\documentclass[journal,onecolumn]{IEEEtran}
\pdfoutput=1
\usepackage{hyperref}
\usepackage{subcaption}
\usepackage{cite}
\usepackage{amsmath,amsfonts,amssymb,amsthm}
\usepackage{algorithmic}
\usepackage{algorithm}
\usepackage{array}
\usepackage{textcomp}
\usepackage{stfloats}
\usepackage{url}
\usepackage{cmll}
\usepackage{verbatim}
\usepackage{balance}
\usepackage{graphicx}
\graphicspath{{./mpx-simulations/}}
\usepackage{float}
\usepackage{hhline}

\usepackage{bytefield}
\usepackage{tikz}

\usepackage{memorygraphs}

\usetikzlibrary{shapes.multipart,graphs}
\tikzset{
    bnode/.style = {   
        text width=1.0em, text height=1.0em, align=center, 
        draw,
        anchor=north,
        rectangle split,
        rectangle split part align=base,
        rectangle split horizontal
    }
}

\DeclareMathOperator*{\argmax}{arg\,max}
\DeclareMathOperator*{\argmin}{arg\,min}

\theoremstyle{definition}
\newtheorem{definition}{Definition}

\newtheorem{proposition}{Proposition}

\newcommand{\alphabet}{\mathcal{A}}
\newcommand{\lang}{\mathcal{L}}
\newcommand*{\prob}{\textbf{p}}

\begin{document}

\title{Parselets: An Abstraction for Fast, General-Purpose Algorithmic Information Calculus}

\author{Fran\c cois Cayre\thanks{Grenoble-INP / Université Grenoble-Alpes / GAIA (hosted by GIPSA-Lab).}}

\maketitle

\begin{abstract}
  This work describes the principled design of a theoretical framework leading to fast and accurate algorithmic information measures on finite multisets of finite strings by means of compression. One distinctive feature of our approach is to manipulate {\em reified}, explicit representations of the very entities and quantities of the theory itself: compressed strings, models, rate-distortion states, minimal sufficient models, joint and relative complexity. 
  To do so, a programmable, recursive data structure called a {\em parselet} essentially provides modeling of a string as a concatenation of parameterized instantiations from sets of finite strings that encode the regular part of the data. This supports another distinctive feature of this work, which is the native embodiment of Epicurus' Principle on top of Occam's Razor, so as to produce both a most-significant and most-general explicit model for the data. This model is iteratively evolved through the Principle of Minimal Change to reach the so-called minimal sufficient model of the data. Parselets may also be used to compute a compression score to any arbitrary hypothesis about the data.
  A lossless, rate-distortion oriented, compressed representation is proposed, that allows immediate reusability of the costly computations stored on disk for their fast merging as our core routine for information calculus. Two information measures are deduced: one is exact because it is purely combinatorial, and the other may occasionally incur slight numerical inaccuracies because it is an approximation of the Kolmogorov complexity of the minimal sufficient model. Symmetry of information is enforced at the bit level. 
  Whenever possible, parselets are compared with off-the-shelf compressors on real data. Some other applications just get enabled by parselets.
\end{abstract}

\begin{IEEEkeywords}
Effective algorithmic rate-distortion theory, information abstract data type. 
\end{IEEEkeywords}

\section{Introduction and overview}

\IEEEPARstart{T}{his} work is a seemingly unique {\em effective} take on Algorithmic Information Theory {\em itself}, targeting the implementation of a few underlying foundational principles and desirable theoretical properties when measuring information quantities contained in individual objects. Our approach theorefore is essentially constructive, and shown to extend naturally to a wide variety of objects by demonstrating support for bytes (default), texts, signals, and images.

Our framework has roots in Kolmogorov complexity~\cite{li:vitaniy:2019}, which is the minimal size of a Turing machine program that will output a given input string. Kolmogorov defined this uncomputable, absolute quantity of information as the limit obtained from the set of the programs that could compress a given string. Yet, our approach is opportunistic: because of the uncomputability of Kolmogorov complexity (we are only about to add one single element in the set of compression programs), we might as well focus on those programs that enable explicit support for faithful evaluations of any quantities of information on finite strings on an actual computer. And we have to lay down a theoretical framework to enable that match between algorithmic information theory and practice. 

By reasoning on strings and on the sets of all their compression programs, algorithmic information theory provides useful and deep relationships between quantities such as the length of a shortest program to ``compress one string given another'', or to ``compress a multiset of strings.'' Of course, the theory has to be agnostic with respect to the actual encodings or the effective means by which such compression operations are conducted (and using off-the-shelf compressors is customary~\cite{cilibrasi:2005}). So if we are to use compression for data analysis, we might as well devise an encoding that natively {\em (i)} represents the objects of the theory, and {\em (ii)} implements the informal specification of compression operations such as the above. Our tool will obviously be encompassed by algorithmic information theory, while its internals seek the most {\em literal} implementation of some important ideas behind it. 

Implementations are often better served by the simplest proofs and arguments, and this work is no exception. This implies that our greatest effort has to be carried out {\em from the representation level}: our end result is stated as a dozen algorithms that collectively provide fast algorithmic information measures with sound numerical behaviour by design. They implement a recursive data structure called a {\em parselet}, that all algorithms use as their explicit representation of the data {\em and of its structure}.

In essence, this work is an inquiry into {\em ``sufficiently elastic, but at the same time convenient and natural, methods of programming''}~\cite[§3]{kolmogorov:1970}, which further enable native support for modern, algorithmic rate-distortion theory~\cite{vereshchagin:2010}. ``Supporting the theory,'' eventually means parselets are our proposal as an abstract data type for information (by ``duck typing,'' in computer science parlance), and we shall provide routines for the evaluation of any information quantities, on any reified entities that are manipulated {\em by the theory itself}---since it is arguably what one may reasonably hope for in practice.

Through literal application of algorithmic information theory, we target the connection to upstream algorithms that consume evaluations of information quantities. On the technical side, this implies some bold decisions: to start a reasonable application programming interface for algorithmic information evaluations, and to impose two-stage working sessions (the data should first be learned by compression, and cached, so it can be used efficiently for all subsequent evaluations). 

Suppose we are given three collections $X$, $Y$ and $Z$ of individual objects: they are not considered elements of a larger set described by a probability distribution---instead we assume the algorithmic (or deterministic) setting. Also suppose we are given a discrete computer to evaluate $I(X:Y\mid Z)$, the quantity of conditional mutual information that is shared between $X$ and $Y$ when $Z$ is known. We say we may delegate the evaluations of these quantities to the computer iff:
\begin{eqnarray}
I(X:Y\mid Z) = I(Y:X\mid Z)\geq 0, \label{eq:info:sym:pos}
\end{eqnarray}
and we understand Eq.~\ref{eq:info:sym:pos} (symmetry and positivity of $I$) as faithful manipulation of Shannon-type, algorithmic information by a discrete computer. 

We identify the collections above with finite multisets of finite strings over finite alphabet $\alphabet=\{\alpha_i\}_i$. Let $\alphabet^\star$ the set of finite strings on $\alphabet$ (string constants are enclosed in double quotes), and denote $\epsilon\in\alphabet^\star$ the empty string. Also let juxtaposition denote concatenation, so $x\in\alphabet^\star\implies \epsilon x = x\epsilon = x$ (and $\epsilon$ is considered a substring of any string).

Let $K(X)\in\mathbb{N}$ a measure of joint algorithmic information contained in ordered string multiset $X$. We may omit the accolades for singletons. By computing a quantity of joint information for $X$, we essentially mean an effective way of compressing, storing together, and decompressing any of the input strings---and the quantity that is measured is the length of the compressed bitstream. This is what we call a ``Solomonoff archiver'' hereafter: a program that uses compression for information calculus, not the converse. For the arguments of $K$, let the comma denote string multiset union such that $K(X,Y)$ denotes joint information quantity in string multisets $X$ and $Y$: this will be our core routine to compute all subsequent quantities. Let $Z$ be a third string multiset. We shall compute relative information quantity (first-order derivative) like:
\begin{eqnarray}
  K(X \mid Z) \triangleq K(X,Z) - K(Z), \label{eq:info:relative}
\end{eqnarray}
and conditional mutual information measure (second-order derivative) like:
\begin{eqnarray}
  I(X:Y \mid Z) \triangleq K(X\mid Z) - K(X\mid Y,Z). \label{eq:info:cond:mutual}
\end{eqnarray}

A popular choice for the axiomatic~\cite{csiszar:2008} construction of $K$ in the deterministic setting, that arguably offers the most satisfying equivalent of Shannon entropy, relies on submodularity~\cite{iyer:2021}. An information measure compatible with Eq.~\ref{eq:info:sym:pos} can be stated following Def.~\ref{def:info:measure} below~\cite{steudel:2010}.

\begin{definition}[Information measure on string multisets]\label{def:info:measure}

  Let $\Omega$ a multiset of observed strings defined on $\mathcal{A}$ (the ``universe'') and multisets $X,Y\subseteq \Omega$. $K(X)\in\mathbb{N}$ is called an information measure on string multiset $\Omega$ iff it is:
  \begin{enumerate}
  \item Normalized: $K(\emptyset)=0$;
  \item Monotone: $K(X)\leq K(\Omega)$;
  \item Submodular: $K(X)+K(Y)\geq K(X,Y) + K(X\cap Y)$.
  \end{enumerate}
\end{definition}

Positivity follows by normalization and monotonicity, strong additivity is encapsulated in Eq.~\ref{eq:info:relative}, expansibility is immediate and implies $K(X)=K(X\mid \emptyset)$, and subadditivity follows by normalization and submodularity. Symmetry of information in Eq.~\ref{eq:info:sym:pos} is immediate. Taking $S=X\cup Z$ and $T=Y\cup Z$ for submodularity implies positivity in Eq.~\ref{eq:info:sym:pos} holds (by developing Eq.~\ref{eq:info:cond:mutual} using Eq.~\ref{eq:info:relative}). In~\cite[Sec. 2]{steudel:2010}, a data processing inequality lemma is eventually obtained (albeit asymmetric).

To keep our data analysis tool as generic as possible, we postulate that our representation of an object should be obtained by application of Occam's Razor (seek a representation that is as short as possible, which is what an off-the-shelf compressor does) and Epicurus' Principle (seek a representation that is as general as possible, which the probabilistic approach natively embodies).

This work delves into what may be obtained when {\em replacing the usual probabilistic description of sets of strings by a programmable representation of finite sets of strings}---that is, in practice, when data is expressed in some regular language. It also implies that we shall get an explicit, non-probabilistic model of the data that is arguably more easily grasped by the human mind---thereby providing some support for explanation of the results, unlike off-the-shelf compressors. 

\subsection{Parameterizing a regular language to describe the data}

Let $\lang_1,\lang_2\subseteq \alphabet^\star$ two languages, and let the usual regular operations on languages (union, concatenation, finite repetitions---Kleene closure is unused, see~\cite[Def.~1.23]{sipser:2012}):

\begin{eqnarray*}
  \lang_1\cup\lang_2     & = & \{x\mid x\in\lang_1\;\lor\;x\in\lang_2\}, \\
  \lang_1\cdot\lang_2    & = & \{xy\mid x\in\lang_1\;\land\;y\in\lang_2\}, \\
  \lang_1^0              & = & \{\epsilon\}, \\
  \forall k>0, \lang_1^k & = & \lang_1\cdot\lang_1^{k-1}. 
\end{eqnarray*}

We shall use the following regular language on $\alphabet$ described by regular expressions $p_L,p_R$ ({\em regexps}, see~\cite[Def.~1.52]{sipser:2012}), so as to enable support for concatenations, alternatives\footnote{The context will always make the meaning of the vertical bar unambiguous (be it for expressing comprehensive set definitions, regexp alternatives, or conditioning in relative quantities).} ($\mbox{\tt |}$ operator), options ($\mbox{\tt ?} $ operator) and a {\em finite} number of repetitions ($\mbox{\tt *}$ operator):

\begin{eqnarray}
  \lang(\epsilon) & = & \{\epsilon\}, \label{eq:regexp:0}\\
  \forall \alpha_i\in\alphabet, \lang(\alpha_i) & = & \{"\alpha_i"\}, \label{eq:regexp:1}\\
  \lang(p_L p_R) & = & \lang(p_L)\cdot\lang(p_R), \label{eq:regexp:2}\\
  \lang(p_L\mbox{\tt |} p_R) & = & \lang(p_L)\cup\lang(p_R), \label{eq:regexp:3}\\
  \lang(p_L\mbox{\tt ?}) & = & \lang(p_L|\epsilon), \label{eq:regexp:4}\\
  \forall k\geq 0, \lang(p_L\mbox{\tt *}) & = & \lang^k(p_L). \label{eq:regexp:5}
\end{eqnarray}

This kernel regexp language models how we shall represent both the data and the regularities extracted from it.

\begin{definition}[Parselets]\label{def:parselet}
In grammar form, the axiom $p$ of our kernel regexp language for describing a finite string over $\alphabet=\{\alpha_i\}_i$ is defined recursively in itself like:
\begin{eqnarray*}
  p & \rightarrow & \mbox{irregular} \mbox{\tt |} \mbox{regular} \\
  \mbox{irregular} & \rightarrow & \mbox{atomic} \mbox{\tt |} \mbox{set} \\
  \mbox{atomic} & \rightarrow & \mbox{``}\alpha_i\mbox{''} \\
  \mbox{set} & \rightarrow & \mbox{option} \mbox{{\tt |}} \mbox{repetition} \\
  \mbox{option} & \rightarrow & p\mbox{\tt ?} \\
  \mbox{repetition} & \rightarrow & p\mbox{\tt *} \\
  \mbox{regular} & \rightarrow & \mbox{conjuction} \mbox{\tt |} \mbox{disjunction} \\
  \mbox{conjunction} & \rightarrow & \mbox{proposition} \; \mbox{proposition} \\
  \mbox{disjunction} & \rightarrow & \mbox{proposition} \mbox{\tt |} \mbox{proposition} \\
  \mbox{proposition} & \rightarrow & p \mbox{\tt |} \mbox{set} 
\end{eqnarray*}

Any elementary, binary-recursive regexp construct $p$ above is called a {\em parselet}, that encodes the representation of a finite set of strings $S_p\subseteq\alphabet^\star$. Such a parselet is identified with $p\mbox{.id}\in\mathbb{N}$. 

A parselet $p$ is equipped with associated core predicates that may be one of $p$.atomic (true whenever $p\mbox{.id}<|\alphabet|$), $p$.option, $p$.repetition, $p$.regular and $p$.conjunction. A parselet also features an integer number of activations $p$.count.

Being essentially a binary data structure, a parselet $p$ is also equipped with associated child parselet accessors $p$.left, and $p$.right whenever $p$.regular holds.

We write $p_0\sim p_1$ whenever parselets $p_0$ and $p_1$ share the same syntactic structure. Otherwise, we write $p_0\nsim p_1$. See App.~\ref{app:serialize:model} for an implementation of the $p_0\sim p_1$ predicate. 
\end{definition}

Other predicates are deduced as $p\mbox{.set}=p\mbox{.option} \lor p.\mbox{repetition}$, $p\mbox{.irregular}=p\mbox{.atomic} \lor p\mbox{.set}$, and $p\mbox{.disjunction}=p.\mbox{regular}\land\lnot p\mbox{.conjunction}$. 

A parselet $p$ may be {\em parameterized} to instantiate one particular string in $S_p$. We will need no parselet parameter for Eqs.~\ref{eq:regexp:0}-\ref{eq:regexp:2}, one bit (as an integer) will be needed to instantiate Eqs.~\ref{eq:regexp:3}-\ref{eq:regexp:4} (to select either alternative or the presence of an option), and one unsigned integer is needed to instantiate the number of repetitions in Eq.~\ref{eq:regexp:5}.

Denote with $v$ the ordered multiset (list) of integer parameters needed to instantiate one particular string from the set $S_p$ described by parselet $p$. Such a particular string in $S_p$ is denoted with $p:v$ (the empty list for atomic parselets could be safely omitted). We represent string $x$ as a concatenation of parameterized ir/regular parselets such that:
\begin{eqnarray*}
x\rightarrow p_{s_1}:v_{s_1}\dots p_{s_l}:v_{s_l},
\end{eqnarray*}
and the runs of {\em string slots} describing the $p_{s_i}:v_{s_i}$ are collectively called {\it string data} hereafter. String slots may be navigated using the {\sc next\_ref} ({\it resp.} {\sc next\_int}) primitive to access the next reference to parselet $p_{s_{i+1}}$ ({\it resp.} the next integer parameter in the list $v_{s_i}$ of the running $p_{s_i}:v_{s_i}$) (see the list of primitives in Tab.~\ref{tab:core:primitives}).

One feature of our design is to provide predicates for ir/regularity of any parselet in string data. For instance, a string acquired in uncompressed form as a sequence of atomic parselets like $x\rightarrow p_1:\emptyset\dots p_{|x|}:\emptyset$ only contains parselets for which the irregular predicate holds. Data in such uncompressed form is written $\underline{x}$.

Let $\mathcal{D}_x=\{p_{i}\}_{i>|\mathcal{A}|}$ a set of canonically ordered, regular parselets for describing substrings of $x$. Denote with $p_1\preceq p_2$ the binary relation on parselets that reads ``$p_1$ may be parameterized to produce a substring of a string in $S_{p_2}$''. Because parselets have a recursive structure, $(\mathcal{D}_x,\preceq)$ forms a lattice with $\bot=\epsilon$ and $\top=\underline{x}$. In effect, the explicit meet/join information of $(\mathcal{D}_x,\preceq)$ shall be encoded as most-significant and most-general parselets added to $\mathcal{D}_x$, which is the {\em dictionary} of $x$ or its {\em model}. Hence, our approach belongs to the flavour of Kolmogorov complexity with two-part codes, which is at the foundation of the Minimum Description Length Principle (MDL~\cite{rissanen:1978}-\cite[Sec.~5.5]{li:vitaniy:2019}).

Alg.~\ref{alg:expand} specifies decoding of string data of $x$, that uses the associated dictionary $\mathcal{D}_x$ in practice to access $p$.left or $p$.right (but we omit it). It also features our evaluation of Bennett's Logical Depth~\cite{bennett:1988}, which is the number of elementary steps to decode the data. We model iteration with terminal recursion. In practice, Alg.~\ref{alg:expand} is very fast and makes our first efficient building block for subsequent algorithmic information calculus algorithms.

\begin{algorithm}[!ht]
\caption{Parselet expansion. Compute $\underline{x}=x\mid\mathcal{D}$.}\label{alg:expand}
\begin{algorithmic}[1]
\STATE {\textsc{dec\_parselet}( $x$, $p$ )}
\STATE \hspace{0.25cm}{$p$.count++ }
\STATE \hspace{0.25cm}{\textbf{if} ( $p$.disjunction )}
\STATE \hspace{0.5cm}{$L_D$++}
\STATE \hspace{0.5cm}{$p\leftarrow$ \textsc{next\_bit( $x$ )} ? $p$.right : $p$.left }
\STATE \hspace{0.5cm}{\textbf{return} \textsc{dec\_parselet}( $x$, $p$ ) }
\STATE 
\STATE \hspace{0.25cm}{occ $\leftarrow$ 1 }
\STATE \hspace{0.25cm}{\textbf{if} ( $p$.left.option ) occ$\leftarrow$ ? \textsc{next\_bit}( $x$ ) }
\STATE \hspace{0.25cm}{\textbf{if} ( $p$.left.repetition ) occ$\leftarrow$ ? \textsc{next\_int}( $x$ ) }
\STATE \hspace{0.25cm}{\textbf{while} ( occ-{}- ) }
\STATE \hspace{0.5cm}{$L_D$++}
\STATE \hspace{0.5cm}{\textbf{if} ( $p$.left.atomic ) \textsc{insert\_after}( $\underline{x}$, $p$.left.id )}
\STATE \hspace{0.5cm}{\textbf{else} \textsc{dec\_parselet}( $x$, $p$.left )}
\STATE \hspace{0.25cm}{\textbf{if} ( $\lnot$ $p$.regular ) \textbf{return}}
\STATE 
\STATE \hspace{0.25cm}{occ $\leftarrow$ 1 }
\STATE \hspace{0.25cm}{\textbf{if} ( $p$.right.option ) occ$\leftarrow$ \textsc{next\_bit}( $x$ ) }
\STATE \hspace{0.25cm}{\textbf{if} ( $p$.right.repetition ) occ$\leftarrow$ \textsc{next\_int}( $x$ ) }
\STATE \hspace{0.25cm}{\textbf{while} ( occ-{}- ) }
\STATE \hspace{0.5cm}{$L_D$++}
\STATE \hspace{0.5cm}{\textbf{if} ( $p$.right.atomic ) \textsc{insert\_after}( $\underline{x}$, $p$.right.id )}
\STATE \hspace{0.5cm}{\textbf{else} \textsc{dec\_parselet}( $x$, $p$.right )}
\STATE 
\STATE {$L_D\leftarrow 0$}
\STATE {$\underline{x}\leftarrow \epsilon$}
\STATE 
\STATE {\textsc{expand}( $x$ )}
\STATE \hspace{0.25cm}{$s$ $\leftarrow$ \textsc{next\_ref}( $x$ ) }
\STATE \hspace{0.25cm}{\textbf{if} ( $s$.EndOfString ) \textbf{return} $\underline{x}$}
\STATE 
\STATE \hspace{0.25cm}{occ $\leftarrow$ 1 }
\STATE \hspace{0.25cm}{\textbf{if} ( $s$.option ) occ$\leftarrow$ \textsc{next\_bit}( $x$ ) }
\STATE \hspace{0.25cm}{\textbf{if} ( $s$.repetition ) occ$\leftarrow$ \textsc{next\_int}( $x$ ) }
\STATE \hspace{0.25cm}{\textbf{while} ( occ-{}- ) }
\STATE \hspace{0.5cm}{$L_D$++}
\STATE \hspace{0.5cm}{\textbf{if} ( $s$.atomic ) \textsc{insert\_after}( $\underline{x}$, $s$.id )}
\STATE \hspace{0.5cm}{\textbf{else} \textsc{dec\_parselet}( $x$, $s$ )}
\STATE 
\STATE \hspace{0.25cm}{\textsc{expand}( $x$ ) }
\end{algorithmic}
\end{algorithm}

\subsection{Compressing data after its minimal sufficient model}

The parselet representation of a string $x$ is well suited to extend to the rate-distortion setting, where $x$ in general should be considered some noisy version of an underlying noise-free string $\hat{x}$. Because we assume a fully algorithmic setting, let us denote $C(\cdot)$ the number of bits needed to write its compressed argument (our computable approximation of the Kolmogorov complexity of the argument). In its most general form, a string $x$ shall be instantiated losslessly from $\hat{x}$ using an amount of bits given by:

\begin{eqnarray}
  C( \mathcal{D}_{\hat{x}},\hat{x}, x ) \triangleq C(\mathcal{D}_{\hat{x}}) + C(\hat{x}\mid\mathcal{D}_{\hat{x}}) + C(\underline{x}\mid\hat{x}). \label{eq:three:part}
\end{eqnarray}

Eq.~\ref{eq:three:part} is the basis of our lossless representation of $x$ on disk as a ``three-part code'' (all serialization issues are put to App.~\ref{app:serialize}): first we write $\mathcal{D}_{\hat{x}}$, and then we write $\hat{x}\mid\mathcal{D}_{\hat{x}}$ (so $\hat{x}$ is decompressed with Alg.~\ref{alg:expand}, that needs $\mathcal{D}_{\hat{x}}$), together with $\underline{x}\mid\hat{x}$ {\em patching information} to eventually recover $x$ losslessly. 

Finding $(\mathcal{D}_{\hat{x}},\hat{x})$ from uncompressed $\underline{x}$ shall rely upon successive applications of the Principle of Minimal Change, and we may eventually specify $x$ in compressed form using its rate-distortion state triplet $(\mathcal{D}_{\hat{x}}, \hat{x}, x)$, which should be read as ``uncompressed string $\underline{x}$ is reconstructed by applying patching data to $\hat{x}$, which is decoded using Alg.~\ref{alg:expand} after $\mathcal{D}_{\hat{x}}$ has been decoded.'' Following the algorithmic rate-distortion setting, $\mathcal{D}_{\hat{x}}$ is called the {\em minimal sufficient model} of $x$, which is deemed to contain all information of statistical interest about $x$~\cite{vereshchagin:2010}. We call $\hat{x}$ our algorithmically, optimally-denoised version of $x$. It shall turn out that the socalled {\em codelength} function $C^\star(\mathcal{D}_{\hat{x}}, \hat{x},x)$ has to be evaluated and minimized in some idealized domain for $C^\star(\underline{x}\mid\hat{x})$, so we let:
\begin{eqnarray}
  C^\star( \mathcal{D}_{\hat{x}},\hat{x}, x ) \triangleq C(\mathcal{D}_{\hat{x}}) + C(\hat{x}\mid\mathcal{D}_{\hat{x}}) + C^\star(\underline{x}\mid\hat{x}). \label{eq:codelength}
\end{eqnarray}

It is an unusual task to embody Epicurus' Principle into a compressor, as keeping track of specific strings in sets of size greater than one will only cost additional bits to be written to disk  and the minimum size of the bitstream is almost always reached by disabling the embodiment of Epicurus' Principle. Therefore, we have no choice but to seek the most complex $\mathcal{D}_{\hat{x}}$ within an idealized generalization ball of admissible radius $\delta\geq 0$ around $(\mathcal{D}^\star, x^\star)$, the sole application of Occam's Razor:
\begin{eqnarray}
\mathcal{D}_{\hat{x}} \triangleq \argmax_{ \mathcal{D}\mbox{ s.t. }|C^\star( \mathcal{D},\hat{x},x )-\min C^\star( \mathcal{D}^\star, x^\star, x )| \leq \delta} C(\mathcal{D}). \label{eq:model}
\end{eqnarray}
Setting $\delta=0$ shall disable the embodiment of Epicurus' Principle, to only keep Occam's Razor as the default epistemological workhorse to minimize the codelength function. Reaching a good candidate in reasonable time for Eq.~\ref{eq:model} involves a heuristic that iterates towards smaller $C(\mathcal{D}_{\hat{x}})+C(\hat{x}\mid\mathcal{D}_{\hat{x}})$, by alternating single-string compression and application of the Principle of Minimal Change to generate the next candidate, until $C^\star(\mathcal{D}_{\hat{x}},\hat{x},x)$ is minimized. Eq.~\ref{eq:model} also has the advantage to discourage seeking empty models.

Our implementation of Eq.~\ref{eq:model} spans from the description of the core compressor in Sec.~\ref{sec:single}, to using it for finding a reasonable solution to the search of the minimal sufficient model in Sec.~\ref{sec:min:suff}.

\subsection{Merging compressed representations}

The parselet representation is also shown to support efficient computation of a compressed bitstream from a multiset of compressed strings $X=\{(\mathcal{D}_{\hat{x}}, \hat{x}, x)\}$, {\em by combining their compressed representation} so they share a common dictionary $\mathcal{D}_{\hat{X}}$. Because the $\mathcal{D}_{\hat{x}}$ are sets, we pay special attention to fast procedures based on:
\begin{eqnarray}
  \mathcal{D}_{\hat{X}} \triangleq \bigcup_{(\mathcal{D}_{\hat{x}},\hat{x},x)\in X} \mathcal{D}_{\hat{x}}, \label{eq:mset:union}
\end{eqnarray}
which gets naturally extended to string multisets with $\mathcal{D}_{\hat{X},\hat{Y}} \triangleq \mathcal{D}_{\hat{X}}\cup\mathcal{D}_{\hat{Y}}$.

Eq.~\ref{eq:mset:union} raises one important point as to how similar information between two string multisets is to be assessed. When one uses a compressor, the concatenation of the strings is preceded by integers that are the size of the next string, except for the last. This formally turns any single-string compressor into an archiver, and compression may take advantage from fainter (or even spurious) similarities that may appear just because of concatenation (which is called ``solid'' compression). 

When combining compressed representations, similarity of information is assessed with any string considered independently of the others (and our approach may need longer data than if using an off-the-shelf compressor). This is one important departure from the usual approach in the literature, and it is our understanding of the difference between a compressor for storage or communications, and a ``Solomonoff archiver'', the core of which is described in Sec.~\ref{sec:joint}.

\subsection{Compression for data analysis}

The main mathematical structure that is constructed from data compression is the lattice of its most-general and most-significant parselets, which is our best hypothesis about the data. Another feature of our design, that is described in Sec.~\ref{sec:arbitrary}, is the ability to compute a compression score for any arbitrary hypothesis we might formulate about the data. 

From Eqs.~\ref{eq:model}-\ref{eq:mset:union}, we get our definitions of compressibility (Def.~\ref{def:compressibility}), and of total ordering of string multisets (Def.~\ref{def:mset:total:order}).

\begin{definition}[In/compressible string multisets]\label{def:compressibility}
  Let $X$ a finite multiset of finite strings over $\mathcal{A}$. $X$ is called incompressible iff $\mathcal{D}_{\hat{X}}=\emptyset$. Otherwise, it is called compressible.
\end{definition}

\begin{definition}[Shannon information lattice on string multisets]\label{def:mset:total:order}
  Let $X,Y$ finite multisets of finite strings over $\mathcal{A}$. We declare:
  \begin{itemize}
  \item $X<Y$ whenever $\mathcal{D}_{\hat{X}}\subset\mathcal{D}_{\hat{Y}}$;
  \item $X=Y$ whenever $\mathcal{D}_{\hat{X}}=\mathcal{D}_{\hat{Y}}$;
  \item $X>Y$ otherwise.
  \end{itemize}

  Following~\cite{shannon:1953}, $(X,\leq)$ defines an information lattice on string multisets with $\bot=\{\emptyset\}$ and $\top=\alphabet^{\star^\star}$, and $Y\geq X$ is called an {\em abstraction} of $X$. 
\end{definition}

Finally, we propose two information measure candidates, namely:
\begin{eqnarray}
K^\star(X) & \triangleq & |\mathcal{D}_{\hat{X}}|, \label{eq:measure:cardinality} \\
K_{\mathcal{D}}(X) & \triangleq & C(\mathcal{D}_{\hat{X}}), \label{eq:measure:complexity}
\end{eqnarray}
which is cardinality {\em vs.} approximation of Kolmogorov complexity of the union of the minimal sufficient models.

Theses measure candidates give rise to our two algorithmic information quantity candidates $I^\star$ and $I_{\mathcal{D}}$. $K^\star$ is clearly an information measure according to Def.~\ref{def:info:measure} so Eq.~\ref{eq:info:sym:pos} holds for $I^\star$, and $I_{\mathcal{D}}$ trades occasional, slight inaccuracies in the positivity of Eq.~\ref{eq:info:sym:pos} for allegedly better modeling capabilities (the accuracies are expected to be as rare and small as possible since our last layer of entropy coding uses maximally compact Huffman codes at all times~\cite{vitter:1987}). Hence, we indulge taking the absolute value of Eqs.~\ref{eq:info:relative}-\ref{eq:info:cond:mutual} so upstream algorithms do not have to handle these corner cases for $K_{\mathcal{\hat{D}}}$. We define syntactic conditional independence from $I^\star$ in Def.~\ref{def:syntactic:indep}.

\begin{definition}[Syntactic conditional independence of string multisets]\label{def:syntactic:indep}
Let $X,Y,Z$ finite multisets of finite strings over $\mathcal{A}$. Syntactic conditional independence is denoted with:
\begin{eqnarray}
X\Perp Y \mid Z \iff I^\star(X:Y\mid Z)=0.
\end{eqnarray}
\end{definition}

\subsection{Contributions}

Symmetry in Eq.~\ref{eq:info:sym:pos} holds when using $K^\star$ because of Eq.~\ref{eq:mset:union}, and it holds when using $K_{\mathcal{D}}$ also because $\mathcal{D}_{\hat{X}}$ is written in canonical order. In fact, we are able to generate bitstreams for lossless storage of $X$ that use the exact same number of bits independently of the order of enumeration of the strings in $X$.

The quantities in Eqs.~\ref{eq:info:relative}-\ref{eq:info:cond:mutual} are eventually used to compute a host of variants for algorithmic information-theoretic distances and algorithmic equivalents of conditional independence statistics (Sec.~\ref{sec:appli}). The relative merits of $K^\star$ and $K_{\mathcal{D}}$ are demonstrated on real data. Further, it is also shown how parselets accomodate reasonable forms of text, signal, and image objects---thereby working on multimodal data using the very same internal representation at all times for all said kinds of objects (Sec.~\ref{sec:single:core} and App.~\ref{app:multimodal}). 

Eventually, we advocate parselets as one possible abstract data type for the algorithmic information contained in string multisets:
\begin{itemize}
\item Parselets implement a rate-distortion based, {\em three-part code} for lossless storage based on an explicit model;
\item Parselets live on a lattice, which enables reliable searches and fast procedures for Eq.~\ref{eq:info:sym:pos};
\item Parselets support optimal, algorithmic denoising based on efficient rate-distortion space exploration;
\item Parselets lead to information measures and quantities with great numerical accuracy, symmetry of information is ensured by design;
\item Parselets enable fully-algorithmic hypothesis testing;
\item Parselets embody exploration of the bounded application of Epicurus' Principle around Occam's Razor.
\end{itemize}

\subsection{Datasets, computer, code instrumentation}

We borrow the following datasets from~\cite{cilibrasi:2005}.\footnote{These two datasets were downloaded from the \texttt{CompLearn.org} page, which is the (now unmaintained) implementation of~\cite{cilibrasi:2005}: the associated code was easily backported for the simulations in this paper.} Their most important characteristics for the present work are reported in Tab.~\ref{tab:datasets}. The dataset \texttt{mammals} contains 34 mitochondrial DNA samples of various species, it is considered an example of highly redundant data on a very small alphabet. The dataset \texttt{languages} contains the Declaration of Human Rights in 49 languages, it is considered an example of less redundant data on a mid-sized alphabet.

In Tab.~\ref{tab:datasets}, we also report the sum of the sizes of the files compressed with four off-the-shelf compressors:
\begin{itemize}
\item \texttt{gzip}~\cite{deutsch:1996} (DEFLATE) is LZ77~\cite{ziv:lempel:1977} with 32KiB sliding window and canonical Huffman entropy-coding;
\item \texttt{xz}~\cite{pavlov:2013} (LZMA2) is LZ77 with 1MiB sliding window, move-to-front context modeling, and range coding;
\item \texttt{bzip2}~\cite{seward:1996} uses the Burrows-Wheeler transform~\cite{burrows:1994} to implement block-sorting (block size of 256KiB), of which the run-length coding of the move-to-front transform is followed by Huffman entropy-coding;
\item \texttt{ppmd}~\cite{shkarin:2002} uses prediction by partial matching, adaptive Markov-chain context modeling and range coding.
\end{itemize}

\begin{table}[!ht]
  \caption{Datasets. DNA bases and glyphs are \textsc{ascii} encoded (one byte for each). The mean file lengths are 16963 bytes (\texttt{mammals}, 34 files) and 11680 bytes (\texttt{languages}, 49 files). Reported lengths are the sums of those of the files compressed separately with four off-the-shelf compressors (in KiB).
    \label{tab:datasets}}
  \centering
  \begin{tabular}{|l||c|c|}\hline
    Dataset                   & \texttt{mammals} (DNA) & \texttt{languages} (Text) \\\hline
    Uncompressed files        &           553.5        &             558.9 \\\hline\hline
    \texttt{gzip} (DEFLATE)   &           166.2        &             205.3 \\\hline
    \texttt{xz} (LZMA2)       &           158.7        &             201.6 \\\hline
    \texttt{bzip2}            &           151.8        &             183.9 \\\hline
    \texttt{ppmd}             &           140.9        &             165.7 \\\hline
  \end{tabular}
\end{table}

We assume a PRAM-CREW machine model. In practice, we target multi-core, 64-bit commodity machines. Our physical instrument for evaluating information quantities is an 8\textsuperscript{th} Gen. Intel Core i7-8565U commodity laptop from 2018. Its main specifications relevant to this work include 4 physical cores (8 logical), 8MiB L3 cache and solid-state storage. We understand logical cores as exploiting free CPU instruction execution slots on-the-fly in physical cores. The performance of this opportunistic execution strategy is quite dependent on the instruction flows executing concurrently in the system. Our computations are spawned asynchronously in a thread pool, and they are all implemented with the simple, obvious batch-scheduling strategy at hand. 

Part of our inquiry is about evaluating logical depth, so we instrument our code to provide the following values:
\begin{enumerate}
\item Execution time, as per the \texttt{CLOCK\_MONOTONIC} system clock, see \texttt{clock\_gettime(2)};
\item Number of CPU instructions (Linux-specific \texttt{PERF\_COUNT\_HW\_INSTRUCTIONS} performance counter);
\item Number of decompression steps.
\end{enumerate}

Monitoring the execution time is the usual and arguably only approach reported in the literature, see {\em e.g.}~\cite{zenil:2012}. We identify counting the number of CPU instructions with the idealized version of the usual approach, because it is immune to CPU frequency-scaling, to using logical cores or not, speculative execution, and to time-sharing in the operating system. For this reason, we shall not bother averaging execution times.

We also identify counting decompression steps, which is our evaluation of logical depth, as the idealized version of counting the number of CPU instructions, because the evaluation now becomes a deterministic function of the sole compressed representation of the data. Hence, we only report the other two for the sake of comparison.

The reference implementation for $|\mathcal{A}|=2^{8}$ is called \texttt{mpx}. It is written in 20k lines of C99 for GNU/Linux. Because counting CPU instructions is part of our inquiry, the compiler becomes part of the experimental setup, and we report figures for \texttt{-O3 -march=native} compilation options as of \texttt{clang-14.0.0}. Usual data structures (hashtables, dynamic arrays) are implemented with macro-templated \texttt{inline} functions (so we avoid constantly dereferencing generic pointers). This is certainly relevant to mention, because aggressively inlined and optimized code is expected to produce flattened plots of CPU instructions count (at least, many function call prologues and epilogues may be saved, also when the compiler optimizes terminal recursion). 

\section{Single-string compression} \label{sec:single}

In the sequel, we assume the primitives and predicates listed in Tab.~\ref{tab:core:primitives} are available. They rely on fully-dynamic, compact data structures (for the string slots and also for an index) that are described in App.~\ref{app:data:structures}. 

\begin{table}[!ht]
  \caption{Core predicates and primitives.
    \label{tab:core:primitives}}
  \centering
  \begin{tabular}{|l|}\hline
    System interface \\ \hline\hline
    $n\leftarrow$\textsc{sys}.written \\
    \hspace{.25cm}Returns the number of bits written so far. The evaluations of $C(\mathcal{D}_{\hat{x}})$ \\
    and $C(\hat{x}\mid\mathcal{D}_{\hat{x}})$ in Eqs.~\ref{eq:three:part}-\ref{eq:codelength} are done by monitoring actual bit count \\
    when writing to {\tt /dev/null} sink. \\ \hline\hline
    Generic primitives \\ \hline\hline
    \textsc{$n\leftarrow$array\_push}( $a$, generic ) \\
    \hspace{.25cm}Push generic object at the end of dynamic array $a$, \\
    returns $n=|a|-1$. \\ \hline\hline
    String slot predicates \\ \hline\hline
    $s$.EndOfString \\
    \hspace{.25cm}True if slot string $s$ encodes the end of string. \\ \hline\hline
    $s$.value \\
    \hspace{.25cm}True if slot string $s$ contains an unsigned integer value. \\ \hline\hline

    String slot primitives \\ \hline\hline
    $s$ ::= $p$ \\
    \hspace{.25cm}Change slot string $s$ contents to parselet $p$, the special value \\
    {\tt Empty} denotes string slot erasure. \\ \hline
    $s$.len \\
    \hspace{.25cm}The number of parselet occurrences in string data to which $s$ belongs. \\ \hline
    $p\leftarrow$\textsc{next\_slot}( $s$ ) \\
    \hspace{.25cm}Returns string slot $s$, advance to next slot. \\ \hline
    $p\leftarrow$\textsc{next\_ref}( $s$ ) \\
    \hspace{.25cm}Consume string slot $s$ as an expected proposition (Def.~\ref{def:parselet}) parselet $p$. \\ \hline
    val$\leftarrow$\textsc{next\_int}( $s$ ) \\
    Consume string slot $s$ as an expected unsigned integer. \\ \hline
    bit$\leftarrow$\textsc{next\_bit}( $s$ ) \\
    \hspace{.25cm}Consume string slot $s$ as an expected bit. \\ \hline
    \textsc{insert\_after}($s$,val) \\
    \textsc{insert\_before}($s$,val) \\
    \hspace{.25cm}Insert new string slot containing unsigned integer val \\
    after/before string slot $s$. \\ \hline
    \textsc{rlc}($\{s_p\}$) \\
    \hspace{.25cm}$\{s\}_p$ is a set of string slot that contain parselet $p$. \\
    First, filter out slots that share the same identifier with the \\
    next left parselet string slot. Then apply run-length coding to the \\
    remaining slots: if the number of successive slots containing \\
    parselet $p$ is greater than one, add a {\tt *} operator, insert number of \\
    repetitions and erase remaining occurrences at the right. This is the \\
    only way to introduce {\tt *} operators. \\ \hline\hline
    Index primitives \\ \hline\hline
    $\{s\}_p,p,\mbox{cost},n\leftarrow$\textsc{most\_promising\_option}($x$,$\mathcal{S}$) \\
    $\{s\}_p,p,\mbox{cost}\leftarrow$\textsc{most\_promising\_disjunction}($x$,$\mathcal{S}$) \\
    $\{s\}_p,p\leftarrow$\textsc{most\_frequent\_conjunction}($x$,$\mathcal{S}$) \\
    \hspace{.25cm}Query index for most promising option, disjunction or the most \\
    frequent conjunction in string data of $x$, forbidding parselet leaves in \\
    the tokens separator set $\mathcal{S}\subseteq\alphabet$. \\
    Returns the set $\{s\}_p$ of string slots to be compiled into regular \\
    parselet $p$, and the generalization cost, together with the number $n$ of \\
    occurrences  in string data that creating option parselet $p$ will remove.  \\ \hline\hline
    Serialization primitives (that update \textsc{sys}.written) \\ \hline\hline
    \textsc{write\_uint}($n$), \textsc{write\_ref}($p$), \textsc{read\_uint}($n$), \textsc{read\_ref}($p$) \\
    \hspace{.25cm}Use dedicated dynamic Huffman coders~\cite{vitter:1987} for unsigned integers and \\
    references to the dictionary. These coders are reset upon request. \\ \hline
    \textsc{write\_const}($c$), \textsc{read\_const}($c$) \\
    \hspace{.25cm}Use $\omega$-codes~\cite{elias:1975} for unbounded unsigned integer values. \\ \hline
    \textsc{write\_bit}($b$), \textsc{read\_bit}($b$) \\
    \hspace{.25cm}Do what the names suggest. \\ \hline
  \end{tabular}
\end{table}

The basic mechanism of our compressor is the iterative inference of the most general regexp language in the vicinity of the shortest representation of the data, and on-the-fly compilation of the data in this language (to instantiate the desired string in a set). 

\subsection{Elementary compilations and searches}

The first technical step is to describe how to compile a set $\{s\}_p$ of string slots that will be represented by parselet $p$.

Let $p\rightarrow p_L\; p_R$ a conjunction and $p_L:v_{p_L}\; p_R:v_{p_R} $  consecutive slots at $p_L\in\{s\}_p$. The conjunction $p$ is compiled in-place into slots $p:\{v_{p_L}, v_{p_R}\}$. One string slot (that of $p_R$) is erased from string data.

Let $p\rightarrow p_{A_0}\mbox{\tt |} p_{A_1}$ a disjunction and $p_{A_{0\mid 1}}:v$  consecutive slots at $p_{A{0\mid 1}}\in\{s\}_p$. The alternative bit should be inserted before parameters $v$ to produce either slots $p:\{0, v\}$ (if $p_{A{0\mid 1}}=p_{A_0}$), or $p:\{1, v\}$ otherwise ($p_{A_{0\mid 1}}=p_{A_1}$). 

Let $p\rightarrow p_L\;p_C\mbox{\tt ?}$ an option and $p_L:v_{p_L}\; p_q:v $  consecutive slots at $p_L\in\{s\}_p$. The option bit should be inserted after parameters $v_{p_L}$ to produce either slots $p:\{v_{p_L}, 1, v\}$ (if $p_q=p_C$), or $p:\{v_{p_L},0\}\;p_q:v$ otherwise. 

The primitives in Tab.~\ref{alg:compile} implement these elementary compilations, that reuse as much existing string slots as possible. 

\begin{algorithm}[!ht]
\caption{Parselet compilation primitives inside string data.}\label{alg:compile}
\begin{algorithmic}[1]
\STATE {\textsc{compile\_conjunction}( $s$, $p$ )}
\STATE \hspace{0.25cm}{$s$ \texttt{:=} $p$}
\STATE \hspace{0.25cm}{$s_R$ $\leftarrow$ \textsc{next\_slot}( $s$ ) }
\STATE \hspace{0.25cm}{\textbf{while} ( $s_R$.value ) } $s_R\leftarrow$ \textsc{next\_slot}( $s_R$ )
\STATE \hspace{0.25cm}{$s_R$ \texttt{:=} \texttt{Empty}}
\STATE 
\STATE {\textsc{compile\_disjunction}( $s$, $p$ )}
\STATE \hspace{0.25cm}{\textsc{insert\_after}( $s$, $s$.id == $p$.right.id )}
\STATE \hspace{0.25cm}{$s$ \texttt{:=} $p$}
\STATE 
\STATE {\textsc{compile\_option}( $s$, $p$ )}
\STATE \hspace{0.25cm}{$s$ \texttt{:=} $p$}
\STATE \hspace{0.25cm}{$s_R$ $\leftarrow$ \textsc{next\_slot}( $s$ ) }
\STATE \hspace{0.25cm}{\textbf{while} ( $s_R$.value ) } $s_R\leftarrow$ \textsc{next\_slot}( $s_R$ )
\STATE \hspace{0.25cm}{\textbf{if} ( $s_R$.id == $p$.right.id ) $s_R$ \texttt{:=}} 1
\STATE \hspace{0.25cm}{\textbf{else} \textsc{insert\_before}( $s_R$, 0 )}
\end{algorithmic}
\end{algorithm}

We shall need to quickly lookup candidate parselet $p$ and the associated locations. Given the index structure outlined in App.~\ref{app:data:structures} and parselets $p_L, p_R$, the index gives $O(|\mathcal{D}|)$ worst-case time access to all locations at which conjunction $p_L\;p_R$ is found. Some search provide an associated generalization cost, which is the sum of the {\em replacement cost} (that of replacing a pattern by a new parselet), and of the {\em incentive} (what may be gained from later compilation into a new conjunction). Let $\forall 0< f\leq 1, l(f)=-\log f$ the idealized bit-length estimate associated to a proportion $f$ of occurrences. 

The primitive \textsc{most\_frequent\_conjunction} terminates in worst-case $O(|\mathcal{D}|^2)$ time (but it performs much better in practice as not all pairs of parselets are used). Suppose it returned $n$ locations. Its replacement cost would be computed like $n(l(n/(x\mbox{.len}-n))-l(p_L\mbox{.count}/x\mbox{.len})-l(p_R\mbox{.count}/x\mbox{.len}))$. 

The primitive \textsc{most\_promising\_option} locates in worst-case $O(|\mathcal{D}|^2)$ time the most numerous locations at which either $p_L\;p_R$ or $p_L\;p_C\;p_R$ holds (with $\lnot$ $p_C$.set), such that they are compiled with parselet $p\rightarrow p_L\;p_C\mbox{\tt ?}$ into incentive locations at which $p\;p_R$ holds. In order to ensure compression termination, we avoid creating parselet $p\rightarrow p_1 p_C\mbox{\tt ?}$ such that $p_1\rightarrow p_2 p_C\mbox{\tt ?}$ already exists. There are $n_{C}$ locations of type $p_L\;p_C\;p_R$ and $n_{\bar{C}}$ locations of type $p_L\;p_R$ that are returned in $\{s\}_p$. The replacement cost is computed like $(n_C+n_{\bar{C}})(1+l((n_C+n_{\bar{C}})/(x\mbox{.len}-n_C))-l(p_L\mbox{.count}/x\mbox{.len}))-n_C l( p_C\mbox{.count}/x\mbox{.len} )$, and the incentive is computed like $(n_C+n_{\bar{C}})(l((n_C+n_{bar{C}})/(x\mbox{.len}-n_C+(n_C+n_{\bar{C}})))-l((n_C+n_{\bar{C}})/(x\mbox{.len}-n_C))-l(p_R\mbox{.count}/(x\mbox{.len}-n_C)))$. 

The primitive \textsc{most\_promising\_disjunction} locates the most numerous locations at which either $p_L\;p_{A_0}$ or $p_L\;p_{A_1}$ holds (stopping the $O(|\mathcal{D}|^3)$ search on the first pair of sets with maximum 10\% imbalance in size for speed-up), such that they are compiled with $p\rightarrow p_{A_0}\mbox{\tt |}p_{A_1}$ into incentive locations at which $p_L\;p$ holds. There are $n_{A_0}$ locations of type $p_L\;p_{A_0}$ and $n_{A_1}$ of type $p_L\;p_{A_1}$ that are returned in $\{s\}_p$. The replacement cost is computed like $(n_{A_0}+n_{A_1})(1+l((n_{A_0}+n_{A_1})/x\mbox{.len}))-n_{A_0}l(p_{A_0}\mbox{.count}/x\mbox{.len})-n_{A_1}l(p_{A_1}\mbox{.count}/x\mbox{.len})$, and the incentive is computed like $(n_{A_0}+n_{A_1})(l((n_{A_0}+n_{A_1})/(x\mbox{.len}-(n_{A_0}+n_{A_1})))-l((n_{A_0}+n_{A_1})/x\mbox{.len})-l(p_L\mbox{.count}/x\mbox{.len}))$.

\subsection{Compressing despite Epicurus}\label{sec:single:core}

Compilation of a conjunction empties one slot in string data: this is our embodiment of Occam's Razor, together with \textsc{rlc}. Compilation of options and disjunctions add one bit each time: this is our embodiment of Epicurus' Principle, which should be done in the hope of future inclusion in a new conjunction---and so we keep in the vicinity of Occam's Razor.

Alg.~\ref{alg:deflate} is a worst-case quadratic-time deflation in the length of string $\underline{x}$, that implements Occam's Razor while favouring application of Epicurus' Principle up to a given admissible amount: generalization with options and disjunctions is subject to maximum admissible costs. The iteration is ultimately governed by application of Occam's Razor: it exits when most frequent conjunctions are no longer frequent enough to be significant. Alg.~\ref{alg:deflate} allocates increasing parselet identifiers starting from $|\alphabet|$, so no cycle may be created. 

Also, Alg.~\ref{alg:deflate} offers native support for tokenization: suppose we know a string encodes ``words'' (tokens) delimited by separators $\mathcal{S}\subseteq\alphabet$. We first perform a first-stage compression, forbidding creation of parselets with leaves in $\mathcal{S}$ to concentrate within tokens, and then we relax the constraint to seek relationships between tokens. If $\mathcal{S}=\emptyset$, tokenization is disabled. Alg.~\ref{alg:deflate} is called with a first round of \textsc{rlc} coding on $\underline{x}$ as \textsc{deflate}( \textsc{rlc}($\{\underline{x}\})$, $\emptyset$, $T_{sig}$, $T_{opt}$, $T_{alt}$, $\mathcal{S}$ ). At the end of Alg.~\ref{alg:deflate}, $x$ is expressed in the regexp language of both its most-significant and most-general parselets within the admissible generalization ball.

Also note that setting $T_{sig}=1$ is equivalent to finding an axiom for $x$, as all of it will end up stored in $\mathcal{D}_x$, and a single parselet in string data (the axiom) will remain, followed by all instantiation parameters (see Sec.~\ref{sec:arbitrary}). Alternatively, increasing $T_{sig}$ will discourage creation of parselets, and eventually our compressor degenerates into entropy coding of single letters (possibly with their repetitions) if no parselet may be used from the model (it is empty when $T_{sig}$ is big enough). 

Empirical evidence suggests that $T_{sig}=6$ is a sound default, and that defaults $T_{opt}=T_{alt}=2$ keep $\delta$ in Eq.~\ref{eq:model} within variations between off-the-shelf compressors themselves, see Tab.~\ref{tab:cache}. We can know by disabling application of Epicurus' Principle with $T_{opt}=T_{alt}=0$. By default, simulations consider no application of Epicurus' Principle and no tokenization. When they do, they are annotated with (g) when $T_{opt}=T_{alt}=2$ generalization is allowed, and with (t) when the default tokenizer is used (which filters out non-alphanumerical characters).

\begin{algorithm}[!ht]
\caption{Deflation by parselets. Get $\mathcal{D},x$ from $\underline{x}$ in worst-case quadratic-time, within generalization ball bounded by $T_{opt},T_{alt}$ on tokens separated by $\mathcal{S}$.} \label{alg:deflate}
\begin{algorithmic}[1]
\STATE {\textsc{deflate\_tok}( $x$, $\mathcal{D}$, $T_{sig}$, $T_{opt}$, $T_{alt}$, $\mathcal{S}$ )}
\STATE \hspace{0.25cm}{$\{s\}_p,p,\mbox{cost},n\leftarrow$ \textsc{most\_promising\_option}($x$,$\mathcal{S}$)}
\STATE \hspace{0.25cm}{\textbf{if} ( $\mbox{cost} < T_{opt}$ ) }
\STATE \hspace{0.5cm}$ p\mbox{.count}\leftarrow |\{s\}_p|$
\STATE \hspace{0.5cm}$ x\mbox{.len}\leftarrow x\mbox{.len}-n$
\STATE \hspace{0.5cm}$ \mathcal{D}\leftarrow \mathcal{D}\cup\{p\}$
\STATE \hspace{0.5cm}{\textbf{foreach} ( $s\in \{s\}_p$ ) \textsc{compile\_option}($s$,$p$)}
\STATE \hspace{0.5cm}{\textsc{rlc}( $\{s\}_p$ ) }
\STATE

\STATE \hspace{0.25cm}{$\{s\}_p,p,\mbox{cost}\leftarrow$\textsc{most\_promising\_disjunction}($x$,$\mathcal{S}$)}
\STATE \hspace{0.25cm}{\textbf{if} ( $\mbox{cost} < T_{alt}$ ) }
\STATE \hspace{0.5cm}$ p\mbox{.count}\leftarrow |\{s\}_p|$
\STATE \hspace{0.5cm}$ \mathcal{D}\leftarrow \mathcal{D}\cup\{p\}$
\STATE \hspace{0.5cm}{\textbf{foreach} ( $s\in \{s\}_p$ ) \textsc{compile\_disjunction}($s$,$p$)}
\STATE \hspace{0.5cm}{\textsc{rlc}( $\{s\}_p$ ) }
\STATE

\STATE \hspace{0.25cm}{$\{s\}_p,p\leftarrow$ \textsc{most\_frequent\_conjunction}($x$,$\mathcal{S}$)}
\STATE \hspace{0.25cm}{\textbf{if} ( $|\{s\}_p| < T_{sig}$ ) \textbf{return} $(\mathcal{D},x)$}
\STATE \hspace{0.25cm}$ p\mbox{.count}\leftarrow |\{s\}_p|$
\STATE \hspace{0.25cm}$ x\mbox{.len}\leftarrow x\mbox{.len}-|\{s\}_p|$
\STATE \hspace{0.25cm}$ \mathcal{D}\leftarrow \mathcal{D}\cup\{p\}$
\STATE \hspace{0.25cm}{\textbf{foreach} ( $s\in \{s\}_p$ ) \textsc{compile\_conjunction}($s$,$p$)}
\STATE \hspace{0.25cm}{\textsc{rlc}( $\{s\}_p$ ) }
\STATE
\STATE \hspace{0.25cm}{\textsc{deflate\_tok}( $x$, $\mathcal{D}$, $T_{sig}$, $T_{opt}$, $T_{alt}$, $\mathcal{S}$ )}
\STATE
\STATE
\STATE {\textsc{deflate}( $x$, $\mathcal{D}$, $T_{sig}$, $T_{opt}$, $T_{alt}$, $\mathcal{S}$ )}
\STATE \hspace{0.25cm}{\textsc{deflate\_tok}( $x$, $\mathcal{D}$, $T_{sig}$, $T_{opt}$, $T_{alt}$, $\mathcal{S}$ )}
\STATE \hspace{0.25cm}{\textsc{deflate\_tok}( $x$, $\mathcal{D}$, $T_{sig}$, $T_{opt}$, $T_{alt}$, $\emptyset$ )}
\end{algorithmic}
\end{algorithm}

\subsection{Related works}

Much of the intuition of parselets as a recursive data type for describing the data was borrowed from Parsing Expression Grammar (\textsc{peg}~\cite{ford:2004}), where modular parsers (implementing a superset of the regexp operators we use) are synthesized into any arbitrary, full-fledged recursive descent parser. See also the discussion at end of Sec.~\ref{sec:appli:proba}.

The above describes a compressor that has a few variants in the literature of grammar-based compression, although ours exhibits a more complex language model. This discussion concentrates on the case $T_{opt}=T_{alt}=0$ (only conjunction parselets are used). Contrary to~\cite{kieffer:yang:2000}, we seek the creation of a reusable dictionary that can be stored apart. This allows to keep the irregular part of the data explicit within string data (which is called a {\em ``skeleton''} in~\cite{storer:1982}). Conjunction parselets are reminiscent of the digram structure used in~\cite{nevill:witten:1997} and of the byte pairs used in~\cite{gage:1994} (which is one method for tokenization in natural language processing), both of which are online, linear-time compressors. 

The very same offline deflation as in Alg.~\ref{alg:deflate} likely was first described by Solomonoff~\cite{solomonoff:1961}, and also more recently in the context of compressing partially ordered strings~\cite[Sec.~3.2]{alur:2003} (the authors acknowledge both inspiration from~\cite{nevill:witten:1997} and tedious implementation). It is analyzed in~\cite{savari:2004}, to which the reader is referred for a discussion on universality. Curiously enough, none of the preceding acknowledged~\cite{solomonoff:1961} as a distant cousin. This has to be an algorithm everyone has to reinvent. 

Another pionneering approach by Chaitin~\cite{chaitin:1987} along the exact same line uses Pure\textsc{lisp} S-expressions as a vehicle for the shortest program representation. This flavour of experimental algorithmic information calculus has later been extended to small Turing machines~\cite{soler:2014} on binary strings up to 12 bits. In this regard, we seek a much less powerful language than \textsc{lisp} or those of Turing machines to describe the data, and hope to handle (much) longer strings more efficiently in reward. 

Nowadays, it is customary to use an off-the-shelf compressor~\cite{cilibrasi:2005} to estimate an upper bound on the quantity of information in an individual object and build subsequent quantities from there. It turns out that the internals of such compressors may introduce unpredictable, spurious numerical artifacts that are not negligible at all (see~\cite{cebrian:2005} for a detailed account). The root causes are left-to-right parsing or buffering fixed-size blocks of data during compression, as well as concatenating input strings when computing joint information quantities. Our compressor uses no internal fixed-size block, it is offline, and only the decompressor uses left-to-right parsing. By construction, our compressor is immune to the numerical artifacts reported for off-the-shelf compressors. The price for this is explicit storage of the dictionary, worst-case quadratic-time and linear space (due to the compressor being offline and also with a high constant due to the index).

Of course, just like any other actual compressor, Alg.~\ref{alg:deflate} is subject to pathological cases. If $x=\mbox{\texttt{ababababab}}$ and $y=\mbox{\texttt{bababababa}}$, then subsequent information measures will miss the similarity because Alg.~\ref{alg:deflate} would not have issued the same parselet. This issue is more likely to occur with small alphabets. Another limitation we face is handling numerical data: assessing similarity at the numerical level by factorizing out chunks of identical digits is not going to help very much in general. However, some hope remains in the area of 8-bit regularly-sampled data (see App.~\ref{app:multimodal}).

\section{Minimal sufficient models} \label{sec:min:suff}

We now finish to describe our heuristic for solving Eq.~\ref{eq:model}. Let $r\triangleq C(\mathcal{D}_r)+C(x_r\mid\mathcal{D}_r)$ the {\em rate} of $x_r$. Our task now is to generate successive $\{(\mathcal{D}_{r_n},x_{r_n})\}_n$ candidates that get closer to solving Eq.~\ref{eq:model}. Since we embodied Epicurus' Principle in Alg.~\ref{alg:deflate} already, we resort to the classical issue of minimizing Eq.~\ref{eq:codelength}.

The first technical step is to be able to evaluate the codelength function in Eq.~\ref{eq:codelength} for any rate-distortion state triplet $(\mathcal{D}_{r}, x_r, x)$ candidate. The second technical step is to evolve the current $(\mathcal{D}_r,x_r)$ candidate to the next. 

\subsection{Evaluating writing costs}

The first term in Eq.~\ref{eq:codelength} is the number of bits that are written during the course of the serialization routine \textsc{write\_model} described in App.~\ref{app:serialize:model}. The other two terms of Eq.~\ref{eq:codelength} are returned by the \textsc{write} primitive in Algs.~\ref{alg:write:rec}-\ref{alg:write}..

The evaluation of $C^\star(\underline{x}\mid x_r)$ cannot rely on monitoring the actual patching mechanism described in App.~\ref{app:serialize:patching}, as it only costs additional bits on disk: exploration of the rate-distortion space would become almost always infeasible. Algorithmic information theory states that this ``shortest program'' approach to patching is formally equivalent to the ``abstract list'' approach, where one counts the number of bits of the compressed index of the desired information in some abstract list that starts at $\hat{x}$.

This is where Shannon entropy keeps useful for us: our abstract list for $\underline{x}\mid x_r$ is the set described by the empirical frequencies of the letter-to-letter differences. In the idealized domain of $C^\star$, each letter in uncompressed $\underline{x}_r$ is followed by a bit indicating whether there is a difference, and an idealized number of bits corresponding to the first-order entropy estimate of the running letter-to-letter difference with $\underline{x}$ if it is non-zero. By using an idealized number of bits, we essentially mean to neutralize the actual convergence time of an entropy coder downstream. 

\begin{algorithm}[!ht]
\caption{Evaluation of writing costs $C(x_r\mid\mathcal{D}_{x_r})$ and $C^\star(\underline{x}\mid x_r)$, recursive part.}\label{alg:write:rec}
\begin{algorithmic}[1]
\STATE {\textsc{write\_parselet}( $x_r$, $\underline{x}$, $p$, diff, $n_{\mbox{diffs}}$ )}
\STATE \hspace{0.25cm}{\textbf{if} ( $p$.disjunction )}
\STATE \hspace{0.5cm}{alt$\leftarrow$\textsc{next\_bit}( $x_r$ ) }
\STATE \hspace{0.5cm}{$p\leftarrow$ alt ? $p$.right : $p$.left }
\STATE \hspace{0.5cm}{\textsc{write\_bit}( alt ) }
\STATE \hspace{0.5cm}{\textbf{return} \textsc{write\_parselet}( $x_r$, $\underline{x}$, $p$, diff, $n_{\mbox{diffs}}$ ) }
\STATE 
\STATE \hspace{0.25cm}{occ $\leftarrow$ 1 }
\STATE \hspace{0.25cm}{\textbf{if} ( $p$.left.option ) } 
\STATE \hspace{0.5cm}{occ $\leftarrow$ \textsc{next\_bit}( $x_r$ ) }
\STATE \hspace{0.5cm}{\textsc{write\_bit}( occ ) }
\STATE \hspace{0.25cm}{\textbf{if} ( $p$.left.repetition ) } 
\STATE \hspace{0.5cm}{occ $\leftarrow$ \textsc{next\_int}( $x_r$ ) }
\STATE \hspace{0.5cm}{\textsc{write\_int}( occ ) }
\STATE \hspace{0.25cm}{\textbf{while} ( occ-{}- ) }
\STATE \hspace{0.5cm}{\textbf{if} ( $p$.left.atomic ) }
\STATE \hspace{0.75cm}{lossless$\leftarrow$\textsc{next\_slot}( $\underline{x}$ ) }
\STATE \hspace{0.75cm}{\textbf{if} ( $p$.left.id $\neq$ lossless.id ) }
\STATE \hspace{1cm}{$n_{\mbox{diffs}}$++ }
\STATE \hspace{1cm}{diff[$p$.left.id-lossless.id]++ }
\STATE \hspace{0.5cm}{\textbf{else} \textsc{write\_parselet}( $x_r$, $\underline{x}$, $p$.left, diff, $n_{\mbox{diffs}}$ )}
\STATE \hspace{0.25cm}{\textbf{if} ( $\lnot$ $p$.regular ) \textbf{return}}
\STATE 
\STATE \hspace{0.25cm}{occ $\leftarrow$ 1 }
\STATE \hspace{0.25cm}{\textbf{if} ( $p$.right.option ) } 
\STATE \hspace{0.5cm}{occ $\leftarrow$ \textsc{next\_bit}( $x_r$ ) }
\STATE \hspace{0.5cm}{\textsc{write\_bit}( occ ) }
\STATE \hspace{0.25cm}{\textbf{if} ( $p$.right.repetition ) }
\STATE \hspace{0.5cm}{occ $\leftarrow$ \textsc{next\_int}( $x_r$ ) }
\STATE \hspace{0.5cm}{\textsc{write\_int}( occ ) }
\STATE \hspace{0.25cm}{\textbf{while} ( occ-{}- ) }
\STATE \hspace{0.5cm}{\textbf{if} ( $p$.right.atomic ) }
\STATE \hspace{0.75cm}{lossless$\leftarrow$\textsc{next\_slot}( $\underline{x}$ ) }
\STATE \hspace{0.75cm}{\textbf{if} ( $p$.right.id $\neq$ lossless.id ) }
\STATE \hspace{1cm}{$n_{\mbox{diffs}}$++ }
\STATE \hspace{1cm}{diff[$p$.right.id-lossless.id]++ }
\STATE \hspace{0.5cm}{\textbf{else} \textsc{write\_parselet}( $x_r$, $\underline{x}$, $p$.right, diff, $n_{\mbox{diffs}}$ )}
\end{algorithmic}
\end{algorithm}

\begin{algorithm}[!ht]
\caption{Evaluation of writing costs $C(x_r\mid\mathcal{D}_{x_r})$ and $C^\star(\underline{x}\mid x_r)$, main driver. Implements Eq.~\ref{eq:codelength} with App.~\ref{app:serialize:model}.}\label{alg:write}
\begin{algorithmic}[1]
\STATE {$C_x\leftarrow$\textsc{sys}.written}
\STATE {$n_{\mbox{diffs}}\leftarrow$ 0 }
\STATE 
\STATE {\textsc{write}( $x_r$, $\underline{x}$, diff )}
\STATE \hspace{0.25cm}{$s$ $\leftarrow$ \textsc{next\_ref}( $x_r$ ) }
\STATE \hspace{0.25cm}{\textsc{write\_ref}( $s$.id ) }
\STATE \hspace{0.25cm}{\textbf{if} ( $s$.EndOfString ) }
\STATE \hspace{0.5cm}{$C_x\leftarrow$\textsc{sys}.written - $C_x$}
\STATE \hspace{0.5cm}{$C^\star\leftarrow - \sum_{d=-|\alphabet|}^{|\alphabet|} \mbox{diff}[d]/\underline{x}\mbox{.len}\times l( \mbox{diff}[d]/\underline{x}\mbox{.len} )$}
\STATE \hspace{0.5cm}{$C^\star\leftarrow \underline{x}\mbox{.len} + n_{\mbox{diffs}} \times C^\star$}
\STATE \hspace{0.5cm}{\textbf{return} $C_x$, $C^\star$}
\STATE 
\STATE \hspace{0.25cm}{occ $\leftarrow$ 1 }
\STATE \hspace{0.25cm}{\textbf{if} ( $s$.option ) } 
\STATE \hspace{0.5cm}{occ $\leftarrow$ \textsc{next\_bit}( $x_r$ ) }
\STATE \hspace{0.5cm}{\textsc{write\_bit}( occ ) }
\STATE \hspace{0.25cm}{\textbf{if} ( $s$.repetition ) } 
\STATE \hspace{0.5cm}{occ $\leftarrow$ \textsc{next\_int}( $x_r$ ) }
\STATE \hspace{0.5cm}{\textsc{write\_int}( occ ) }
\STATE 
\STATE \hspace{0.25cm}{\textbf{while} ( occ-{}- ) }
\STATE \hspace{0.5cm}{\textbf{if} ( $s$.atomic ) }
\STATE \hspace{0.75cm}{lossless$\leftarrow$\textsc{next\_slot}( $\underline{x}$ ) }
\STATE \hspace{0.75cm}{\textbf{if} ( $p$.id $\neq$ lossless.id ) }
\STATE \hspace{1cm}{$n_{\mbox{diffs}}$++ }
\STATE \hspace{1cm}{diff[$s$.id-lossless.id]++ }
\STATE \hspace{0.5cm}{\textbf{else} \textsc{write\_parselet}( $x_r$, $\underline{x}$, $s$, diff, $n_{\mbox{diffs}}$ )}
\STATE 
\STATE \hspace{0.25cm}{\textsc{write}( $x_r$, $\underline{x}$, diff ) }
\end{algorithmic}
\end{algorithm}

Now we can evaluate the codelength function of Eq.~\ref{eq:codelength} efficiently for any $(\mathcal{D}_{r}, x_r, x)$ rate-distortion triplet candidate.

\subsection{Towards simpler, lossy models}

The only missing ingredient is the generation of a suite of $\{(\mathcal{D}_{r_n},x_{r_n})\}$ candidates. Aiming at the minimization of Eq.~\ref{eq:codelength}, we seek $r_{n+1}\leq r_n$. To make the process efficient, proposed changes should occur right in the dictionary $\mathcal{D}_{r_n}$, but their effects are ultimately to be assessed in uncompressed form {\em in advance}, so we avoid actually committing the proposed change. Now let $d(x,y)$ a string distortion measure between uncompressed strings $\underline{x}$ and $\underline{y}$. We seek the least change that may be performed from model space towards a simpler $x_{r_{n+1}}$:

\begin{eqnarray}
\mathcal{D}_{r_{n+1}} = \argmin_{\mathcal{D}_r\mbox{ s.t. }r\leq r_n} d(x_{r_{n+1}}\mid\mathcal{D}_r,x_{r_n}\mid\mathcal{D}_{r_n}), \label{eq:min:change}
\end{eqnarray}
which is our embodiment of the Principle of Minimal Change, so simplification stops when $\mathcal{D}_{r_{n+1}}=\mathcal{D}_{r_n}$. 

In Artificial Intelligence, removal from a knowledge base is known as a \textit{contraction} operation and is often implemented using the Principle of Minimal Change~\cite{ribeiro:2013}. Denote with $\searrow$ such a contraction operation governed by minimal change. Successive contractions of the model shall form the converging suite $(\mathcal{D}_{r_0},x_{r_0}) \searrow (\mathcal{D}_{r_1},x_{r_1}) \searrow \dots \searrow (\hat{\mathcal{D}}_,\hat{x})$, starting from $(\mathcal{D}_{r_0},x_{r_0})=(\mathcal{D},x)$ obtained from Alg.~\ref{alg:deflate}. For simplicity, our contraction operation will only need $(\mathcal{D}_{r_n}, x_{r_n})$ to generate the next rate-distortion state $(\mathcal{D}_{r_{n+1}},x_{r_{n+1}})$ in the suite, which we understand as a {\em local} contraction.

Due to the $(\mathcal{D}_{r_n},\preceq)$ lattice structure, a local contraction (removal from model space) may only involve one parselet in $\mathcal{D}_{r_n}$, so the search space of Eq.~\ref{eq:min:change} is tractable. Only being allowed to prune one parselet from $\mathcal{D}_{r_n}$ leaves no other choice but to {\em replace} it by a remaining one. For obvious efficiency reasons, we seek keeping decoding with the same instantiation parameter values in string data.  This further restrict the search space to candidate parselets having an available replacement. Eventually, our search will stop when the model only contains parselets with different syntactic structures.

Parselet replacement by an existing one is admissible: $C(\mathcal{D}_{r_{n+1}})<C(\mathcal{D}_{r_n})$ because we only have one less parselet to write, and $C(x_{r_{n+1}}\mid\mathcal{D}_{r_{n+1}}) \leq C(x_{r_{n}}\mid\mathcal{D}_{r_{n}})$ because $x_{r_{n+1}}$ uses one less identifier and the remaining one appears more often (which a final-stage entropy coder will take advantage of)---so $r_{n+1} < r_n$. 

Hence, we choose an existing parselet $\hat{p}_{n}$ as the \textit{replacement parselet} for $p_{n}$, such that $\mathcal{D}_{r_{n+1}}\cup\{p_n\}=\mathcal{D}_{r_{n}}$, and occurrences of $p_{n}$ in $x_{r_n}$ get replaced by $\hat{p}_{n}$ to form $x_{r_{n+1}}$. In practice, we only need the \textsc{rename} primitive to change occurrences of $p_{n}$.id by $\hat{p}_{n}$.id in data structures $(\mathcal{D}_{r_n},x_{r_n})$. Parselet replacement means that $p_n$.count occurrences of $p_n$ will incur a modification.

Extend the string distortion measure $d$ to parselets $p_n\sim\hat{p}_n$ such that $d(p_n,\hat{p}_n)$ is computed on the strings of the letter leaves of $p_n$ and $\hat{p}_n$ listed in some common arbitrary order (we use that of Alg.~\ref{alg:canonical}), or $d(p_n,\hat{p}_n)=\infty$ whenever $p_n\nsim\hat{p}_n$. Our proxy for solving Eq.~\ref{eq:min:change} through local contractions relies on determining the first pair of replacement parselets $(p_n,\hat{p}_n)$, when it exists, such that:

\begin{eqnarray*}
(p_n,\hat{p}_n) \triangleq \argmin_{(p,\hat{p})\in\mathcal{D}_{r_n}^2\mbox{ s.t. } p\neq\hat{x}\land p \sim \hat{p}} p\mbox{.count } d( p, \hat{p} ).
\end{eqnarray*}

It may seem tempting to consider a fast $O(|\mathcal{D}|)$ procedure by letting $\mathcal{D}_{r_{n+1}}\leftarrow\mathcal{D}\backslash \{p_{n}\}$. Alas that would defeat the purpose, since this would constrain the search space to the noisy structures of $\mathcal{D}$ that had been constructed from $x$ in the first place. Hence, we should instead use Alg.~\ref{alg:deflate} after Alg.~\ref{alg:decompress} to regenerate $\mathcal{D}_{r_{n+1}}$ from the uncompressed form of $\underline{x}_{r_{n+1}}$: the simpler structures in $x_{r_{n+1}}$ should be best reflected in $\mathcal{D}_{r_{n+1}}$ as well. 

Under the assumptions above, we are finally able to describe the actual compressor for one string $x$ in Alg.~\ref{alg:compress}. The single-string compressor is called with \textsc{compress}( $\underline{x}$, $\underline{x}$, $T_{sig}$, $T_{alt}$, $T_{opt}$, $\infty$ ). Observe that Alg.~\ref{alg:deflate} maintains $p$.count values, so they are available for line 3. If no parselet replacement is possible, Alg.~\ref{alg:compress} stops after line 3 (eventually the model would only keep a single representant for different syntactic structures). Alg.~\ref{alg:compress} eventually implements our heuristic for solving Eq.~\ref{eq:model} efficiently. 

\begin{algorithm}[!ht]
\caption{Single-string lossy compression: minimization of Eq.~\ref{eq:codelength} by parselet contraction. Implements Eq.~\ref{eq:model}: get $\mathcal{D}_{\hat{x}},\hat{x}$ from $\underline{x}$.}\label{alg:compress}
\begin{algorithmic}[1]
\STATE {\textsc{compress}( $\underline{x}$, $\underline{x}_{r_n}$, $T_{sig}$, $T_{opt}$, $T_{alt}$, $\mathcal{S}$, codelen\_min ) }
\STATE \hspace{0.25cm}{$\mathcal{D}_{r_n},x_{r_n}$ $\leftarrow$ \textsc{deflate}( \textsc{rlc}($\{\underline{x}_{r_n}\}$), $\emptyset$, }
\STATE \hspace{3.75cm}{$T_{sig}$, $T_{opt}$, $T_{alt}$, $\mathcal{S}$ )}
\STATE \hspace{0.25cm}{$p_{n},\hat{p}_{n}$ $\leftarrow \argmin_{p\neq\hat{p}, p\sim\hat{p}} p\mbox{.count }d(p,\hat{p})$ }
\STATE \hspace{0.25cm}{$\mathcal{D}_{r_{n+1}},x_{r_{n+1}}$ $\leftarrow$ \textsc{rename}( $\mathcal{D}_{r_n}$, $x_{r_n}$, $p_{n}$, $\hat{p}_{n}$ ) }
\STATE 
\STATE \hspace{0.25cm}{$C_{\mathcal{D}}$ $\leftarrow$\textsc{write\_model}( $\mathcal{D}_{r_{n+1}}$, $x_{r_{n+1}}$ ) }
\STATE \hspace{0.25cm}{$C_x$,$C^\star$ $\leftarrow$ \textsc{write}( $x_{r_{n+1}}$, $\underline{x}$, [] ) }
\STATE \hspace{0.25cm}{\textbf{if} ( $C_{\mathcal{D}}$+$C_x$+$C^\star$ $>$ codelen\_min ) \textbf{return} $\mathcal{D}_{r_n}$, $x_{r_n}$ }
\STATE \hspace{0.25cm}{\textbf{else} codelen\_min = $C_{\mathcal{D}}$+$C_x$+$C^\star$}
\STATE 
\STATE \hspace{0.25cm}{\textsc{compress}( $\underline{x}$, \textsc{expand}( $x_{r_{n+1}}$ ), }
\STATE \hspace{2.1cm}{$T_{sig}$, $T_{opt}$, $T_{alt}$, $\mathcal{S}$, codelen\_min )}
\end{algorithmic}
\end{algorithm}

The maximum number of iterations depends on the syntactic complexity of the data, and we allow for more iterations than is sketched in Alg.~\ref{alg:compress} to have a chance to escape from local minima. By default, we explore the next $L=250$ iterations after the last minimum of the codelength function has been found. Setting $L=0$ enables purely lossless compression as no parselet replacement is attempted. Setting $L=\infty$ enables scanning all reachable rate-distortion states of the data.

The suite $\{(\mathcal{D}_{r_n},x_{r_n})\}_n$ as a whole may occasionally contain states where a few previous local contractions up to $(\mathcal{D}_{r_n}, x_{r_n})$ suddenly resolve in dramatic changes in $\mathcal{D}_{r_{n+1}}$ after recompression at line 2. The actual suite of states reached during rate-distortion space exploration is called the rate-distortion {\em profile}, and the rate-distortion function would be computed from it~\cite{derooij:2012}. This explains why we only report rate-distortion profiles, and why they may contain such dramatic incidents. 

\subsection{Illustration: rate-distortion profiles}\label{sec:rd:profiles}

\begin{figure}[ht]
  \centering
  \includegraphics[scale=.65]{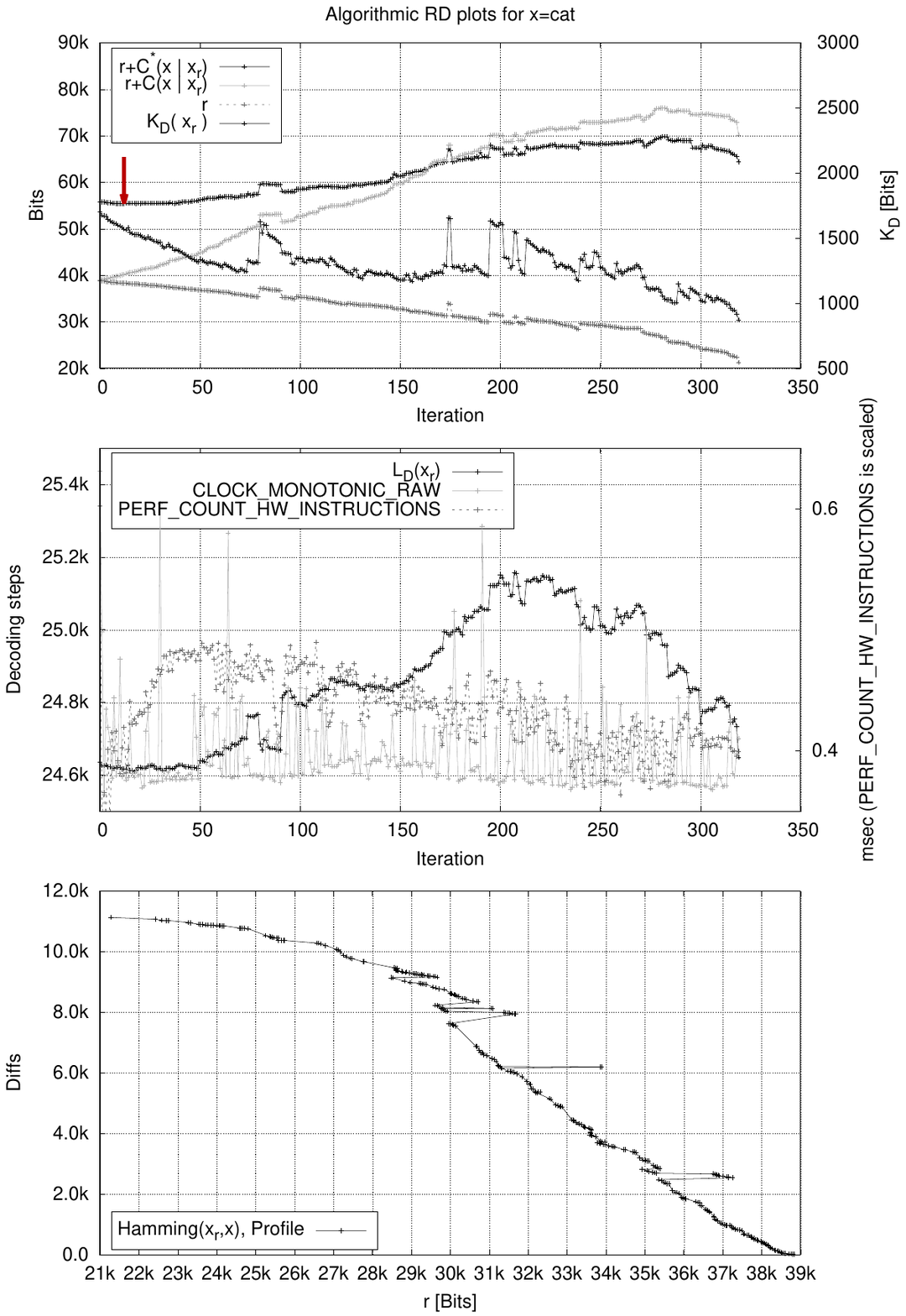}
  \caption{Rate-distortion plots of the mitochondrial DNA sample \texttt{cat} (17009 bases). Lossless model: 188 parselets. Minimal sufficient model: 150 parselets at $r$ = 37.2kb. Last model: 87 parselets. Wall-clock time: 4.8s (full plot).}
  \label{fig:rd:cat}
\end{figure}

\begin{figure}[!ht]
  \centering
  \includegraphics[scale=.65]{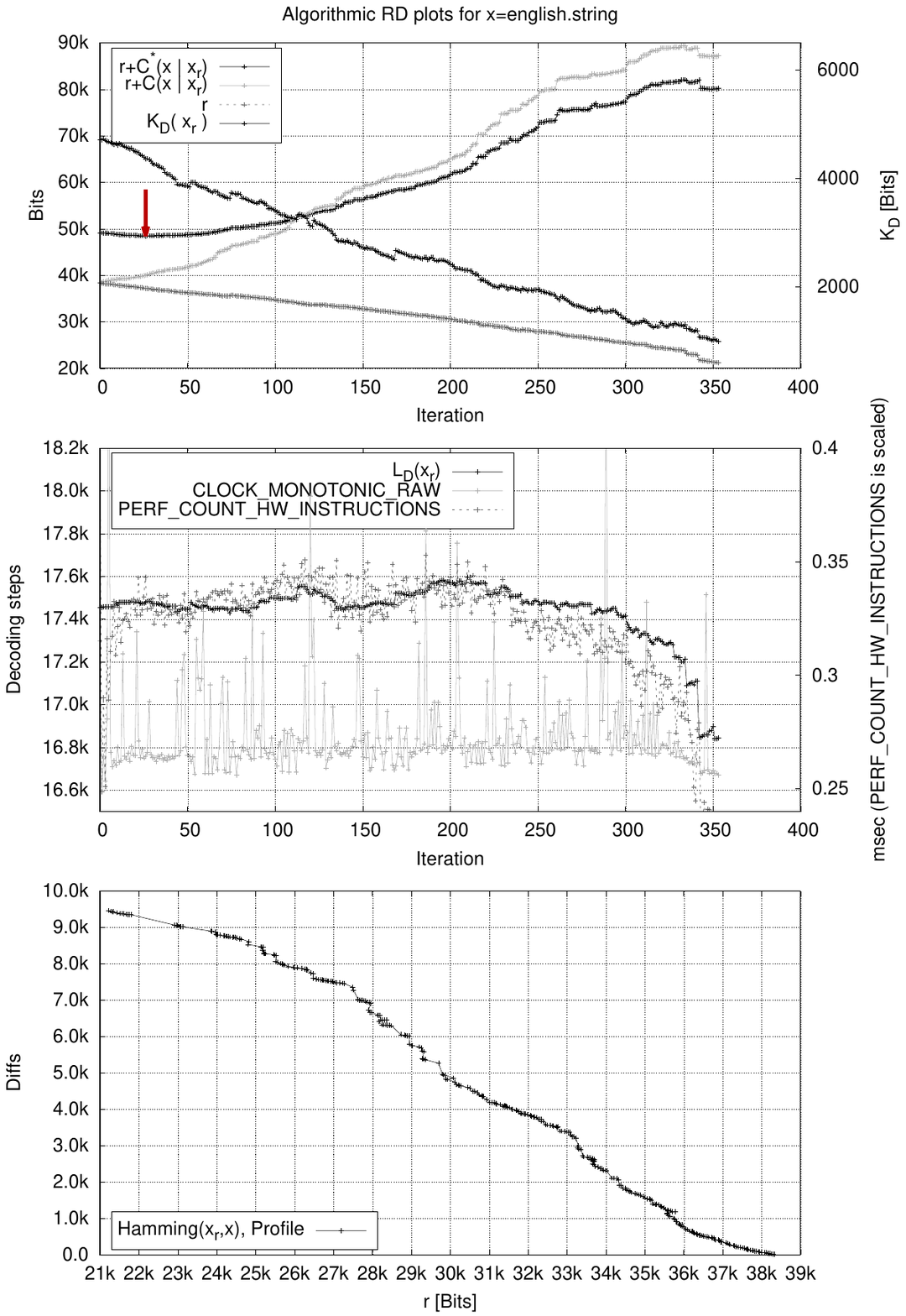}
  \caption{Rate-distortion plots of the \texttt{english} string (10843 letters). Lossless model: 315 parselets. Minimal sufficient model: 259 parselets at $r$ = 36kb. Last model: 80 parselets. Wall-clock time: 4.7s (full plot).}
  \label{fig:rd:english}
\end{figure}

For each example input string, we shall present the results in a systematic fashion and assuming $T_{opt}=T_{alt}=0$ so a separate discussion on generalization shall be in order. Black plots pertain to a value of interest in the algorithmic information-theoretic setting. From top to bottom in Figs.~\ref{fig:rd:cat}-\ref{fig:rd:english}, we shall depict:
\begin{itemize}
  \item The codelength function $r+C^\star(\underline{x}\mid x_r)$ (Eq.~\ref{eq:codelength}, the vertical arrow corresponds to the minimal sufficient model), the true size of the patched lossy representation on disk $r+C(\underline{x}\mid x_r)|$ (Eq.~\ref{eq:three:part}), and the size $r$ of the lossy string on disk;
  \item Logical depth estimates: the number of CPU instructions is mapped linearly to an arbitrary time scale so the correlations between the three, if any, get visible;
  \item Rate-distortion profile of the input string: this is the plot along the states that are reached by Alg.~\ref{alg:compress} -- the ``glitches'' are expected from the theory (see above).
\end{itemize}

We depict in Fig.~\ref{fig:rd:cat} the parselet-based rate-distortion plots of a DNA sample. For DNA, the rate-distortion profiles at the bottom show more catastrophic events during the first iterations on alphabets of small size. Also, the codelength function is quite flat, indicating that Alg.~\ref{alg:compress} mostly fails to build significantly better models than the lossless one. 

For English data in Fig.~\ref{fig:rd:english}, the codelength function visibly fits much better the expected behaviour for less redundant data~\cite{derooij:2012}. By design, we generate successive candidates for $\hat{x}$ by pruning models, so the last iterations work on models only remotely related to the data and the codelength function reaches a plateau. This is the major departure from the expected behaviour, but it appears long after the minimal sufficient model has been found. 

Our understanding of the logical depth estimates is as follows: first, monitoring the running time externally (\texttt{CLOCK\_MONOTONIC\_RAW}) clearly requires a few experiments to mitigate the side-effects of multi-tasking and other system integration issues. This is consistent with the recommendations for best practices in the literature~\cite{zenil:2012}.

On the contrary, the other two estimates should be considered deterministic and much closer to the data. The occasional discrepancies between the two are due to the way the compiler renders the different code paths of our implementation, and to our implementation techniques and data structures (see App.~\ref{app:data:structures}).

This ``compiler noise'' is completely removed from $L_{\mathcal{D}}(x_r)$, because it is computed at the programmatic level. Hence, experiments using the logical depth computed on parselets are expected to be more reliable and easier to implement because of determinism. 

Sudden variations in $L_{\mathcal{D}}(x_r)$ account for massive changes in $\mathcal{D}_{r_{n+1}}$ from $\mathcal{D}_{r}$, as it is connected to the expected length of the parselets. Figs.~\ref{fig:rd:cat}-\ref{fig:rd:english} suggest that the natural move for our algorithm is to start producing parselets that instantiate longer substrings for noisy data. 

\begin{figure}[!ht]
  \centering
  \includegraphics[trim={800mm 0 2300mm 0},clip,scale=.2]{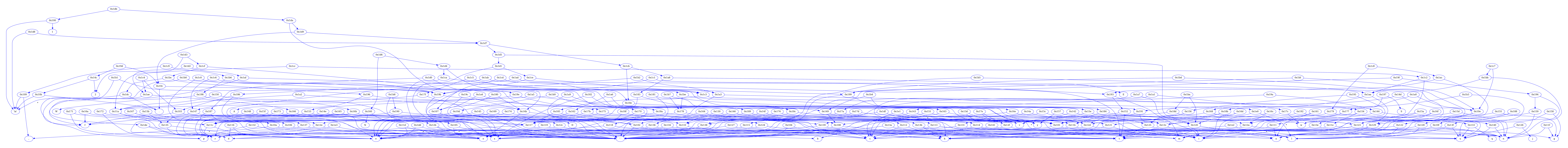}
  \caption{Excerpt of the minimal sufficient model built for \texttt{english}, in canonical order (206 parselets). Regexp operators appear on the edges (look at the small star on incoming edges to letter \texttt{'s'} below parselets \texttt{0x105} and \texttt{0x1dd} for instance). This is the Hasse diagram of lattice $(\mathcal{D}_{\hat{x}},\preceq)$ (albeit with inverted arrows, and omitting $\top=\underline{x}$ and $\bot=\epsilon$).}
  \label{fig:minsuff:english}
\end{figure}

We report in Fig.~\ref{fig:english:denoised} the beginning of the uncompressed string associated to the minimal sufficient model extracted on Fig.~\ref{fig:rd:english}, of which we depict an excerpt on Fig.~\ref{fig:minsuff:english}. The comparison with~\cite{derooij:2012} may suggest parselet-based algorithmic denoising is slightly more aggressive. This is because we operate from model space to select the next lossy version, while \cite{derooij:2012} operates with the help of a compressor oracle on the output of an external (genetic) model. We believe the much lower running times (seconds with parselets {\em vs.} hours for~\cite{derooij:2012}) compensate for the slightly cruder exploration of the local rate-distortion space. 

\begin{figure}[!ht]
\texttt{\tiny
eniversal Declaration of Human Rights\\
\\
Preanble\\
\\
Whereas recognition of the inherent tignity and of the equal and inalienacle rights of all menbers of the human fanily is the foundation of freedom, justice and peace in the world,\\
Whereas disregard and contenpt for human rights have resulted in barbarous asts which have outraged the conscience ou mankind, and the advent of a world in which human beings shall enjoy freedom of speech and delief and freedom from fear and want has been proclained as the highest aspisation of the common punple,\\
Whereas it is essential, if man is nou to be compelled to have recourse, as a mast resort, to redellion ahainst tyranny and oppression, that human rights should be proterted by the rule ou maw,\\
}
\caption{Beginning of the algorithmically, optimally denoised \texttt{english} string (decompression of the lossy string associated to the minimal sufficient model of Fig.~\ref{fig:rd:english}).}
  \label{fig:english:denoised}
\end{figure}

The directed acyclic graph (DAG) of Fig.~\ref{fig:minsuff:english} encodes the local join/meet information of the lattice $(\mathcal{D}_{\hat{x}},\preceq)$. Like some other well-known hierarchies of connected nodes, we observe less connections on the higher layers that capture more complex (and less frequent) regularities.

Unlike off-the-shelf compressors, we can provide such an explicit model of the data learned by compression, and it is the basis for the fast routines described in the next section. Whether this DAG may be used for generation is out of the scope of this work.

\section{String multiset compression} \label{sec:joint}

Now that we have a single-string compressor, we could just follow the literature and use concatenation to formally provide archiving capabilities (to approximate joint Kolmogorov complexity). Since Alg.~\ref{alg:deflate} runs in quadratic-time and it is called numerous times by Alg.~\ref{alg:compress}, one is more likely interested in the much faster procedures resulting from Eq.~\ref{eq:mset:union}.

This means that we should be able to describe a procedure to merge compressed representations $(\mathcal{D}_{\hat{x}},\hat{x},x)$ and $(\mathcal{D}_{\hat{y}},\hat{y},y)$ into $(\mathcal{D}_{\hat{x},\hat{y}},\{\hat{x},\hat{y}\},\{x,y\})$. Technically, the problem is twofold: we shall describe below how to ensure proper decoding of $\hat{x}$ and $\hat{y}$ using $\mathcal{D}_{\hat{x},\hat{y}}=\mathcal{D}_{\hat{x}}\cup\mathcal{D}_{\hat{y}}$, but tackling the marginal redundancy that may appear at this stage is addressed as a serialization issue in App.~\ref{app:serialize:marginal}.

\subsection{Efficient string merging}

Our main problem here is that identical string sets may be encoded in $\mathcal{D}_{\hat{x}}$ with different identifiers than in $\mathcal{D}_{\hat{y}}$, as different compression histories for the two would allow. Obviously, we can start with $\mathcal{D}_{\hat{x},\hat{y}}=\mathcal{D}_{\hat{x}}$, and only the identifiers in string data of $\hat{y}$ would need be changed (which is called {\em transcoding} hereafter). Alg.~\ref{alg:merge} follows the compression history of $\hat{y}$ to either find a matching identifier along the compression history of $\mathcal{D}_{\hat{x}}$ that should be reused, or add the currently missing parselet to $\mathcal{D}_{\hat{x},\hat{y}}$, such that $\hat{y}$ string data may still be decoded.

\begin{algorithm}[!ht]
\caption{Merging dictionaries $\mathcal{D}_{\hat{x}}$ and $\mathcal{D}_{\hat{y}}$ into $\mathcal{D}_{\hat{x},\hat{y}}$, transcoding $y$.}\label{alg:merge}
\begin{algorithmic}[1]
\STATE {\textsc{merge}( $\mathcal{D}_{\hat{x}}$, $\mathcal{D}_{\hat{y}}$, $y$ ) }
\STATE \hspace{0.25cm}$\mathcal{D}_{\hat{x},\hat{y}}$ $\leftarrow$ $\mathcal{D}_{\hat{x}}$
\STATE \hspace{0.25cm}y\_mapping $\leftarrow$ $[0\dots|\mathcal{D}_{\hat{y}}|-1]$
\STATE \hspace{0.25cm}
\STATE \hspace{0.25cm}\textbf{for} ( $i$ $\leftarrow$ $|\mathcal{A}|$ ; $i$ $<$ $|\mathcal{D}_{\hat{y}}|$ ; $i$++ )
\STATE \hspace{0.5cm}$p$ $\leftarrow$ $\mathcal{D}_{\hat{y}}[i]$
\STATE \hspace{0.5cm}$p$.left.id $\leftarrow$ y\_mapping[ $p$.left.id ]
\STATE \hspace{0.5cm}\textbf{if} ( $p$.regular ) $p$.right.id $\leftarrow$ y\_mapping[ $p$.right.id ]
\STATE 
\STATE \hspace{0.5cm}\textbf{for} ( $j$ $\leftarrow$ $|\mathcal{A}|$ ; $j$ $<$ $|\mathcal{D}_{\hat{x}}|$ ; $j$++ )
\STATE \hspace{0.75cm}\textbf{if} ( $p$ == $\mathcal{D}_{\hat{x}}[j]$ ) 
\STATE \hspace{1cm}y\_mapping[ $i$ ] $\leftarrow$ $j$
\STATE \hspace{1cm}\textbf{break}
\STATE \hspace{0.5cm}\textbf{if} ( j == $|\mathcal{D}_{\hat{x}}|$ )
\STATE \hspace{0.75cm}$\mathcal{D}_{{\hat{x}},\hat{y}}$ $\leftarrow$ $\mathcal{D}_{\hat{x},\hat{y}}\cup\{p\}$
\STATE \hspace{0.75cm}y\_mapping[ $i$ ] $\leftarrow$ $|\mathcal{D}_{\hat{x},\hat{y}}|$
\STATE
\STATE \hspace{0.25cm}\textbf{while} ( $\lnot$ $y$.EndOfString ) 
\STATE \hspace{0.5cm}$y$.id $\leftarrow$ y\_mapping[ $y$.id ]
\STATE \hspace{0.5cm}\textsc{next\_ref}( $y$ )
\STATE
\STATE \hspace{0.25cm}\textbf{return} $\mathcal{D}_{\hat{x},\hat{y}}$
\end{algorithmic}
\end{algorithm}

Our {\em recursive macro scheme} is acyclic: it works on the lattice $(\mathcal{D}_{\hat{x}},\preceq)$, therefore it features the socalled {\em topological recursion} property~\cite[Sec.~2]{storer:1982}, which enables repurposing any sorting algorithm to find any permutation of identifiers (this is instrumental to writing the model in canonical order). It also enables finding a mapping of identifiers in any arbitrary order, not just by compression history like in Alg.~\ref{alg:merge}. In practice, our implementation of Alg.~\ref{alg:merge} uses a hashtable instead of the $O(|\mathcal{D}_{\hat{x}}|)$ search in the inner loop (see App.~\ref{sec:impl:dictionary}). 

Alg.~\ref{alg:merge} lays all the basic ideas for our core \textsc{union} primitive, that merges any number of strings $\{(\mathcal{D}_{\hat{x}_i},\hat{x}_i,x_i)\}_i$ into $\{(\mathcal{D}_{\hat{X}},\hat{x}_i,x_i)\}_i$. It is implemented along divide-and-conquer, in bottom-up, parallelized batch executions of merging models using hashtables, and transcoding (half) sets of strings is only needed if we must output an actual archive: evaluations of information quantities based on the sole minimal sufficient model may skip it. In turn, the (parallelized) \textsc{union} primitive described above is used into Alg.~\ref{alg:kompress}, which is the core of our Solomonoff archiver (the \textbf{spawn} keyword annotates work done in parallel, and the remaining details for actual archival storage are in App.~\ref{app:serialize:marginal}). 

\begin{algorithm}[!ht]
  \caption{Parallel Solomonoff archiver core for joint information calculus. Implements Eq.~\ref{eq:mset:union}: get $\{(\mathcal{D}_{\hat{X}},\hat{x}_i,x_i)\}_i$ from $\{(\mathcal{D}_{\hat{x}_i},\hat{x}_i,x_i)\}_i$.}\label{alg:kompress}
\begin{algorithmic}[1]
\STATE {\textsc{kompress}( X, $T_{sig}$, $T_{opt}$, $T_{alt}$, $\mathcal{S}$ ) }
\STATE \hspace{0.25cm}\textbf{spawn} $\mathcal{D}_{\hat{x}_i},\hat{x}_i$ $\leftarrow$ {\textsc{compress}}( $\underline{x}_i$, $\underline{x}_i$, 
\STATE \hspace{4.75cm}$T_{sig}$, $T_{opt}$, $T_{alt}$, $\mathcal{S}$, $\infty$ )
\STATE \hspace{0.25cm}$\mathcal{D}_{\hat{X}}$ $\leftarrow$ \textbf{spawn} \textsc{union}( $[\mathcal{D}_{\hat{x}_i}]$, $[\hat{x}_i]$ )
\STATE \hspace{0.25cm}\textbf{return} $\mathcal{D}_{\hat{X}}$, $[\hat{x}_i]$
\end{algorithmic}
\end{algorithm}

In practice, the calls to \textsc{compress} check for cached, compressed versions on disk to make our implementation of Eq.~\ref{eq:mset:union} only ever need one single call to Alg.~\ref{alg:compress} per string---so scanning directories allows to build on the filesystem to define and store string multisets for certain named classes/concepts described extensively, ready for fast evaluations once the cache has been populated.

\subsection{Illustration: populating the cache on disk}

In a typical working session, the first step is to transform the data into the compressed, patched parselets representation that uses the minimal sufficient model. Tab.~\ref{tab:cache} reports the execution performance with default $L=250$ additional contractions for safety. We report the wall-clock time. Speed-up due to multi-threading scales almost linearly on this machine but maximum memory consumption keeps very high.

For comparison with Tab.~\ref{tab:datasets}, Tab.~\ref{tab:cache} includes results for lossless compression ($\hat{x}=x$). DNA data is decently compressed by \texttt{mpx} (159.1KiB, less than the 166.2KiB of \texttt{gzip} in Tab.~\ref{tab:datasets}), but text data gives much longer bitstreams (244.4KiB).

\begin{table}[!ht]
  \caption{Computation of the models. Focus on generalization (g) and tokenization (t). Maximum memory consumption varied between 55MiB and 83MiB. In the last column, we report the sum of compressed file sizes as reported by the filesystem.\label{tab:cache}}
  \centering
  \begin{tabular}{|l|c|c|c|}\hline
    Lossless                 &  Time (s) & MT speed-up & Size (KiB) \\\hline\hline
    \texttt{mammals}         &      0.15 &        x7.1 & 159.1 \\\hline
    \texttt{mammals} (g)     &      0.25 &        x7.5 & 159.2 \\\hline\hline
    \texttt{languages}       &      0.23 &        x6.6 & 244.4 \\\hline
    \texttt{languages} (t)   &      0.25 &        x6.7 & 250.0 \\\hline
    \texttt{languages} (g)   &      0.50 &        x6.7 & 275.2 \\\hline
    \texttt{languages} (g,t) &      0.61 &        x7.1 & 265.0 \\\hline\hline
    Minimal sufficient       &  Time (s) & MT speed-up & Size (K) \\\hline\hline
    \texttt{mammals}         &      22.1 &        x7.5 & 166.7 \\\hline
    \texttt{mammals} (g)     &      57.7 &        x7.5 & 171.1 \\\hline\hline
    \texttt{languages}       &      38.0 &        x7.6 & 256.9 \\\hline
    \texttt{languages} (t)   &      43.8 &        x7.6 & 265.6 \\\hline
    \texttt{languages} (g)   &      78.5 &        x7.4 & 279.2 \\\hline
    \texttt{languages} (g,t) &     121.5 &        x7.7 & 275.4 \\\hline
  \end{tabular}
\end{table}

Several observations are in order. First, the relative increases due to generalization or tokenization keep in par with, for instance, that of \texttt{gzip} {\em vs.} \texttt{ppmd} (see Tab.~\ref{tab:datasets}). Secondly, instead of the lossless model, storing the data in the language of its minimal sufficient model incurs a $10\%$ increase on disk usage (the formatting details to lossless reconstruction are in App.~\ref{app:serialize:patching}).

\section{Arbitrary models}\label{sec:arbitrary}

So far, we miss a way to formulate hypotheses about the data, in the form of an arbitrary parselet model $\mathcal{D}$. This is intrumental to interfacing with hypothetical other procedures for manipulating and producing new models, or to integrate prior knowledge that had been acquired by other means (if the data were structured according to some known regular grammar, one should certainly devise a compiler to produce the model).

The first technical step is to describe how an arbitrary parselet model $\mathcal{D}$ is to be applied on arbitrary string $\underline{x}$, and the second technical step is to explain how to generate a workable model from arbitrary prior knowledge. 

\subsection{Conditional deflation}

Let $(\mathcal{D}_{\hat{x}}, \hat{x}, \underline{x})$ obtained from $\underline{x}$ and Alg.~\ref{alg:compress}, and let $\underline{\hat{x}}$ the uncompressed version of $\hat{x}$. Our conditional representation of $\underline{x}$ using given model $\mathcal{D}$ is computed by finding the subset of $\mathcal{D}$ that describes $\hat{x}$ as compactly as possible, and we denote:

\begin{eqnarray}
  C(\hat{x}\Vert\mathcal{D}) \triangleq \argmin_{\mathcal{D}_x\subseteq\mathcal{D}} C(\mathcal{D}_x) + C(\underline{\hat{x}}\mid\mathcal{D}_x).\label{eq:cond:defl}
\end{eqnarray}
We borrow the double-bar to indicate using a different model than the appropriate one, like in the Kullback-Leibler divergence notation. The reason we work on $\hat{x}$ is because we seek to make the Fundamental Inequality holds, that links Bayes' Rule to the MDL~\cite[Sec.~5.5.4]{li:vitaniy:2019}. 

We now describe a procedure to implement Eq.~\ref{eq:cond:defl}. Note that we do not need to take tokenization into account, as that would have been done upon creation of $\mathcal{D}$. Obviously we seek compression of $\underline{\hat{x}}$, yet under the constraint of being able to explore all subsets of $\mathcal{D}$. Our rationale therefore is to keep Occam's Razor as in Alg.~\ref{alg:deflate}, but to accept any options or alternatives of $\mathcal{D}$ that are contained in $\hat{x}$ and referenced by another parselet in $\mathcal{D}$.

The proposed procedure for Eq.~\ref{eq:cond:defl} is depicted in Alg.~\ref{alg:cond:deflate}. It works by trying to reuse as much parselets from $\mathcal{D}$ as possible. Several iterations are needed so all parselets in $\mathcal{D}$ get a chance to be used to describe $\hat{x}$ (parselets may come in any order in $\mathcal{D}$). When the parselets in $\mathcal{D}$ are listed in canonical order, we do not need more than 6 iterations over $\mathcal{D}$ (one per type of regular parselet).

\begin{algorithm}[!ht]
\caption{Conditional deflation by parselets (implements Eq.~\ref{eq:cond:defl}). Get $(\mathcal{D}_x,x)$ from $\underline{\hat{x}}$ and $\mathcal{D}$ such that $\mathcal{D}_x\subseteq\mathcal{D}$ and $C(\mathcal{D}_x)+C(\underline{\hat{x}}\mid\mathcal{D}_x)$ is minimized.} \label{alg:cond:deflate}
\begin{algorithmic}[1]
\STATE {\textsc{deflate\_cond}( $\hat{x}$, $\mathcal{D}$, $T_{sig}$ )}
\STATE \hspace{0.25cm}{\textbf{foreach} ( $p\in\mathcal{D}$ )} $p$.count$\leftarrow$0
\STATE \hspace{0.25cm}{$\mathcal{D}_x\leftarrow\emptyset$}
\STATE \hspace{0.25cm}{reused$\leftarrow$1}
\STATE \hspace{0.25cm}{\textbf{while} ( reused $>$ 0 )}
\STATE \hspace{0.5cm}{reused$\leftarrow$0}
\STATE \hspace{0.5cm}{\textbf{foreach} ( $p\in\mathcal{D}$ )}
\STATE \hspace{0.75cm}{\textbf{if} ( $p$.count $>$ 0 ) \textbf{continue}}
\STATE \hspace{0.75cm}{\textbf{if} ( $p$.disjunction )}
\STATE \hspace{1cm}{$\{s\}_p\leftarrow$\textsc{index\_find\_disjunction( $\hat{x}$, $p$ )}}
\STATE \hspace{1cm}{\textbf{if} ( $|\{s\}_p|$ $>$ 0 )}
\STATE \hspace{1.25cm}{reused++}
\STATE \hspace{1.25cm}{$p$.count$\leftarrow |\{s\}_p|$}
\STATE \hspace{1.25cm}$ \mathcal{D}_x\leftarrow \mathcal{D}_x\cup\{p\}$
\STATE \hspace{1.25cm}{\textbf{foreach} ( $s\in\{s\}_p$ ) }
\STATE \hspace{1.5cm}{\textsc{compile\_disjunction}( $s$, $p$ )}
\STATE \hspace{1.25cm}{\textsc{rlc}( $\{s\}_p$ ) }
\STATE \hspace{0.75cm}{\textbf{else if} ( $p$.right.option )}
\STATE \hspace{1cm}{$\{s\}_p\leftarrow$\textsc{index\_find\_option( $\hat{x}$, $p$ )}}
\STATE \hspace{1cm}{\textbf{if} ( $|\{s\}_p|$ $>$ 0 )}
\STATE \hspace{1.25cm}{reused++}
\STATE \hspace{1.25cm}{$p$.count$\leftarrow |\{s\}_p|$}
\STATE \hspace{1.25cm}$ \mathcal{D}_x\leftarrow \mathcal{D}_x\cup\{p\}$
\STATE \hspace{1.25cm}{\textbf{foreach} ( $s\in\{s\}_p$ ) }
\STATE \hspace{1.5cm}{\textsc{compile\_option}( $s$, $p$ )}
\STATE \hspace{1.25cm}{\textsc{rlc}( $\{s\}_p$ ) }
\STATE \hspace{0.75cm}{\textbf{else}}
\STATE \hspace{1cm}{$\{s\}_p\leftarrow$\textsc{index\_find\_conjunction( $\hat{x}$, $p$ )}}
\STATE \hspace{1cm}{\textbf{if} ( $|\{s\}_p|$ $\geq$ $T_{sig}$ )}
\STATE \hspace{1.25cm}{reused++}
\STATE \hspace{1.25cm}{$p$.count$\leftarrow |\{s\}_p|$}
\STATE \hspace{1.25cm}$ \mathcal{D}_x\leftarrow \mathcal{D}_x\cup\{p\}$
\STATE \hspace{1.25cm}{\textbf{foreach} ( $s\in\{s\}_p$ ) }
\STATE \hspace{1.5cm}{\textsc{compile\_conjunction}( $s$, $p$ )}
\STATE \hspace{1.25cm}{\textsc{rlc}( $\{s\}_p$ ) }
\STATE \hspace{0.25cm}{\textsc{trim}( $\mathcal{D}_x$, $\hat{x}$ )}
\STATE \hspace{0.25cm}{\textbf{return} $\mathcal{D}_x,\hat{x}$}
\end{algorithmic}
\end{algorithm}

In Alg.~\ref{alg:cond:deflate}, we used several primitives that we only outline here for brevity:
\begin{itemize}
\item The \textsc{trim} primitive, that reallocates parselet indices of $\mathcal{D}_x$ starting from $|\mathcal{A}|$ and subsequently transcodes $\hat{x}$ (so it is only about getting the right permutation of indices and following the lines of Alg.~\ref{alg:merge});
\item The \textsc{index\_find\_*} primitives, that use the index to locate the occurrences $\{s\}_p$ of a given parselet $p$ in $\hat{x}$ (options and alternatives should be referenced by another parselet in $\mathcal{D}$ as well). Searching for known parselets is much easier (and faster) than for the \textsc{most\_*} primitives.
\end{itemize}

\subsection{Building a model from prior knowledge}

Our reference implementation shall eventually feature a textual input reader for models and strings. Yet, it is easy to trick Alg.~\ref{alg:compress} into producing workable models from prior knowledge expressed as a string. 

Let prior knowledge be encoded in string $\underline{z}$. All of it should be available to describe string $\underline{\hat{x}}$ using a conditional deflation, so we let $(\mathcal{D},z)\leftarrow\mbox{\textsc{compress}}( \underline{z}, \underline{z}, 1, T_{opt}, T_{alt}, \mathcal{S}, \infty )$. Setting $T_{sig}=1$ ensures all of $\underline{z}$ will be stored in $\mathcal{D}$. Since conditional deflation will filter out unused parselets in $\mathcal{D}$ to describe $\hat{x}$, this is not an issue in practice. Alternatively, we may well reuse any models we have at hand, as we illustrate below. 

\subsection{Illustration: Scoring hypotheses}\label{sec:appli:hypo}

\begin{figure}[!ht]
  \centering
  \includegraphics[width=\columnwidth]{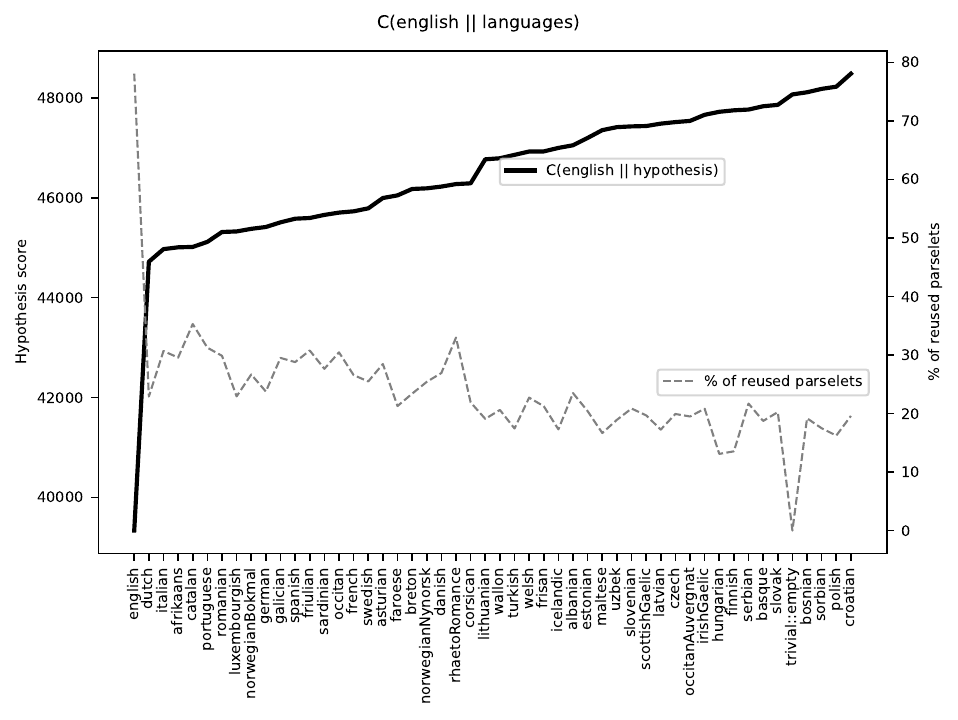}
  \caption{Applying Alg.~\ref{alg:cond:deflate} to \texttt{english} with minimal sufficient models from \texttt{languages}. The value of $C(\mbox{\texttt{english}}\Vert\emptyset)$ is reported for reference. Wall-clock time: 0.142s (parallelization: x5.8).}
  \label{fig:hypothesis:testing}
\end{figure}

To illustrate the behaviour of Alg.~\ref{alg:cond:deflate} in solving Eq.~\ref{eq:cond:defl}, we report in Fig.~\ref{fig:hypothesis:testing} the (ranked) values of $C(\mbox{\texttt{english}}\Vert\mbox{\texttt{languages}})$ and $C(\mbox{\texttt{english}}\Vert\emptyset)$, along with the proportion of parselets that were reused from the minimal sufficient model of the proposed hypothesis. The MDL principle tells us to select the best hypothesis on the data as $\argmin_i C(\mbox{\texttt{english}}\Vert\mbox{\texttt{languages}}_i)$. 

The left-most value of $C(\mbox{\texttt{english}}\Vert\mbox{\texttt{english}})$ is reported in order to assess how Alg.~\ref{alg:cond:deflate} may reuse parselets created with Alg.~\ref{alg:compress} on the very same input data. We cannot reuse all of them because they were loaded from canonical order, which is different from that of compression history, and the sets of locations that may be compiled to the same parselet cannot be expected to be the same depending on the order they are searched. Nevertheless, as much as 78\% of parselets from Alg.~\ref{alg:compress} could be reused. Alg.~\ref{alg:cond:deflate} produced a bitstream for the lossy \texttt{english} string that is 2000 bits longer (+5.3\%) than that produced by Alg.~\ref{alg:compress}. Including the slightly larger patch, Alg.~\ref{alg:cond:deflate} takes 5284 bytes on disk, while Alg.~\ref{alg:compress} only needs 5008.

The value of $C(\mbox{\texttt{english}}\Vert\emptyset)$ is located to the right, where zero parselets could be reused. It is among the highest, in the middle of mostly Eastern-European languages hypotheses. The empty hypothesis may serve to set a threshold for non-informative hypotheses. 

Slightly better scores are obtained for Western-European languages hypotheses, yet the gap from $C(\mbox{\texttt{english}}\Vert\mbox{\texttt{english}})$ to  $C(\mbox{\texttt{english}}\Vert\mbox{\texttt{catalan}})$ suggests the value of the latter is still in the higher range.

Also, one may consider that not using the same order for parselet creation between Alg.~\ref{alg:compress} and Alg.~\ref{alg:cond:deflate}, makes the minimal sufficient model of \texttt{english} an admissible hypothesis about the data (not one that had been learned exactly the same way on the data, see~\cite[Sec.~5.5.1]{li:vitaniy:2019}). In this perspective, the gap after $C(\mbox{\texttt{english}}\Vert\mbox{\texttt{english}})$ illustrates more how a good model for the data will be discriminated from others. 

\section{Applications}\label{sec:appli}

In this section, we first connect parselets to Jaynes' system of probabilities. Eventually, we are able to compute algorithmic information-theoretic distances and conditional independence measures. They are all illustrated on real data.

\subsection{Abstract probabilities}\label{sec:appli:proba}

Quite different from (uncomputable) Solomonoff's algorithmic probabilities, parselets allow to assign a probability to any string multiset $X=\{(\mathcal{D}_{\hat{x}},\hat{x},x)\}$ against a given compressible string multiset $\Omega$, see Def.~\ref{def:proba}.

Because $\mathcal{D}_{\hat{\Omega}}$ is finite, $(\mathcal{D}_{\hat{\Omega}}, 2^{\mathcal{D}_{\hat{\Omega}}}, \mathcal{D}\mapsto |\mathcal{D}|/|\mathcal{D}_{\hat{\Omega}}|)$ is a probability space following Kolmogorov's probability axioms. Under syntactic independence of two string multisets ($X\Perp Y\mid\emptyset$ after Def.~\ref{def:syntactic:indep}), so is $(\Omega, 2^{\mathcal{D}_{\hat{\Omega}}}, X\mapsto |\mathcal{D}_{\hat{X}}|/|\mathcal{D}_{\hat{\Omega}}|)$. In practice however, we would like to relax the constraint that $X\subseteq\Omega$, so we may evaluate the probability of any string multiset $X$ against a compressible one $\Omega$. Therefore, our probability evaluations in Def.~\ref{def:proba} embed a projection on $\mathcal{D}_{\hat{\Omega}}$ to the numerator. 

\begin{definition}[Abstract probabilities]\label{def:proba}
  Let current knowledge defined by compressible finite multiset $\Omega$ (the ``universe'') of finite strings defined over $\mathcal{A}$. We compute abstract probabilities against $\Omega$ for two string multisets $Y$ and $X=\{(\mathcal{D}_{\hat{x}},\hat{x},x)\}$ in compressed form like:
  \begin{eqnarray}\label{eq:prob}
    \prob(X) & \triangleq & \frac{|\left(\bigcap_{x\in X}\mathcal{D}_{\hat{x}}\right)\cap\mathcal{D}_{\hat{\Omega}}|}{|\mathcal{D}_{\hat{\Omega}}|},\\
    \prob(X\mid Y) & \triangleq & \frac{\prob(X,Y)}{\prob(Y)}, \forall\;\prob(Y)>0.
  \end{eqnarray}
\end{definition}

\begin{proposition}[$(\Omega, 2^{\mathcal{D}_{\hat{\Omega}}}, \prob)$ is a probability space]
  Eq.~\ref{eq:prob} defines a normalized, well-defined, and finitely additive probability measure. 
  \begin{proof}
    $2^{\mathcal{D}_{\hat{\Omega}}}$ is a $\sigma$-algebra and $\forall\;X, \prob(X)\geq 0$ ($\prob$ is normalized). By Eq.~\ref{eq:prob}, $\forall\;X, \prob(X)\leq\prob(\Omega)=1$ ($\prob$ is well-defined). Let $X\Perp Y\mid\emptyset$ two syntactically independent finite multisets of finite strings. Then $\mathcal{D}_{\hat{X}}\cap\mathcal{D}_{\hat{Y}}=\emptyset$ and $|\mathcal{D}_{\hat{X},\hat{Y}}\cap\mathcal{D}_{\hat{\Omega}}|=|\mathcal{D}_{\hat{X}}\cap\mathcal{D}_{\hat{\Omega}}|+|\mathcal{D}_{\hat{Y}}\cap\mathcal{D}_{\hat{\Omega}}|$ ($\prob$ is finitely additive).
  \end{proof}
\end{proposition}

While Kolmogorov's system of probability builds on measure theory (like we used in our exposition), Jaynes' starts from a fully computational setting~\cite{jaynes:2003}, where an automaton in charge of faithfully manipulating plausibility of logical propositions, is used to formally derive the (very same) rules of probability calculus. In~\cite[Ch.~2]{jaynes:2003}, Jaynes states that for his theory to be rightfully modeled with Venn diagrams (which is not required at all however), their {\em ``points must represent some ultimate elementary propositions $\omega_i$ into which [logical conjunction] A can be resolved. Of course, consistency then requires us to suppose that [logical conjunctions] B and C can also be resolved into these same propositions $\omega_i$.''} Ironically, we became aware of Jaynes' work only a few weeks after we implemented abstract probabilities.

Later on, Jaynes~\cite[App. A]{jaynes:2003} makes explicit the formal equivalence with Kolmogorov's system in the important case of discrete finite set-theory, concluding that both systems reach {\em ``essentially complete agreement''} in this setting. It was quite likely that probabilities would arise as a close by-product of an implementation of a combinatorial information measure (our underlying quantity of information here is the size of regular parselet sets). Kolmogorov advocated that information should be efficiently measured in discrete machines by using combinatorial information measures~\cite{kolmogorov:1970}, but his illustrative example could not easily be contrived into a workable implementation.

Let us now speculate a bit. Jaynes discusses at length why logical conjunctions are almost always a better default choice than disjunctions for expressing logical theories. People who would have started the very same work as ours from Jaynes' intuition, would also likely have implemented disjunctive parselets only for the sake of completeness, but they would equally likely have missed the \texttt{*} and \texttt{?} regexp operators in the first place (which Jaynes does not need to mention because the full regexp language would be akin to syntactic sugar for his matter). This suggests that the resulting algorithms would have been very close to ours and only marginally simpler without compression in the picture. But the whole design would have been driven first and foremost by compression, so the integration of \textsc{rlc} through a \texttt{*} operator seems inevitable at some point and, as suggested by our pseudo-codes, supporting \texttt{?} in addition to \texttt{*} is almost free.

In retrospect, our attempt at defining (abstract) probabilities coincides with Jaynes' approach on two fundamental points, namely:
\begin{itemize}
\item Ensuring faithful manipulation of the data by the machine at all times, as we enforce Eq.~\ref{eq:info:sym:pos}, and the lossless strings may be recovered at any time during computations because of using joint compression as the core building block;
\item We adequately model string data as one big logical conjunction of parselets, which have the required recursive type to populate sound Venn diagrams (the parselet model seen as a set).
\end{itemize}

In a sense, parselets revisit Jaynes' work by enforcing faithfulness at the lower level of an underlying, combinatorial information measure.

In Jaynes' system, which started out from modeling what kind of logical reasoning could be reliably delegated to the machine, of which our introductory material is reminiscent, the minimal sufficient model would encode the shortest defining logical statement about the class to which the data belongs. Computing unions of parselet sets is how the minimal language encompassing each piece of data is constructed, to contain the necessary elementary propositions for logical reasoning about the data as a whole.

Hopefully the transparent Jaynesian background firmly establishes abstract probabilities as little more than a serendipity hack from language theory. The fundamental intuition of parselets as a generic recursive type to describe data should obviously be credited to Jaynes: we gladly acknowledge we stumbled upon it much later indeed, and deprived of such profound views. 

Assuming all data is observed (otherwise dedicated, external procedures to evolve the data are needed, of which Alg.~\ref{alg:compress} is an example in the compressed domain), abstract probabilities readily extend the reach of any algorithm expressed in this language to the deterministic setting. 

\subsection{Illustration: abstract probability universe}

In case the language of abstract probabilities is used by upstream algorithms, one may want to abstract the union of parselets into an archive without strings. This shall illustrate the parallel performance of the \textsc{union} primitive and reading back the costly computations above from the cache. Tab~\ref{tab:universe} shows some decent speed-up can still be obtained by parallel execution, resulting in a very fast operation overall.

Tab.~\ref{tab:universe} allows us to reflect on our general strategy for implementing Epicurus' Principle by favoring the augmentation of the size of the dictionary under some allowance in compressed length increase (as reported in Tab.~\ref{tab:cache}). Generalization is mostly useless on DNA. On textual data, tokenization has a less important influence than generalization on the size of the dictionary. Recalling Jaynes, this provides some experimental insight as to ``how much'' expressing logical theories about the world should be expressed primarily with conjunctions: only textual data used alternatives significantly, for less than 15\% of the total number of parselets.

\begin{table}[!ht]
  \caption{Computing abstract probability universes from the cache. We annotate generalization with (g), and the number of options and alternatives is provided, and tokenization is annotated with (t). Increases are relative to the baseline (no generalization and no tokenization). Average speed-up due to multi-threading was x3.9, maximum running time was 112ms.\label{tab:universe}}
  \centering
  \begin{tabular}{|l|c|c|c|c|}\hline
    Lossless                 & $\sum_{x} K^\star(x)$ & \begin{tabular}{c} $K^\star(X)$ \\ (opt,alt) \end{tabular} & $K_{\mathcal{D}}(X)$ \\\hline\hline
    \texttt{mammals}         &   5153 &  796  &  9971 \\\hline
    \texttt{mammals} (g)     &   5161 & \begin{tabular}{c} 807\\ (0,6) \end{tabular}  & 10263 \\\hline\hline
    \texttt{languages}       &  13729 &  7769  & 202193 \\\hline
    \texttt{languages} (t)   &  15623 &  8716  & 229044 \\\hline
    \texttt{languages} (g)   &  13958 & \begin{tabular}{c} 11381\\ (43,2007) \end{tabular}  & 335025 \\\hline
    \texttt{languages} (g,t) &  16058 & \begin{tabular}{c} 10841\\ (146,985) \end{tabular}  & 304838 \\\hline\hline
    Minimal sufficient       & $\sum_{x} K^\star(x)$ & \begin{tabular}{c} $K^\star(X)$ \\ (opt,alt) \end{tabular} & $K_{\hat{\mathcal{D}}}(X)$ \\\hline\hline
    \texttt{mammals}         &   4531 &  877  & 11324 \\\hline
    \texttt{mammals} (g)     &   4557 & \begin{tabular}{c} 882\\ (0,3) \end{tabular}  & 11520 \\\hline\hline
    \texttt{languages}       &  12583 &  7642 & 199740 \\\hline
    \texttt{languages} (t)   &  14033 &  8470 & 222867 \\\hline
    \texttt{languages} (g)   &  13242 & \begin{tabular}{c} 10740\\ (39,1612) \end{tabular}  & 311188 \\\hline
    \texttt{languages} (g,t) &  14794 & \begin{tabular}{c} 10281\\ (149,780) \end{tabular}  & 286274 \\\hline
  \end{tabular}
\end{table}

\begin{figure}[!ht]
  \centering
  \includegraphics[scale=.8]{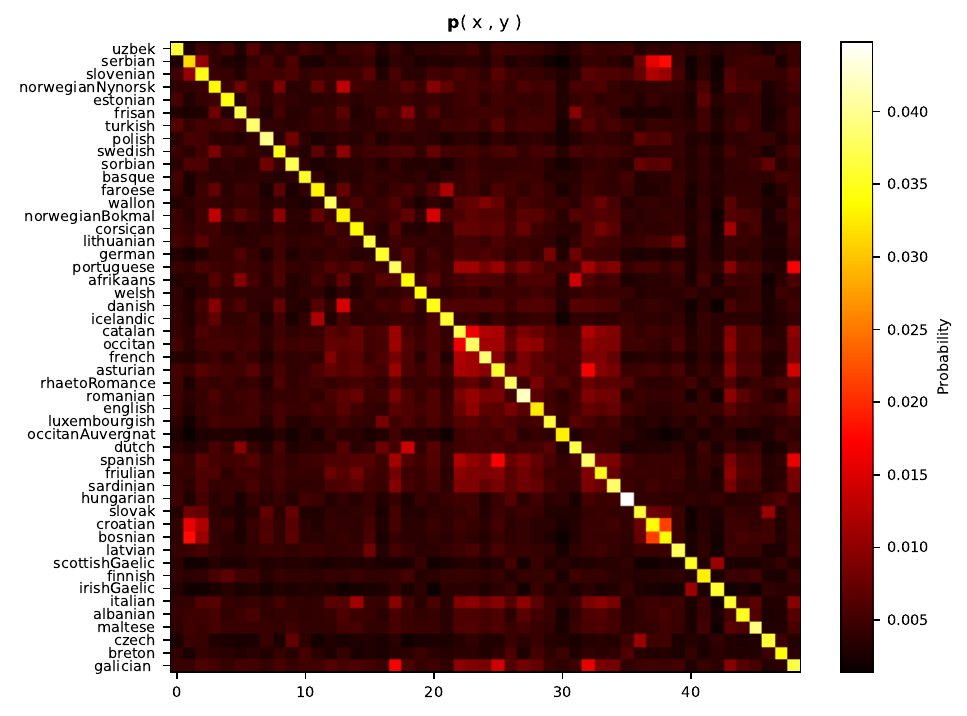}\\
  \includegraphics[scale=.8]{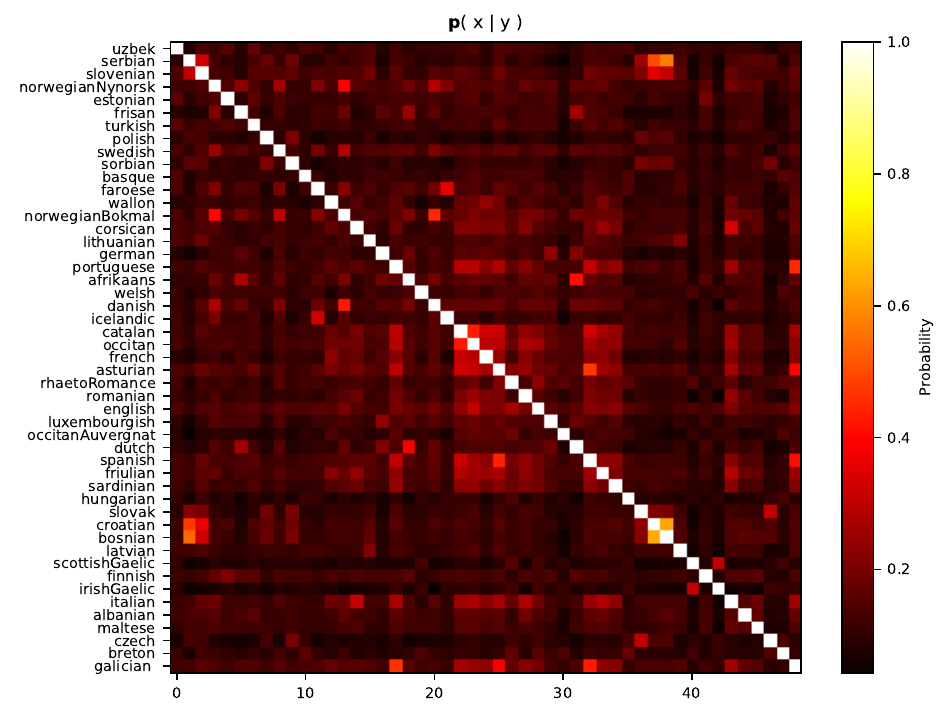}
  \caption{\prob( $x , y$ ) and \prob( $x\mid y$ ) on $\Omega=\texttt{languages}$.\label{fig:prob:cond}}
\end{figure}

\subsection{Information-theoretic distances}\label{sec:appli:dist}

For the last two applications, we write $K$ to denote either information measure above ($K^\star$ or $K_{\mathcal{D}}$). Our reference implementation supports four information-theoretic distances on collections of objects, starting with the Shannon distance~\cite{shannon:1953}:

\begin{equation*}
D_S(X,Y) \triangleq 2K(X,Y)-K(X)-K(Y), 
\end{equation*}
and the Information Distance~\cite{li:chen:2004}:
\begin{equation*}
\mbox{ID}(X,Y) \triangleq \max\left\{K(X\mid Y),K(Y\mid X)\right\}. 
\end{equation*}
Our default distance is the Normalized Information Distance (NID \cite{li:chen:2004}), which is known to minimize a host of other distances as well, and to be more accurate for objects of different sizes:
\begin{equation*}
\mbox{NID}(X,Y) \triangleq \frac{\max\left\{K(X\mid Y),K(Y\mid X)\right\}}{\max\left\{K(X),K(Y)\right\}}. 
\end{equation*}

For greater simplicity in the interface, we assume the user actually wants to compute distance matrices between pairs of files in set $X$ (and $Y$ if provided). Of course, the above can be easily implemented at a higher-level by compressing sets $X$ and $Y$ in separate files first if a distance between two datasets is needed. 

When one uses an off-the-shelf compressor, the NID has to be approximated by the Normalized Compression Distance~\cite{cilibrasi:2005}, parameterized by compressor $C$, that uses the concatenation $xy$ of two strings $x$ and $y$:
\begin{equation*}
\mbox{NCD}(x,y) \triangleq \frac{C(xy)-\min\left\{C(x),C(y)\right\}}{\max\left\{C(x),C(y)\right\}}. 
\end{equation*}

It is an interesting feature of our design that it supports native implementation of the NID, and that we can also implement the NCD on top of it. This allows for a seamless continuum of illustration regarding the computation of information-theoretic distances with parselets.

For fair comparison with off-the-shelf compressors, our NCD is computed using Eq.~\ref{eq:three:part} on the lossless models ($\hat{X}=X$), without generalization. Observe that in this case, we cannot guarantee symmetry of the distances anymore and our NCD becomes subject to small numerical discrepancies due to concatenation when computing Eq.~\ref{eq:three:part} for $xy$ {\em vs.} $yx$. 

\subsection{Illustration: Distance matrices}

Clustering is the only application to fairly compare with off-the-shelf compressors because the NCD above only needs concatenating two strings to use them~\cite{cilibrasi:2005}. We report in Tab.~\ref{tab:abstract:nid} the time to produce a distance matrix. Depending on the applicative workflow, one should sum the running times listed in Tabs.~\ref{tab:cache},~\ref{tab:abstract:nid} to fairly compare with off-the-shelf compressors. In any case, it would be the time of computing the first distance matrix from uncompressed data (and Tab.~\ref{tab:abstract:nid} is an approximate upper bound for subsequent computations in any subset of the same). 

Because we ensure symmetry of all the distances but the NCD, we can safely compute half of the distance matrix (and marginals are computed once). All these computations are spawned in parallel. The reference code for the NCD is not parallelized, so the figures reported for off-the-shelf compressors should be divided by 8 (the number of logical cores in our computer) to be more in line with our built-in parallelization. Computing NID/$K^\star$ matrices is the fastest that may be computed, while computing NID/$K_{\mathcal{\hat{D}}}$ adds up the time to write the compressed dictionaries in canonical order (except for caching or archiving, writing is only simulated in memory so we also save actual disk access). 

For the sake of uniformity, we used hierarchical clustering with Ward linkage to compare all compressors on all datasets. As already reported in~\cite{cilibrasi:2005}, all off-the-shelf compressors produced sound clusters on both datasets. In our case, Tab.~\ref{tab:abstract:nid} shows how \texttt{mpx} would make a terrible off-the-shelf compressor in practice (because of its quadratic-time deflation). 

\begin{table}[!ht]
  \caption{Wall-clock times for Normalized Information Distance matrices computations (figures averaged over models of Tab.~\ref{tab:cache}) {\em vs.} Normalized Compression Distance~\cite{cilibrasi:2005}. Average speed-up due to multi-threading was x6.8 for \texttt{mpx}: the code of the NCD should be also parallelized for fair comparison, even if \texttt{xz} activated multi-threading (x1.9) for \texttt{mammals}. Maximum Resident Set Size was 36MiB for \texttt{mpx} and 20MiB for NCD/\texttt{xz}.\label{tab:abstract:nid}}
    
  \centering
  \begin{tabular}{|l|c|c|}\hline
    Time (s) per dataset        & \texttt{mammals} & \texttt{languages} \\\hline\hline
    NCD/\texttt{gzip}           &  11.7            &   1.6 \\\hline
    NCD/\texttt{xz}             &  12.7            &  29.5 \\\hline
    NCD/\texttt{bzip2}          &   3.3            &   5.2 \\\hline
    NCD/\texttt{ppmd}           &   4.0            &  12.1 \\\hline
    NCD/\texttt{mpx}            &  33.6            &  64.4 \\\hline\hline
    NID/$K^\star$               &   0.1            &   0.5 \\\hline
    NID/$K_{\hat{\mathcal{D}}}$ &   0.3            &   1.2 \\\hline
  \end{tabular}
\end{table}

\begin{figure}[!ht]
  \centering
  \includegraphics[scale=.5]{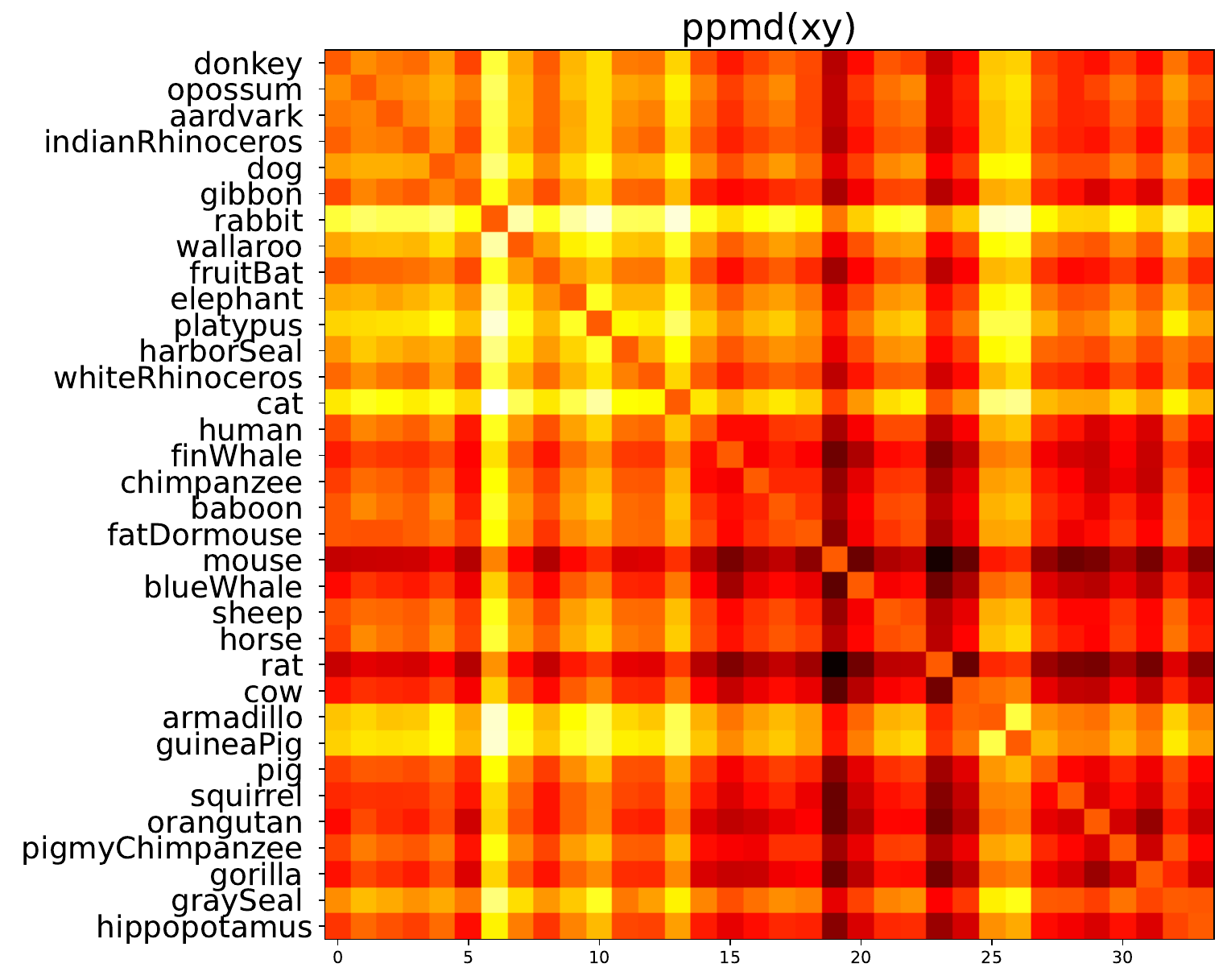}
  \includegraphics[scale=.5]{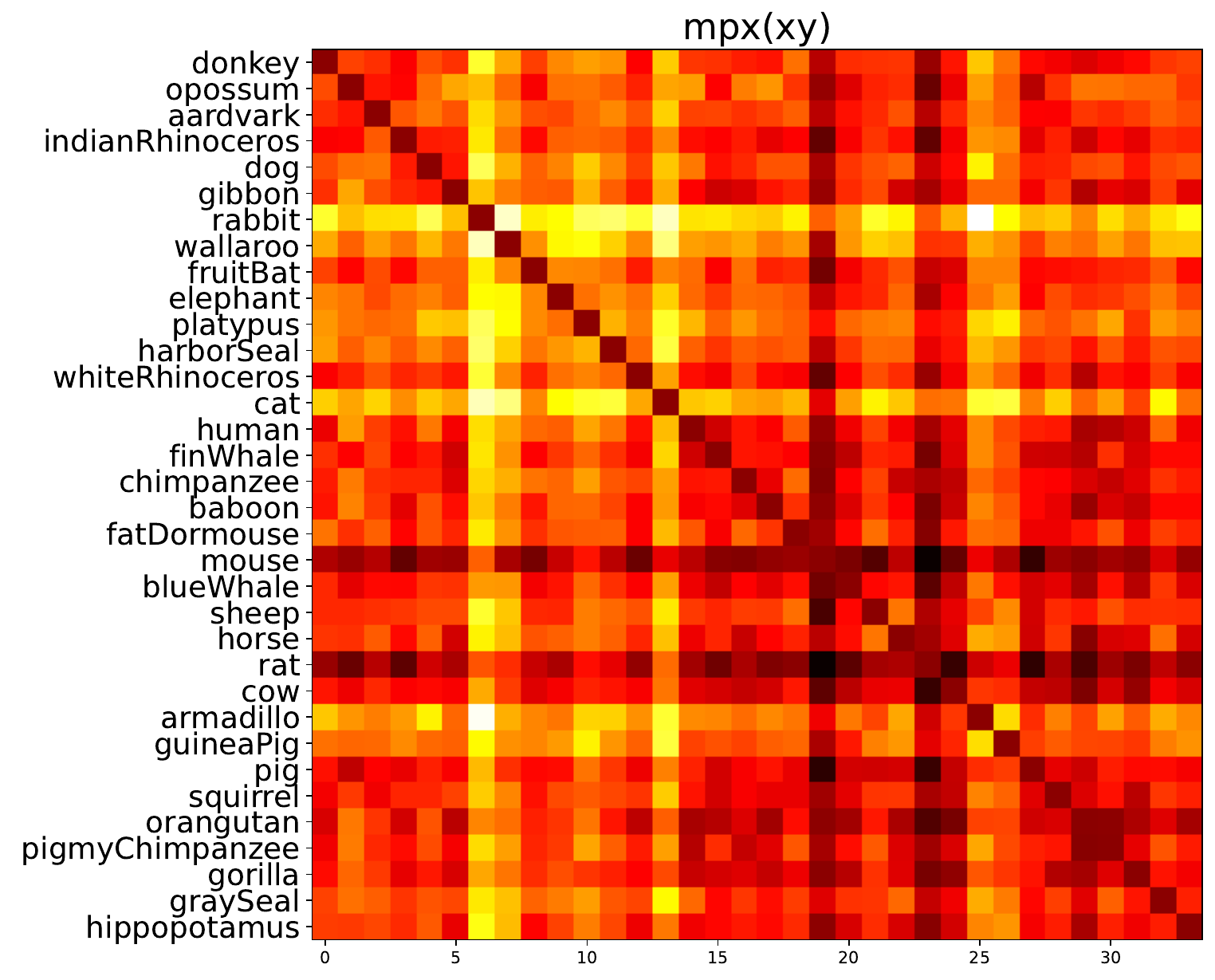}
  \caption{On \texttt{mammals}, \texttt{ppmd} and \texttt{mpx} learned the same structure for $C(xy)$, yet with much more noise than other compressors (their matrices look like those of Fig.~\ref{fig:dist:languages:NID}). The scale is different for both and the diagonal has been set to the mean value for more accurate rendering. \texttt{ppmd} had just enough data for clustering, while \texttt{mpx} needed more.\label{fig:Cxy:mammals}}
\end{figure}

As for \texttt{mpx}, the clustering quality was well below average for \texttt{mammals}, using any distance (which is supported by the flat codelength function of Fig.~\ref{fig:rd:cat}). Fig.~\ref{fig:Cxy:mammals} reports the matrices $C(xy)$ of the compressed concatenations for \texttt{mpx} and \texttt{ppmd} (as used in the NCD), which gave the noisiest distance matrix among off-the-shelf compressors. Both compressors learned the same structure for these matrices: \texttt{ppmd} likely had just enough data for sound clustering, while \texttt{mpx} needed some more. At last, this is all empirical evidence we can provide to support this hypothesis. 

Each compressor will learn the data differently: even if \texttt{mpx} compressed DNA better (see Tab.~\ref{tab:datasets} {\em vs.} Tab.~\ref{tab:cache}) than \texttt{gzip} on average (that was also verified for the concatenations $C(xy)$ when computing the NCD), closer inspection of the results showed \texttt{gzip} produced more significantly smaller bitstreams for $C(xy)$ than \texttt{mpx} (its distribution of $C(xy)$ showed no such outliers to the left). Given the internal architecture of \texttt{gzip} (the compressor with its LZ77 sliding window of 32KiB as the smallest internal state), this is indication that our design may need more data to better learn the objects separately (since it cannot rely on concatenation to make fainter similarities appear and be captured).

\begin{figure}[!ht]
  \centering
  \includegraphics[scale=.8]{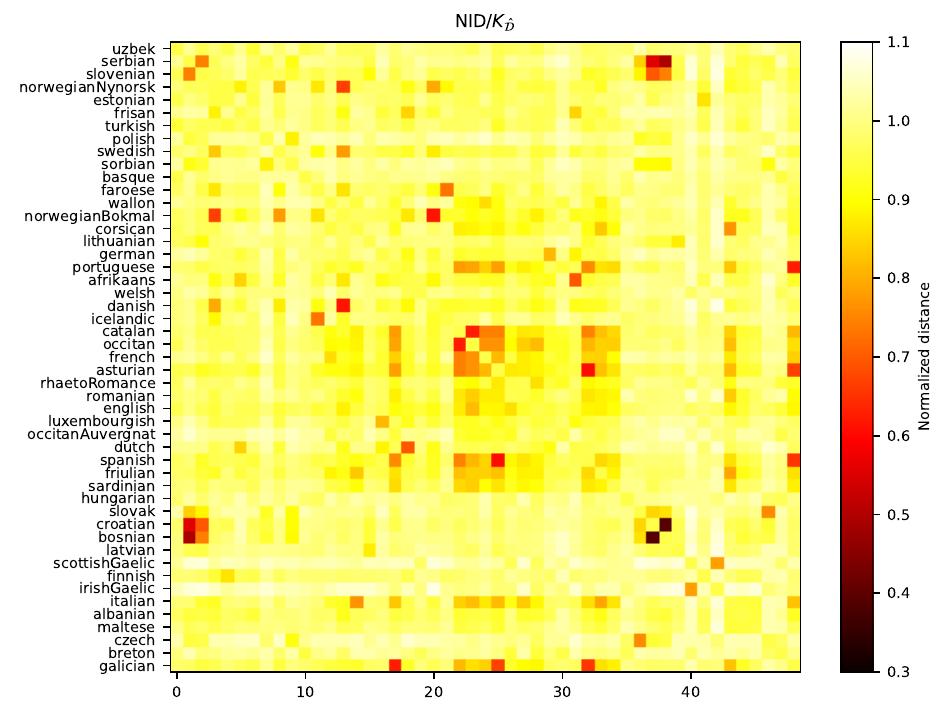}
  \includegraphics[scale=.8]{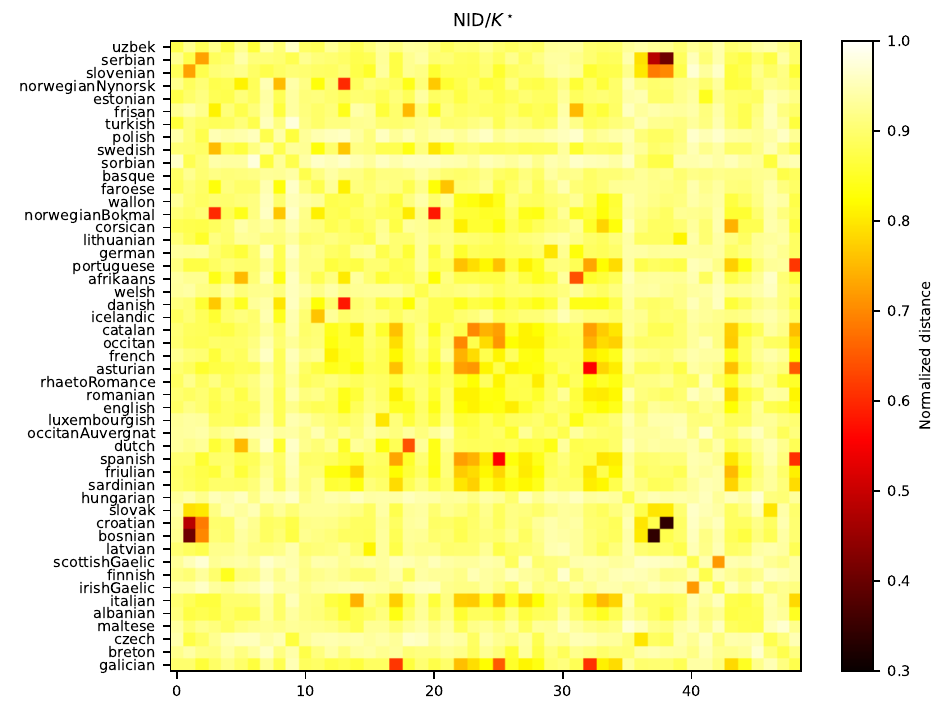}
  \caption{Top: NID/$K_{\mathcal{D}}$ on the lossless models of \texttt{languages}. Bottom: NID/$K^\star$ on the minimal sufficient models. We have indication that {\em (i)} string data is eventually mostly random (both captured the same structures), {\em (ii)} that the lossless models contain the significant information about the data, and {\em (iii)} that the minimal sufficient models retain most of it. Compared to the (unreported) distance matrix of NCD/\texttt{gzip}, parselets tend to produce more detailed differences.\label{fig:dist:languages:NID}}
\end{figure}

The situation is quite different with the \texttt{languages} dataset: \texttt{mpx} had enough data to learn it. We report in Fig.~\ref{fig:dist:languages:NID} the distance matrices for two information measures the NID may use on \texttt{languages}, all of which are truly symmetric (unlike the NCD). Only NID/$K^\star$ is normalized due to actual entropy-coding of the other. We compare two extreme regimes: NID/$K_{\mathcal{D}}$, arguably the most sensitive measure, on the lossless models {\em vs.} NID/$K^\star$, arguably the crudest of the two, on the minimal sufficient models. Empirical evidence suggests both setups capture the same structures.

\begin{figure}[!ht]
  \centering
  \includegraphics[scale=.9]{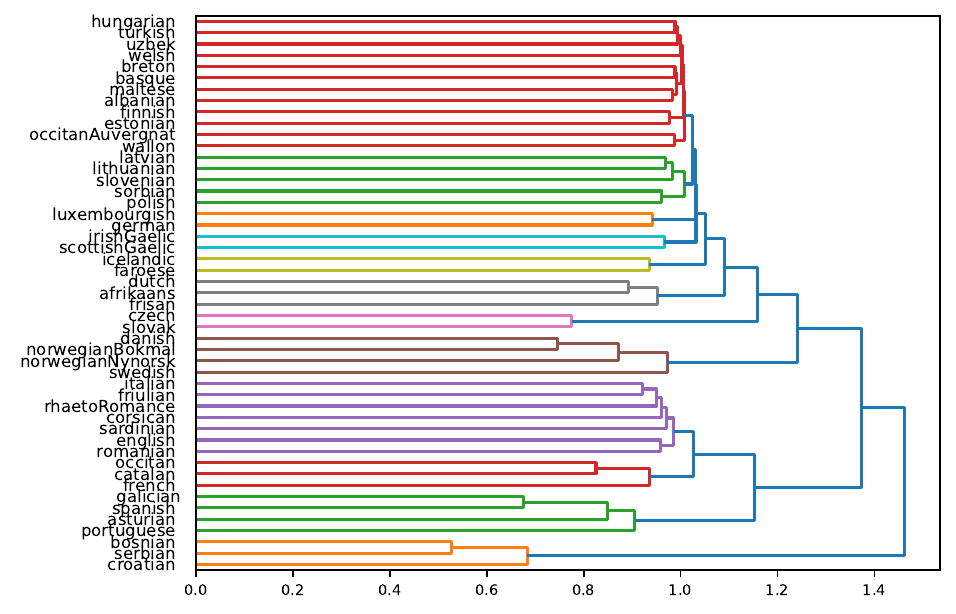}
  \includegraphics[scale=.9]{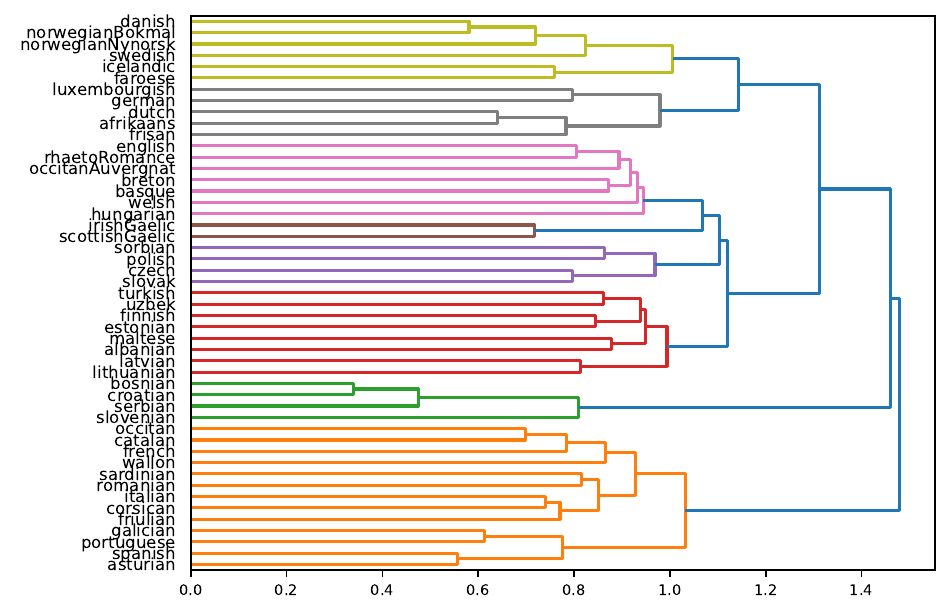}
  \caption{Hierarchical clusterings of \texttt{languages}, NCD/\texttt{gzip} (the best contender for off-the-shelf compressors here) {\em vs.} NID/$K^\star$ on the minimal sufficient models, to finally check we could safely discard both (insignificant) string data and noise in the model, even if we used the crudest information measure (distance matrix from bottom of Fig.~\ref{fig:dist:languages:NID}). This is the net effect of optimal, fully-algorithmic denoising combined with a purely combinatorial information measure on the lattice of most-significant parselets.\label{fig:languages:clustering:NCD:NID}}
\end{figure}

Finally, the clustering of NCD/\texttt{gzip} is compared with that of NID/$K^\star$ computed on the minimal sufficient models in Fig.~\ref{fig:languages:clustering:NCD:NID}. The cost of computing the minimal sufficient models is high, but computing the NID/$K^\star$ is the cheapest we have and it serves to illustrate the default behaviour when computing a normalized distance matrix.

Inspection of variants ($K_{\mathcal{D}}$ or $K^\star$) of the NID on all models (lossless minimal sufficient, with or without tokenization or generalization) showed no significant difference with Fig.~\ref{fig:languages:clustering:NCD:NID}. 

\subsection{Conditional independence statistics}\label{sec:appli:cond}

The information quantities are interfaced to the language of statistics so as to provide a host of conditional independence measures suitable for data analysis by compression. 

\begin{definition}[Conditional independence statistics]
  \begin{equation*}\label{eq:norm:cond:ind}
    I_{nd}(X;Y\mid Z) \triangleq \frac{I(X;Y\mid Z)}{\mbox{denom}(X,Y)},
  \end{equation*}
  where $\mbox{denom}(X,Y)$ encodes algorithmic equivalents of usual quantities in statistics, and may be:
\begin{itemize}
\item $K(X,Y)$ (dual correlation);
\item $K(X)+K(Y)$ (Pearson's correlation coefficient);
\item $\min\{K(X),K(Y)\}$ (total correlation);
\item $\sqrt{K(X)+K(Y)}$ (redundancy, by default).
\end{itemize}
\end{definition}

Just like for the NID, only $K^\star$ leads to numerically pristine values that are free from entropy-coding artifacts. 

\subsection{Illustration: inference of causal relationships}

When no intervention on the data is possible, the PC algorithm is the standard method of inferring causal relationships in data~\cite{colombo:2014}, based on a Markovian hypothesis for the causal structures. In this setting, the use of algorithmic information measures has been justified~\cite{janzing:2010} on the ground that Lauritzen semi-graphoid axioms hold provided the information measure is submodular~\cite{steudel:2010}.

The PC algorithm runs in cubic-time of the number of strings, by successively collapsing edges of an initial complete graph. The remaining edges may be oriented afterwards. The first step is called building the skeleton (undirected graph information in this context), and requires a conditional independence measure. This measure will involve more and more strings in the (conditioning) Markovian neighborhood under scrutiny, as the data is harder to decipher. The evaluation of the conditional independence measure is compared to threshold $\eta$ in order to declare independence.

\begin{figure}[!ht]
  \centering
  \includegraphics[scale=.3]{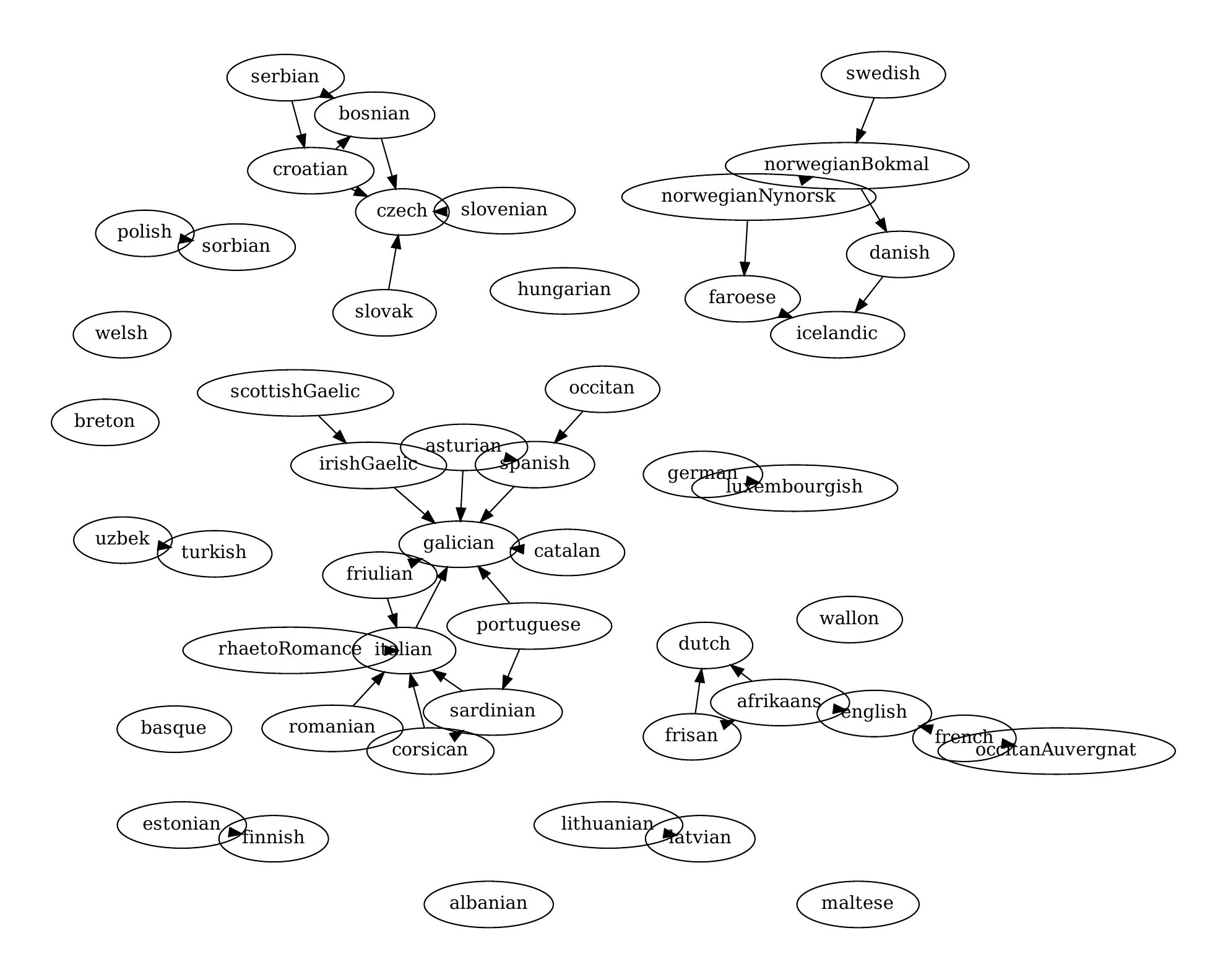}
  \includegraphics[scale=.3]{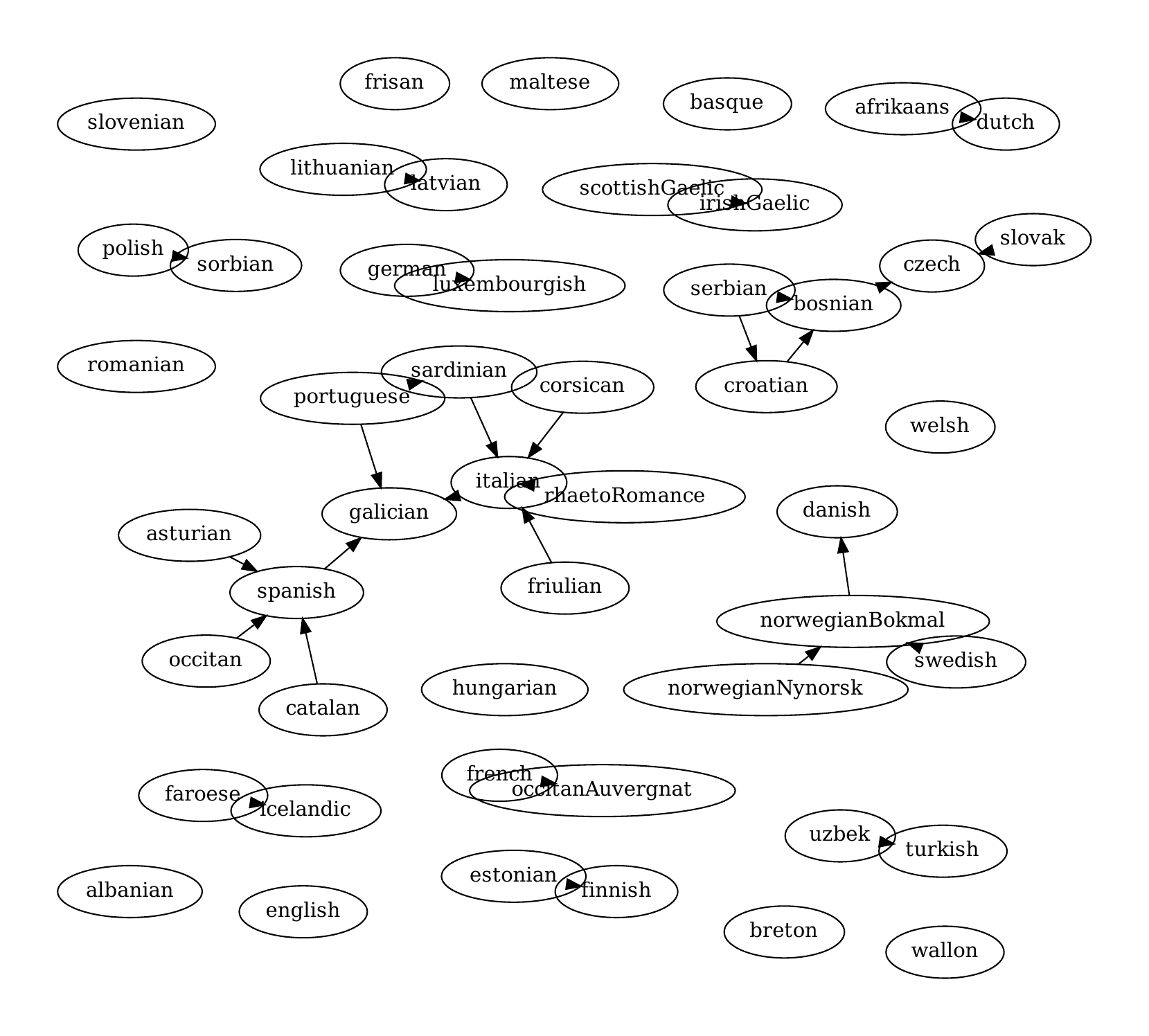}
  \includegraphics[scale=.3]{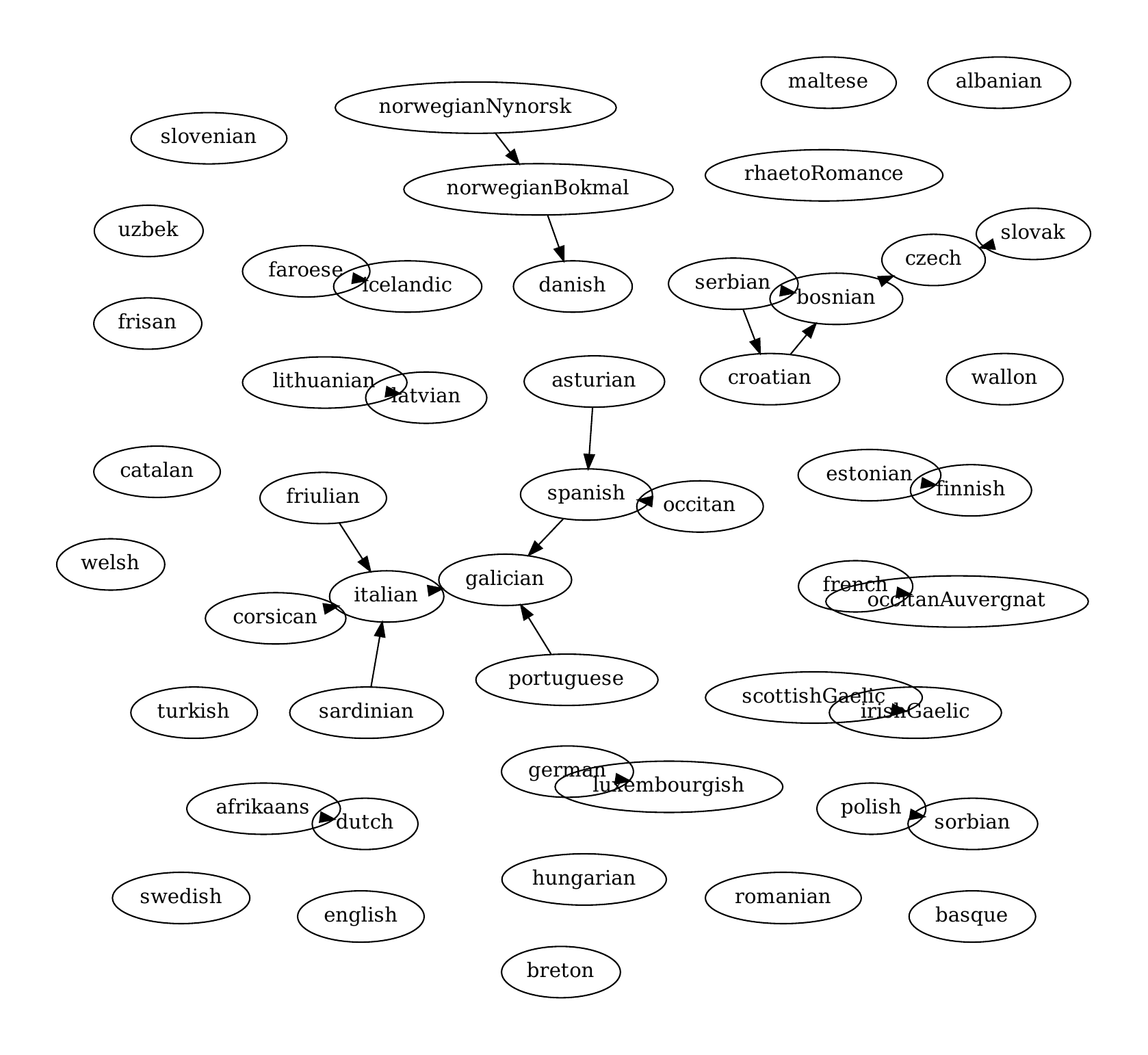}
  \caption{One-shot learning of Markovian causal dependencies on \texttt{languages} with $K_{\hat{\mathcal{D}}}$ (Kolmogorov complexity estimate of the tokenized, minimal sufficient models). Top: $\eta=2\%$ (11.4s), bottom-left: $\eta=3\%$ (8.7s), bottom-right: $\eta=4\%$ (6.7s).}
  \label{fig:pc:languages:KD}
\end{figure}

Of course, different kinds of data or parameter values will need different values for $\eta$. From top to bottom of Fig.~\ref{fig:pc:languages:KD}, we shall increase the value of $\eta$, so the middle graphs are arguably the best we may obtain by manual inspection of $\eta$, and the surrounding graphs provide a glimpse of the evolution around the sweet spot. We take this last opportunity to verify that tokenization does not do harm. On the \texttt{languages} dataset, our implementation of the PC algorithm gives less smooth results and it hardly parallelizes with $I^\star$ (only small Markovian neighborhoods need be checked), while $I_{\hat{\mathcal{D}}}$ caused deeper inspection of the data on larger Markovian neighborhoods so the overall parallelization speed-up was x1.8, and it is arguably the most sensitive measure of the two. The PC algorithm is one application of Algorithmic Information Theory that gets unlocked by parselets, on a seemingly larger scale than~\cite{steudel:2010}.

\subsection*{Acknowledgements}

The author is indebted to I. Sivignon and N. Le Bihan for their patience and comments on the early versions of this work. 

\bibliography{mpx}
\bibliographystyle{IEEEtran}

\appendices

\section{Basic multimodal support} \label{app:multimodal}

Technically, support for textual data is native through the use of tokens separators in Alg.~\ref{alg:compress}. Here, we show how to add support for signals and images. Because we are seeking an actual implementation, we must describe a reasonable interface with actual, regularly-sampled data stored in uncompressed form.

As far as compression is concerned, physical measures are rarely adequately represented in the time domain: compressing frequency-based representations is more natural for such data, as it fits better the capabilities of biological perception and the correlations, if any, in the data. We now describe two simple extensions of our compressor to support 1D and 2D regularly-sampled data: our parselet compressor is just about to replace the last entropy-coding stage of a traditional signal/image coder. Due to the limitation of the signal dynamic to 8 bits, the 2D case is expected to approximately match the behaviour of a true image coder, however the 1D case may only be used as a generic, 8-bit PCM input/output capability: the world of high-definition audio will stay out of reach.

From the algorithmic information point of view, such frequency-transforms are incorporated in the program (the compressed bitstream) as storing an optional set of pre/post-processing operations in a dedicated string header. We detect well-known file formats: \texttt{wav} for 8-bit signals and \texttt{pgm}/\texttt{ppm} for gray-level and color images. Everything else does not trigger pre/post-processing operations.

For the case of 1D, we may only expect a generic signal in time domain, properly pre-processed regarding acquisition and formatting. In the case of 2D, we assume a natural scene, bitmap image. Vector-graphics images obviously have to be treated as tokenized data, and palette images as plain string of bytes in order to take advantage of RLC coding in Alg.~\ref{alg:compress}.

\subsection{Selecting the transform}

We should find a way to reversibly map a signal or an image onto letters so we can use Alg.~\ref{alg:compress} directly. These transform letters should keep some kind of magnitude information (in the range $[0\dots|\mathcal{A}|-1]$). This discards most classical signal / image compression formats, since this magnitude information is either gained from a floating-point transform, or the output range of integer transforms expands well over the numerical input domain. 

To the best of our knowledge, there is only one suitable transform for our task in the literature: the socalled Piecewise-Linear Haar-like transform (PLHaar~\cite{senecal:2004}), which is separable, reversible and $n$-bit-to-$n$-bit. It implicitly and bijectively remaps the output range of a Haar transform onto the input range (so it may even be implemented as a fast static lookup table). In order to accomodate data from any sizes, we implemented 1D and 2D PLHaar block-transforms (with zero-padding). Our compressor therefore senses the physical world from a block-based, Haar-like representation with block size of 32 samples in any dimensions.

\subsection{Generic encoding of sampled data}

There are some fine lessons to learn from actual multimedia coding: it is generally a good idea to separate coarse and finer signal information inside the bitstream. While traditional compressors use this to achieve better compression and/or scalable content delivery, ours will mostly use it so that parselets are most likely created on parts of frequency-data with the same range of physical significance.

Hence, we implement the following simple and reversible pre/post-processing: the coarsest low-pass PLHaar coefficient of all blocks (1D or 2D) are grouped at the start of the byte string, and all other high-pass coefficients follow in time (1D) or raster-scan (2D) order, block after block. Other possible 2D scanning orders of wavelet subbands include zig-zag\footnote{The zig-zag scan in traditional \textsc{jpeg} addresses the different purpose of serializing samples along increasing spatial frequencies. We only need a space-filling curve for serialization, that preferably visits neighboring samples.} order, and following the Hilbert curve. We did not implement them however. 

In the specific case of color images, we first map pixels to $YC_bCr$ chromatic representation, and default lossy pre-processing starts with subsampling of the two $C_b,C_r$ chroma channels by a factor 2. This is the only occasion our compressor introduces irrecoverable loss of data. 

Our string header for describing the type of data associated to a compressed string in the bitstream is then:

\begin{itemize}
\item \texttt{0}: unstructured/textual data as bytes, no pre/post-processing operations;
\item \texttt{100 chans-1 freq length}: signal of \texttt{chans} channels of \texttt{length} samples at \texttt{freq} Hz ; 
\item \texttt{101 chans-1 width height}: 8-bit grayscale or color image.
\end{itemize}

The data type prefix codes above take provision for future extension to other types of data. Using the same parselet representation for bytes, texts, signals and images, likely makes our compressor one of the first computer program to implement a form of synesthesia (``homoiconicity'' was already coined for manipulating programs as data, and we only work at the string level). For instance, we could decode a signal as a text, and a distance between a book and a picture would be computed in a common syntactic representation for the two. 

\subsection{Illustration: image compression, lossless to lossy}

As for lossless ($\mathcal{D}_{\hat{x}}=\mathcal{D}_x$) image compression reported in Tab.~\ref{tab:lossless:lena}, decorrelation of the PLHaar transform allows \texttt{mpx} to catch up despite its average coding performance (disabling the PLHaar transform leads to a 62KiB bitstream). Rather surprisingly, using the \texttt{--delta=dist=3 --lzma2=pb=0} converter option of \texttt{xz} on Lena (to enable delta prediction coding on PCM bytes) gave a 50KiB bitstream.

\begin{table}[!ht]
  \caption{Lossless compression: \texttt{gzip} {\em vs.} \texttt{xz} {\em vs.} \texttt{mpx} (lengths in bytes as reported by the filesystem). Focus on generalization (g): our embodiment of Epicurus' Principle on Lena could not generalize within the default bounds. Maximum Resident Set Size (memory consumption): 9.3M (for 65536 8-bit pixels).\label{tab:lossless:lena}}
  \centering
  \begin{tabular}{|l||c|c||c|c|c|}\hline
    Name                   & \texttt{gzip} & \texttt{xz} & \texttt{mpx} &   sec & $|\mathcal{D}_x|$ \\\hline\hline
    \texttt{lena256}       &         58268 &       47960 &        48370 &  0.26 &       806         \\\hline
    \texttt{lena256} (g)   &           --- &         --- &        48370 &  0.46 &       806         \\\hline
  \end{tabular}
\end{table}

\begin{figure}[!ht]
  \centering
  \includegraphics[scale=.5]{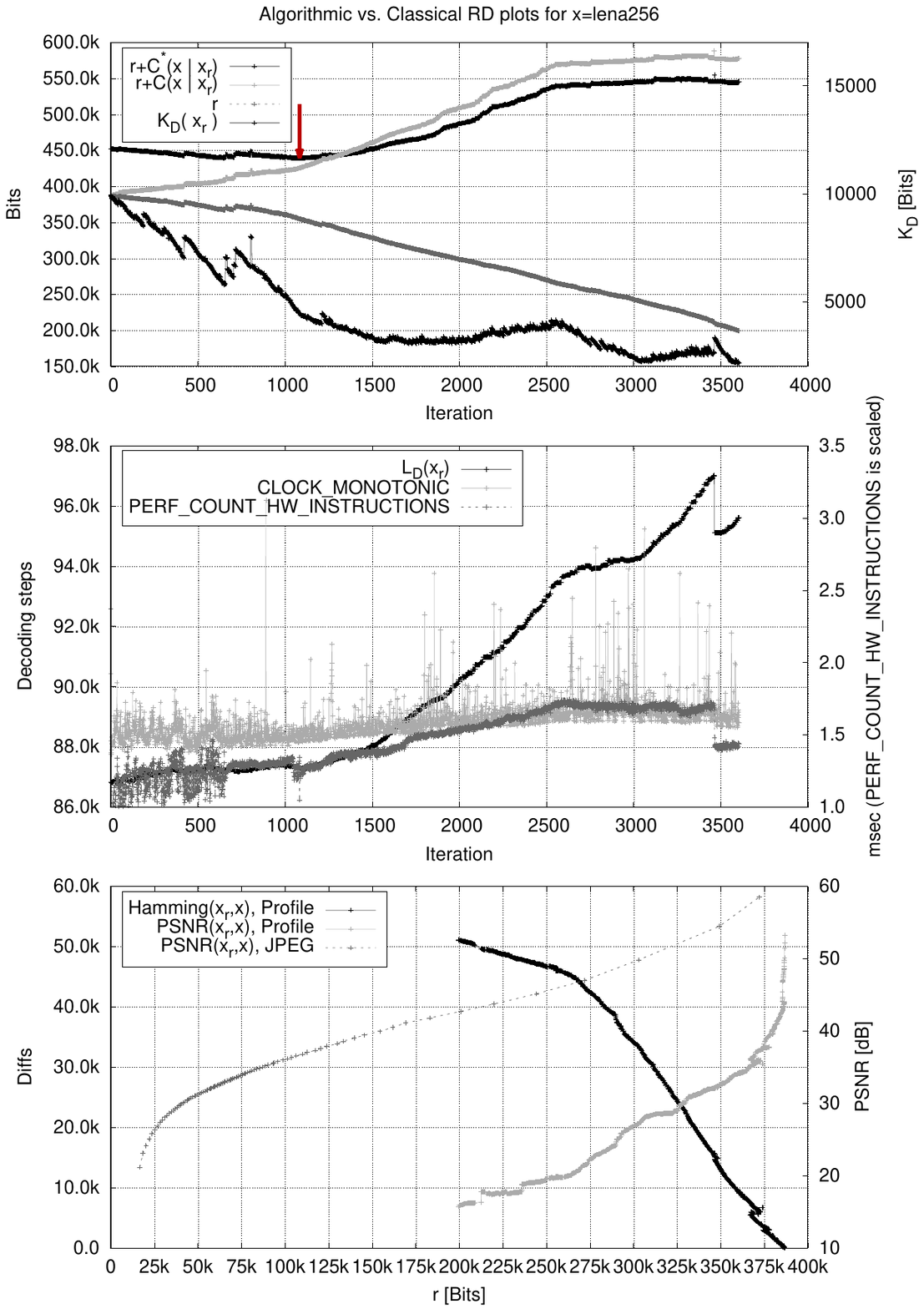}
  \caption{Rate-distortion plots for the \texttt{lena256} image (65536 graymap pixels): compression by parselets {\em vs.} \textsc{jpeg}. Lossless model: 806 parselets. Minimal sufficient model: 351 parselets at rate $r$ = 353.3kb (PSNR=33.28dB). Last model: 172 parselets (PSNR=15.73dB at $r$ = 198.2kb). Wall-clock time: 7m56s (full plot).}
  \label{fig:rd:lena}
\end{figure}

Now, to assess the behaviour of lossy image compression (by nearest-parselet syntactic quantization of Haar-like coefficients), we add the classical rate-PSNR plot for \texttt{lena256} in Fig.~\ref{fig:rd:lena}, including a comparison with standard, lossy \textsc{jpeg}. As an image lossy compressor, \texttt{mpx} does not apply psycho-visual masking, and it generally lacks many of the refinements of a true lossy image coder. However, its sense of the minimal sufficient model (optimal algorithmic denoising) is at PSNR = 33.28dB---compared to 35.38dB for \textsc{jpeg} at the default QF=75 quality factor value, which should be considered \textsc{jpeg}'s minimal sufficient model. In the lossy case, \texttt{mpx} would make a terrible traditional image coder. 

However, the classical rate-distortion plot confirms that even if image lossy compression is driven purely algorithmically, it also explores the classical rate-distortion space in some perceptually meaningful way (this is due to the PLHaar transform and the default $L_2$ norm to find the replacement parselet in Alg.~\ref{alg:compress}). Further work in this area should certainly be devoted to devise a lossy strategy on the PLHaar coefficients that is more in line with traditional lossy image compression strategies. 

\begin{figure}[!ht]
  \centering
  \includegraphics[scale=.55]{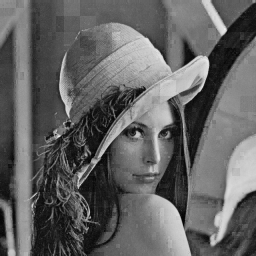}\\
  \vspace*{0.25cm}
  \includegraphics[scale=.55]{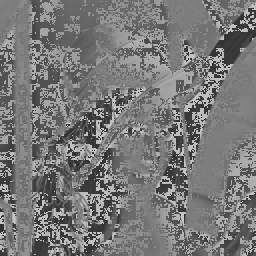}\\
  \vspace*{0.25cm}
  \includegraphics[scale=.55]{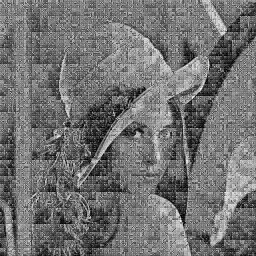}
  \caption{Top: Optimal, fully-algorithmic denoising of \texttt{lena256}: the PSNR with the original is at 36.19dB. \texttt{mpx} contains no psychovisual masking, so visible artifacts may appear (top of hat and shoulder). Middle: Denoising residual (magnified). Bottom: Image decompressed from the last iteration (PSNR=15.74dB) in Fig.~\ref{fig:rd:lena}. Alg.~\ref{alg:compress} could not generate a version with a simpler syntactic model (172 parselets).}
  \label{fig:rd:lena:visuals}
\end{figure}

In Fig.~\ref{fig:rd:lena:visuals}, we depict the magnified residual of the optimal, fully-algorithmic image denoising, and the last lossy image that Alg.~\ref{alg:compress} can reach to the bottom. Alg.~\ref{alg:compress} learned the $32\times 32$-pixel PLHaar block structure and blocking artifacts appear. Perceptually significant information has mostly been left untouched. Unreported experiments suggests that this is also the case with audio data. Even if the ear finds the discrepancies more noticeable and unpleasant than the eye, subjective perception agrees on close levels of discomfort for the two modalities.

\section{Serialization} \label{app:serialize}

This Appendix lists the last steps for serialization, in the order in which they are performed. They cover actually reading/writing Solomonoff archive $(\mathcal{D}_{\hat{X}}, \{\hat{x}_i\}, \{x_i\})$ obtained by Alg.~\ref{alg:kompress} from/to disk. 

\subsection{Parselet flushing} \label{app:serialize:flushing}

In full generality, string data of $\hat{x}$ may contain references to $\mathcal{D}_{\hat{X}}$ with \texttt{?} or \texttt{*} regexp operators, and we want to avoid explicit encoding of the regexp operator associated to each reference to $\mathcal{D}_{\hat{X}}$ in string data.

So we first have to flush residual occurrences of such operators in string slots. This is done by creating irregular parselets (with the relevant operator to flush) in $\mathcal{D}_{\hat{X}}$ and updating the references in the string slots. Now a string slot that is a reference to $\mathcal{D}_{\hat{X}}$ only contains an integer less than $|\hat{\mathcal{D}}|$. And string data now only contains unsigned integers.

Upon reading a compressed archive, such irregular parselets are detected, replaced on-the-fly, and eventually removed from $\mathcal{D}_{\hat{X}}$ so it only contains regular parselets, as expected.

\subsection{Coding the model in canonical order} \label{app:serialize:model}

We may now leverage the topological recursion property to rewrite $\mathcal{D}_{\hat{X}}$ in a canonical order that allows a compact representation of the model on disk.

First, we write regular (conjunctions, then disjunctions) parselets, followed by flushed (irregular) parselets. Disjunctions, conjunctions and flushed parselets are ordered by their left regexp operator (with the right regexp operator as secondary key if it exists), and ultimately by their leaves (which makes our canonical order encompass lexicographical order between parselets $p_0\sim p_1$).

\begin{algorithm}[!ht]
  \caption{Canonical comparison function for parselets $p_0,p_1$. Also provides an implementation of predicate $p_0\sim p_1$.}\label{alg:canonical}
\begin{algorithmic}[1]
\STATE {\textsc{canonical\_order}( $p_0$, $p_1$, lex ) }
\STATE \hspace{0.25cm}{\textbf{if} ( $\lnot$ $p_0$.regular $\land$ $p_1$.regular ) \textbf{return} 1}
\STATE \hspace{0.25cm}{\textbf{if} ( $p_0$.regular )}
\STATE \hspace{0.5cm}{\textbf{if} ( $\lnot$ $p_1$.regular ) \textbf{return} -1}
\STATE \hspace{0.5cm}{\textbf{if} ( $p_0$.conjunction $\land$ $\lnot$ $p_1$.conjunction ) \textbf{return} -1}
\STATE \hspace{0.5cm}{\textbf{if} ( $\lnot$ $p_0$.conjunction $\land$ $p_1$.conjunction ) \textbf{return} 1}
\STATE
\STATE \hspace{0.25cm}{\textbf{if} ( $\lnot$ $p_0$.left.set $\land$ $p_1$.left.set ) \textbf{return} -1}
\STATE \hspace{0.25cm}{\textbf{if} ( $p_0$.left.set ) }
\STATE \hspace{0.5cm}{\textbf{if} ( $\lnot$ $p_1$.left.set ) \textbf{return} 1}
\STATE \hspace{0.5cm}{\textbf{if} ( $p_0$.left.option $\land$ $p_1$.left.repetition ) \textbf{return} -1}
\STATE \hspace{0.5cm}{\textbf{if} ( $p_0$.left.repetition $\land$ $p_1$.left.option ) \textbf{return}  1}
\STATE
\STATE \hspace{0.25cm}{\textbf{if} ( $\lnot$ $p_0$.regular ) }
\STATE \hspace{0.5cm}{\textbf{if} ( $\lnot$ $p_0$.left.atomic $\land$ $p_1$.left.atomic ) \textbf{return}  1 }
\STATE \hspace{0.5cm}{\textbf{if} ( $p_0$.left.atomic ) }
\STATE \hspace{0.75cm}{\textbf{if} ( $\lnot$ $p_1$.left.atomic ) \textbf{return} -1 }
\STATE \hspace{0.75cm}{\textbf{if} ( lex ) }
\STATE \hspace{1cm}{\textbf{if} ( $p_0$.left.id $>$ $p_1$.left.id ) \textbf{return}  1 }
\STATE \hspace{1cm}{\textbf{if} ( $p_0$.right.id $<$ $p_1$.right.id ) \textbf{return} -1 }
\STATE \hspace{0.75cm}{\textbf{return} 0 }
\STATE
\STATE \hspace{0.25cm}{left\_diff$\leftarrow$\textsc{canonical\_order}( $p_0$.left, $p_1$.left, lex )}
\STATE \hspace{0.25cm}{\textbf{if} ( $\lnot$ $p$.regular $\lor$ left\_diff ) \textbf{return} left\_diff}
\STATE
\STATE \hspace{0.25cm}{\textbf{if} ( $\lnot$ $p_0$.right.set $\land$ $p_1$.right.set ) \textbf{return} -1}
\STATE \hspace{0.25cm}{\textbf{if} ( $p_0$.right.set ) }
\STATE \hspace{0.5cm}{\textbf{if} ( $\lnot$ $p_1$.right.set ) \textbf{return} 1}
\STATE \hspace{0.5cm}{\textbf{if} ( $p_0$.right.option $\land$ $p_1$.right.repetition ) \textbf{return} -1}
\STATE \hspace{0.5cm}{\textbf{if} ( $p_0$.right.repetition $\land$ $p_1$.right.option ) \textbf{return}  1}
\STATE
\STATE \hspace{0.25cm}{\textbf{if} ( $\lnot$ $p_0$.right.atomic $\land$ $p_1$.right.atomic ) \textbf{return}  1 }
\STATE \hspace{0.25cm}{\textbf{if} ( $p_0$.right.atomic ) }
\STATE \hspace{0.5cm}{\textbf{if} ( $\lnot$ $p_1$.right.atomic ) \textbf{return} -1 }
\STATE \hspace{0.5cm}{\textbf{if} ( lex ) }
\STATE \hspace{0.75cm}{\textbf{if} ( $p_0$.right.id $<$ $p_1$.right.id ) \textbf{return} -1 }
\STATE \hspace{0.75cm}{\textbf{if} ( $p_0$.right.id $>$ $p_1$.right.id ) \textbf{return}  1 }
\STATE \hspace{0.5cm}{\textbf{return} 0 }
\STATE \hspace{0.25cm}{\textbf{return} \textsc{canonical\_order}( $p_0$.right, $p_1$.right, lex )}
\STATE
\STATE {\textsc{parselet\_sim}( $p_0$, $p_1$ ) }
\STATE \hspace{0.25cm}{\textbf{return} $\lnot$ \textsc{canonical\_order}( $p_0$, $p_1$, 0 ) }
\STATE
\STATE {\textsc{parselet\_compare}( $p_0$, $p_1$ ) }
\STATE \hspace{0.25cm}{\textbf{return} \textsc{canonical\_order}( $p_0$, $p_1$, 1 ) }
\end{algorithmic}
\end{algorithm}

After the relevant indices permutation to canonical order has been found (by repurposing a standard sorting function with the comparison function in Alg.~\ref{alg:canonical}), string data $\{\hat{x}_i\}$ eventually gets transcoded one last time. This is done by the \textsc{canonical\_sort}( $\mathcal{D}_{\hat{X}}$, $\{\hat{x}_i\}$ ) primitive, which is omitted for brevity. 

In canonical order, parselets are listed as $2\times 3\times 3 + 2$ groups of identical syntactic structure of increasing complexity, so we only need listing their group size and identifier(s). We anticipe that writing down a zero size will cost one zero bit. Adding two zero bits for zero conjunctions and disjunctions, to the $2\times 9$ zero bits for each empty syntactic group of conjunctions and disjunctions, requires subtracting 20 bits to enforce $C({\mathcal{D}_{\hat{X}}})=0$ bits in Eqs.~\ref{eq:three:part}-\ref{eq:codelength}-\ref{eq:measure:complexity} if $\mathcal{D}_{\hat{X}}=\emptyset$. 

\begin{algorithm}[!ht]
  \caption{Model serialization. Returns evaluation of $C(\mathcal{D}_{\hat{X}})$ for Eqs.~\ref{eq:three:part}-\ref{eq:codelength}-\ref{eq:measure:complexity}. }\label{alg:write:model}
  \begin{algorithmic}[1]
    \STATE {\textsc{write\_flushed}( $\mathcal{D}_{\hat{X}}$, pred ) }
    \STATE \hspace{0.25cm}{$j$ $\leftarrow$ $n$ }
    \STATE \hspace{0.25cm}{\textbf{while} ( $\mathcal{D}_{\hat{X}}[j]$.left.pred ) $j$++ }
    \STATE \hspace{0.25cm}{\textsc{write\_const} ( $j-n$ ) }
    \STATE \hspace{0.25cm}{\textbf{while} ( $n$ $<$ $j$ ) }
    \STATE \hspace{0.5cm}{\textsc{write\_int} ( $\mathcal{D}_{\hat{X}}[n]$.left.id ) }
    \STATE \hspace{0.5cm}{$n$++ }
    \STATE
    \STATE {\textsc{write\_reg}( $\mathcal{D}_{\hat{X}}$, pred0, pred1 ) }
    \STATE \hspace{0.25cm}{$j$ $\leftarrow$ $n$ }
    \STATE \hspace{0.25cm}{\textbf{while} ( $\mathcal{D}_{\hat{X}}[j]$.left.pred0 $\land$ $\mathcal{D}_{\hat{X}}[j]$.right.pred1 ) $j$++ }
    \STATE \hspace{0.25cm}{\textsc{write\_const} ( $j-n$ ) }
    \STATE \hspace{0.25cm}{\textbf{while} ( $n$ $<$ $j$ ) }
    \STATE \hspace{0.5cm}{\textsc{write\_int} ( $\mathcal{D}_{\hat{X}}[n]$.left.id ) }
    \STATE \hspace{0.5cm}{\textsc{write\_int} ( $\mathcal{D}_{\hat{X}}[n]$.right.id ) }
    \STATE \hspace{0.5cm}{$n$++ }
    \STATE
    \STATE {\textsc{write\_groups\_reg}( $\mathcal{D}_{\hat{X}}$ ) }
    \STATE \hspace{0.25cm}{\textsc{write\_reg} ( $\mathcal{D}_{\hat{X}}$, $\lnot$.set, $\lnot$.set ) }
    \STATE \hspace{0.25cm}{\textsc{write\_reg} ( $\mathcal{D}_{\hat{X}}$, $\lnot$.set, .option ) }
    \STATE \hspace{0.25cm}{\textsc{write\_reg} ( $\mathcal{D}_{\hat{X}}$, $\lnot$.set, .repetition ) }
    \STATE \hspace{0.25cm}{\textsc{write\_reg} ( $\mathcal{D}_{\hat{X}}$, .option, $\lnot$.set ) }
    \STATE \hspace{0.25cm}{\textsc{write\_reg} ( $\mathcal{D}_{\hat{X}}$, .option, .option ) }
    \STATE \hspace{0.25cm}{\textsc{write\_reg} ( $\mathcal{D}_{\hat{X}}$, .option, .repetition ) }
    \STATE \hspace{0.25cm}{\textsc{write\_reg} ( $\mathcal{D}_{\hat{X}}$, .repetition, $\lnot$.set ) }
    \STATE \hspace{0.25cm}{\textsc{write\_reg} ( $\mathcal{D}_{\hat{X}}$, .repetition, .option ) }
    \STATE \hspace{0.25cm}{\textsc{write\_reg} ( $\mathcal{D}_{\hat{X}}$, .repetition, .repetition ) }
    \STATE
    \STATE {$n\leftarrow|\alphabet|$ }
    \STATE
    \STATE {\textsc{write\_model}( $\mathcal{D}_{\hat{X}}$, $\{\hat{x}_i\}$ ) }
    \STATE
    \STATE \hspace{0.25cm}{\textsc{canonical\_sort} ( $\mathcal{D}_{\hat{X}}$, $\{\hat{x}_i\}$ ) }
    \STATE
    \STATE \hspace{0.25cm}{$C\leftarrow$ \textsc{sys}.written }
    \STATE \hspace{0.25cm}{\textsc{write\_const} ( $\mathcal{D}_{\hat{X}}$.conjunctions ) }
    \STATE \hspace{0.25cm}{\textsc{write\_groups\_reg} ( $\mathcal{D}_{\hat{X}}$ ) }
    \STATE
    \STATE \hspace{0.25cm}{\textsc{write\_const} ( $\mathcal{D}_{\hat{X}}$.disjunctions ) }
    \STATE \hspace{0.25cm}{\textsc{write\_groups\_reg} ( $\mathcal{D}_{\hat{X}}$ ) }
    \STATE \hspace{0.25cm}{$C\leftarrow$ \textsc{sys}.written - $C$ - 20 }
    \STATE
    \STATE \hspace{0.25cm}{\textsc{write\_const} ( $\mathcal{D}_{\hat{X}}$.flushed ) }
    \STATE \hspace{0.25cm}{\textsc{write\_flushed} ( $\mathcal{D}_{\hat{X}}$, .option ) }
    \STATE \hspace{0.25cm}{\textsc{write\_flushed} ( $\mathcal{D}_{\hat{X}}$, .repetition ) }
    \STATE
    \STATE \hspace{0.25cm}{\textbf{return } $C$ }
  \end{algorithmic}
\end{algorithm}

\subsection{Coding patching information} \label{app:serialize:patching}

We tackle the algorithmic treatment of $\underline{x}\mid\hat{x}$ for Eq.~\ref{eq:three:part}. This is needed for actual, lossless storage of the data but not for evaluations of information quantities, so it is really a serialization issue. To keep in the spirit of this work, we describe a ``shortest program'' approach to patching, the ``abstract list'' approach was needed for Eq.~\ref{eq:codelength}. 

We have to extend our representation to support the idea that an input noisy string $x$ should be programmatically represented with $r$ bits (the rate) for its approximation by some $x_r$, augmented with some $\underline{x}\mid x_r$ patching information that allows to reconstruct $\underline{x}$ from $x_r$ (and $\mathcal{D}_r$). This is where we nest the third code of our representation to actually implement Eq.~\ref{eq:three:part}.

The fact that the representation of $x_r$ is augmented with nested patching information, allows to monitor (during reading or writing) the number of bits that are used for either part of the representation of $x$, be it the $r$ bits of $x_r$ or those of $\underline{x}\mid x_r$. Eventually, we are able to generate either $\underline{x}_r$ or $\underline{x}$ from the compressed bitstream of $x$.

The first technical step is to describe how to encode the patching information on disk. The problem at hand is called string-to-string correction, and the patching information should exploit the recursive structure of parselets to enforce a minimum-length principle.

\begin{figure}[!ht]
  \centering
  \begin{tikzpicture}[
      level 1/.style={sibling distance=40mm},
      level 2/.style={sibling distance=40mm},
      level 2/.style={sibling distance=21mm},
      level distance=1cm,
    ]

    \node[matrix] (dict) {
      \node{\texttt{0x4}}; & \node{$\rightarrow$}; & \node{\texttt{a* b}};\\
      \node{\texttt{0x5}}; & \node{$\rightarrow$}; & \node{\texttt{c* d}};\\
      \node{\texttt{0x6}}; & \node{$\rightarrow$}; & \node{\texttt{0x4* 0x5}};\\
    };

    \node[matrix,right=of dict,xshift=-.5cm] (strings) {
      \node{$\underline{x}$}; & \node{=}; & \node{\texttt{aabaacbccd}};\\
      \node{$\underline{x}_r$}; & \node{=}; & \node{\texttt{aabaaabccd}};\\
      \node{$x_r$}; & \node{$\rightarrow$}; & \node{\texttt{0x6} : 2 2 3 2};\\
    };

    \node[below=of dict.east] (tree) { \texttt{0x6} \textit{\underline{1}} }
    child{ node {\texttt{0x4*} [\textit{\underline{0}},\textit{\underline{1}}]}
      child{ node (bottom) {\texttt{a*} [[\textit{0},\textit{0}],[\textit{\underline{0}},\textit{\underline{0}},\textit{\underline{1}}]] } }
      child{ node {\texttt{b} [[\textit{0}],[\textit{\underline{0}}]]} }
    }
    child{ node {\texttt{0x5} [\textit{\underline{0}}]}
      child{ node {\texttt{c*} [[\textit{0}],[\textit{0}]]} }
      child{ node {\texttt{d} [[\textit{0}]]} }
    };

    \node[below=of bottom,anchor=west] {
      $x\rightarrow$\texttt{0x6} : \textit{\underline{1}} 2 \textit{\underline{0}} 2 \textit{\underline{1}} 3 \textit{\underline{0}} \textit{\underline{0}} \textit{\underline{1}} \texttt{\underline{c-a}} \textit{\underline{0}} 2};

  \end{tikzpicture}
\caption{Patching $x_r$ with diff information to reconstruct $\underline{x}$. Diff label bits are typeset in italic roman font to the right of nodes, brackets indicate nested levels during parselet nodes traversal. Underlined bits is all that needs be written to disk to locate letter differences. Other zero bits are implicit because their parent nodes have a zero diff bit. Observe that also changing the second \texttt{b} in $x$ would benefit from factorizing the diff location information during parselet traversal. \label{fig:patching}}
\end{figure}
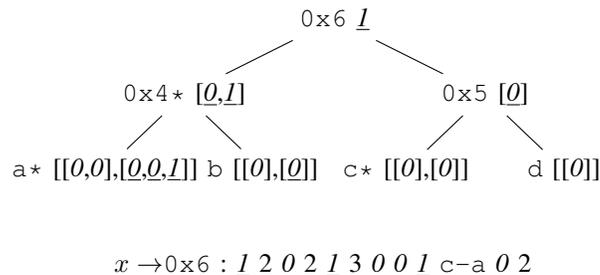

As a serialization step, patching is to be applied upon reading/writing compressed data. As a consequence, we do not need to actually include patching information within string slots sequence: it is computed/decoded on-the-fly. When we specify the full decoder however, we must assume patching information is interspersed within the bitstream. Also, we need to save patching information associated to each parselet, for it is used in the very last encoding step described in Sec.~\ref{app:serialize:marginal}. Fig.~\ref{fig:patching} depicts a one-parselet example of how previous syntactic elements of the bitstream (that generate $x_r$) are interspersed with (underlined) patching information (so $\underline{x}$ may eventually be decompressed). 

When seen as a binary tree (see Fig.~\ref{fig:patching}), every node of a parselet traversed during decoding of $x_r$ is labeled with a bit indicating whether any of its left or right child encodes a different letter than that in the current decoding position of $\underline{x}$. As a minimum-length principle, we seek to avoid encoding diff bits for identical subtrees in $x_r$ and $\underline{x}$, which we know they are when we read their zero parent diff bit during depth-first traversal of the dictionary. Recalling an exact occurrence of a parselet (assuming the same regexp operator values) therefore only costs an additional zero bit. Disjunctive parselets do not need a diff bit (only their selected alternative). The same applies to unselected \texttt{?} regexp options. 

We may now specify the actual decoder, that works on \textit{patched parselets}. Alg.~\ref{alg:decompress} is the pseudo-code for the decompressor of patched parselets, which describe the data following a general, rate-distortion-based representation.

Not depicted is a final run-length coding of zero bits to the next diff bit set, along the occurrences of references in string data. When there is no patching information ($x_r=x$, which is how our compressed format encompasses a purely lossless representation of the data), we only code an integer that is the number of occurrences of references in string data. When a run of zero diff bits has been exhausted, the decoder expects the size of the next run of zero diff bits before proceeding to actual decoding. Describing this last level of coding in Alg.~\ref{alg:decompress} would obfuscate the exposition too much. The length of the first run is discarded when monitoring reading/writing, so the patch counts for zero bit when $\hat{x}=x$ (lossless model).

The decompressor is called with \textsc{decompress}( $\epsilon$, $x_r$, lossy ) on compressed $x_r$, where the last argument selects either uncompressed $\underline{x}_r$ or $\underline{x}$ as output.

\begin{algorithm}[!ht]
\caption{Patched parselet decompressor. Compressed patched input $x_r$, $\underline{x}\mid x_r$ is either read from disk or memory to reconstruct either uncompressed $\underline{x}$ or $\underline{\hat{x}}$ depending on parameter lossy.}\label{alg:decompress}
\begin{algorithmic}[1]
\STATE {\textsc{dec\_patched}( $\underline{x}$, $x_r$, $p$, diff, lossy )}
\STATE \hspace{0.25cm}{\textbf{if} ( $p$.disjunction )}
\STATE \hspace{0.5cm}{$p\leftarrow$ \textsc{next\_bit( $x_r$ )} ? $p$.right : $p$.left }
\STATE \hspace{0.5cm}{\textbf{return} \textsc{dec\_patched}( $\underline{x}$, $x_r$, $p$, diff ) }
\STATE 
\STATE \hspace{0.25cm}{occ $\leftarrow$ 1 }
\STATE \hspace{0.25cm}{\textbf{if} ( $p$.left.option ) occ$\leftarrow$ \textsc{next\_bit}( $x_r$ ) }
\STATE \hspace{0.25cm}{\textbf{if} ( $p$.left.repetition ) occ$\leftarrow$ \textsc{next\_int}( $x_r$ ) }
\STATE \hspace{0.25cm}{\textbf{while} ( occ-{}- ) }
\STATE \hspace{0.5cm}{diff $\leftarrow$ diff ? \textsc{next\_bit}( $x_r$ ) : 0 }
\STATE \hspace{0.5cm}{\textbf{if} ( $p$.left.atomic ) }
\STATE \hspace{0.75cm}{d $\leftarrow$ diff ? \textsc{next\_int}( $x_r$ ) : 0 }
\STATE \hspace{0.75cm}{\textsc{insert\_after}( $\underline{x}$, $p$.left.id + ( lossy ? 0 : d ) )}
\STATE \hspace{0.5cm}{\textbf{else} }
\STATE \hspace{0.75cm}{\textsc{dec\_patched}( $\underline{x}$, $x_r$, $p$.left, diff, lossy )}
\STATE \hspace{0.25cm}{\textbf{if} ( $\lnot$ $p$.regular ) \textbf{return}}
\STATE 
\STATE \hspace{0.25cm}{occ $\leftarrow$ 1 }
\STATE \hspace{0.25cm}{\textbf{if} ( $p$.right.option ) occ$\leftarrow$ \textsc{next\_bit}( $x_r$ ) }
\STATE \hspace{0.25cm}{\textbf{if} ( $p$.right.repetition ) occ$\leftarrow$ \textsc{next\_int}( $x_r$ ) }
\STATE \hspace{0.25cm}{\textbf{while} ( occ-{}- ) }
\STATE \hspace{0.5cm}{diff $\leftarrow$ diff ? \textsc{next\_bit}( $x_r$ ) : 0 }
\STATE \hspace{0.5cm}{\textbf{if} ( $p$.right.atomic ) }
\STATE \hspace{0.75cm}{d $\leftarrow$ diff ? \textsc{next\_int}( $x_r$ ) : 0 }
\STATE \hspace{0.75cm}{\textsc{insert\_after}( $\underline{x}$, $p$.right.id + ( lossy ? 0 : d ) )}
\STATE \hspace{0.5cm}{\textbf{else} }
\STATE \hspace{0.75cm}{\textsc{dec\_patched}( $\underline{x}$, $x_r$, $p$.right, diff, lossy )}
\STATE 
\STATE {\textsc{decompress}( $\underline{x}$, $x_r$, lossy ) }
\STATE \hspace{0.25cm}{$s$ $\leftarrow$ \textsc{next\_ref}( $x_r$ ) }
\STATE \hspace{0.25cm}{\textbf{if} ( $s$.EndOfString ) \textbf{return} $\underline{x}$}
\STATE 
\STATE \hspace{0.25cm}{occ $\leftarrow$ 1 }
\STATE \hspace{0.25cm}{\textbf{if} ( $p$.option ) occ$\leftarrow$ \textsc{next\_bit}( $x_r$ ) }
\STATE \hspace{0.25cm}{\textbf{if} ( $p$.repetition ) occ$\leftarrow$ \textsc{next\_int}( $x_r$ ) }
\STATE \hspace{0.25cm}{\textbf{while} ( occ-{}- ) }
\STATE \hspace{0.5cm}{diff $\leftarrow$ \textsc{next\_bit}( $x_r$ ) }
\STATE \hspace{0.5cm}{\textbf{if} ( $s$.atomic ) }
\STATE \hspace{0.75cm}{diff $\leftarrow$ diff ? \textsc{next\_int}( $x_r$ ) : 0 }
\STATE \hspace{0.75cm}{\textsc{insert\_after}( $\underline{x}$, $s$.id + ( lossy ? 0 : diff ) ) }
\STATE \hspace{0.5cm}{\textbf{else} \textsc{dec\_patched}( $\underline{x}$, $x_r$, $s$, diff, lossy )}
\STATE 
\STATE \hspace{0.25cm}{\textsc{decompress}( $\underline{x}$, $x_r$, lossy ) }
\end{algorithmic}
\end{algorithm}

We now explain how to generate patching information $\underline{x}\mid x_r$. Such information is comprised, for each reference to a parselet in the string slots of $x_r$, of {\em (i)} a bit array to store the diff label bits, including those for internal nodes and those that shall become implicit, and {\em (ii)} an array to store relevant differences of letters. This is the complete patching information that is saved for App.~\ref{app:serialize:marginal}. 

Like in a closed-loop, prediction/correction coder, we simulate decoding of the current reference in $x_r$ (prediction done during the top-bottom phase) and then fill the diff bits array with $\underline{x}$ in the bottom-up phase (to construct correction information). The diff bits array is implemented as an array of integers, so {\em (i)} the value $-1$ may be used as a temporary placeholder, and {\em (ii)} we may store within the call stack the location of the bit to be updated during the bottom-up phase by ORing its children diff bits. We define the \textsc{array\_push} primitive that pushes its argument at the end of an array and returns the size of the array minus one. We use it for any kind of data, including differences of letters. 

Alg.~\ref{alg:diff:labels} uses $\underline{x}$ in uncompressed form to implement the construction of the diff bits array for the parselet $p$ describing lossy data at string slot $x_r$. 

\begin{algorithm}[!ht]
\caption{Construction of $\underline{x}\mid x_r$ patching information (diff bits and letters differences), starting with one parselet $p$ of $x_r$ describing data at $\underline{x}$.}\label{alg:diff:labels:rec}
\begin{algorithmic}[1]
\STATE {\textsc{patch}( $x_r$, $p$, $\underline{x}$, diff, delta )}
\STATE \hspace{0.25cm}{\textbf{if} ( $p$.disjunction )}
\STATE \hspace{0.5cm}{$p\leftarrow$ \textsc{next\_bit( $x_r$ )} ? $p$.right : $p$.left }
\STATE \hspace{0.5cm}{\textbf{return} \textsc{patch}( $x_r$, $p$, $\underline{x}$, diff, delta )}
\STATE \hspace{0.25cm}{idx $\leftarrow$ \textsc{array\_push}( diffs, -1 )}
\STATE 
\STATE \hspace{0.25cm}{occ $\leftarrow$ 1 }
\STATE \hspace{0.25cm}{\textbf{if} ( $p$.left.option ) occ$\leftarrow$ ? \textsc{next\_bit}( $x_r$ ) }
\STATE \hspace{0.25cm}{\textbf{if} ( $p$.left.repetition ) occ$\leftarrow$ ? \textsc{next\_int}( $x_r$ ) }
\STATE \hspace{0.25cm}{left\_diff $\leftarrow$ 0 }
\STATE \hspace{0.25cm}{\textbf{while} ( occ-{}- ) }
\STATE \hspace{0.5cm}{\textbf{if} ( $p$.left.atomic ) }
\STATE \hspace{0.75cm}{lossless$\leftarrow$\textsc{next\_slot}( $\underline{x}$ ) }
\STATE \hspace{0.75cm}{\textbf{if} ( $p$.left.id != lossless.id ) }
\STATE \hspace{1.0cm}{\textsc{array\_push}( delta, lossless.id - $p$.left.id )}
\STATE \hspace{0.75cm}{left\_diff \(|\)= $p$.left.id != lossless.id }
\STATE \hspace{0.5cm}{\textbf{else} }
\STATE \hspace{0.75cm}{left\_diff \(|\)= \textsc{patch}( $x_r$, $p$.left, $\underline{x}$, diff, delta )}
\STATE \hspace{0.25cm}{\textbf{if} ( $\lnot$ $p$.regular ) \textbf{return} diff[ idx ] = left\_diff }
\STATE 
\STATE \hspace{0.25cm}{occ $\leftarrow$ 1 }
\STATE \hspace{0.25cm}{\textbf{if} ( $p$.right.option ) occ$\leftarrow$ ? \textsc{next\_bit}( $x_r$ ) }
\STATE \hspace{0.25cm}{\textbf{if} ( $p$.right.repetition ) occ$\leftarrow$ ? \textsc{next\_int}( $x_r$ ) }
\STATE \hspace{0.25cm}{right\_diff $\leftarrow$ 0 }
\STATE \hspace{0.25cm}{\textbf{while} ( occ-{}- ) }
\STATE \hspace{0.5cm}{\textbf{if} ( $p$.right.atomic ) }
\STATE \hspace{0.75cm}{lossless$\leftarrow$\textsc{next\_slot}( $\underline{x}$ ) }
\STATE \hspace{0.75cm}{\textbf{if} ( $p$.right.id != lossless.id ) }
\STATE \hspace{1.0cm}{\textsc{array\_push}( delta, lossless.id - $p$.right.id )}
\STATE \hspace{0.75cm}{right\_diff \(|\)= $p$.right.id != lossless.id }
\STATE \hspace{0.5cm}{\textbf{else} }
\STATE \hspace{0.75cm}{right\_diff \(|\)= \textsc{patch}( $x_r$, $p$.right, $\underline{x}$, diff, delta )}
\STATE \hspace{0.25cm}{\textbf{return} diff[ idx ] = left\_diff \(|\) right\_diff }
\end{algorithmic}
\end{algorithm}

\begin{algorithm}[!ht]
\caption{Construction of $\underline{x}\mid x_r$ patching information (diff bits and letters differences), main driver.}\label{alg:diff:labels}
\begin{algorithmic}[1]
\STATE {\textsc{diff}( $\underline{x}$, $x_r$, diffs, deltas )}
\STATE \hspace{0.25cm}{$s$ $\leftarrow$ \textsc{next\_ref}( $x_r$ ) }
\STATE \hspace{0.25cm}{\textbf{if} ( $s$.EndOfString ) \textbf{return} diffs, deltas}
\STATE 
\STATE \hspace{0.25cm}{diff $\leftarrow$ $\emptyset$ }
\STATE \hspace{0.25cm}{delta $\leftarrow$ $\emptyset$ }
\STATE \hspace{0.25cm}{occ $\leftarrow$ 1 }
\STATE \hspace{0.25cm}{\textbf{if} ( $s$.option ) occ$\leftarrow$ \textsc{next\_bit}( $x_r$ ) }
\STATE \hspace{0.25cm}{\textbf{if} ( $s$.repetition ) occ$\leftarrow$ \textsc{next\_int}( $x_r$ ) }
\STATE \hspace{0.25cm}{\textbf{while} ( occ-{}- ) }
\STATE \hspace{0.5cm}{\textbf{if} ( $s$.atomic ) }
\STATE \hspace{0.75cm}{lossless$\leftarrow$\textsc{next\_slot}( $\underline{x}$ ) }
\STATE \hspace{0.75cm}{\textbf{if} ( $s$.id != lossless.id ) }
\STATE \hspace{1.0cm}{\textsc{array\_push}( diff, 1 )}
\STATE \hspace{1.0cm}{\textsc{array\_push}( delta, lossless.id - $s$.id )}
\STATE \hspace{0.75cm}{\textbf{else}}
\STATE \hspace{1.0cm}{\textsc{array\_push}( diff, 0 )}
\STATE \hspace{0.5cm}{\textbf{else}}
\STATE \hspace{0.75cm}{\textsc{patch}( $x_r$, $\underline{x}$, $x$, diff, delta ) }
\STATE
\STATE \hspace{0.25cm}{\textsc{array\_push}( diffs , diff ) }
\STATE \hspace{0.25cm}{\textsc{array\_push}( deltas, delta ) }
\STATE \hspace{0.25cm}{\textsc{diff}( $\underline{x}$, $x_r$, diffs, deltas )}
\end{algorithmic}
\end{algorithm}

The last step is to write the data, omitting implicit diff information. The diff bits and letters differences arrays may now be accessed sequentially during parselet traversal, as Alg.~\ref{alg:write:patched} shows. The \textsc{write\_*} primitives return the integer value they wrote and update the number of written bits during the course of Alg.~\ref{alg:write:patched}. The (possibly negative) minimal letters difference is stored in the string header, and letters differences are entropy-coded unsigned offsets from this value.

\begin{algorithm}[!ht]
\caption{Writing compressed $x_r\mid\mathcal{D}_r$ with $\underline{x}\mid x_r$ patch, starting at parselet $p$ in $x_r$.}\label{alg:write:patched:rec}
\begin{algorithmic}[1]
\STATE {\textsc{enc\_patched}( $x_r$, $p$, diff, $b$, delta, $d$, has\_diff )}
\STATE \hspace{0.25cm}{\textbf{if} ( $p$.disjunction )}
\STATE \hspace{0.5cm}{alt$\leftarrow$ \textsc{next\_bit}( $x_r$ )}
\STATE \hspace{0.5cm}{\textsc{write\_bit}( alt )}
\STATE \hspace{0.5cm}{$p\leftarrow$ alt ? $p$.right : $p$.left }
\STATE \hspace{0.5cm}{\textsc{enc\_patched}( $x_r$, $p$, diff, $b$, delta, $d$, has\_diff )}
\STATE \hspace{0.5cm}{\textbf{return} }
\STATE 
\STATE \hspace{0.25cm}{occ $\leftarrow$ 1 }
\STATE \hspace{0.25cm}{\textbf{if} ( $p$.left.option ) }
\STATE \hspace{0.5cm}{occ$\leftarrow$ \textsc{next\_bit}( $x_r$ ) }
\STATE \hspace{0.5cm}{\textsc{write\_bit}( occ )}
\STATE \hspace{0.25cm}{\textbf{if} ( $p$.left.repetition ) }
\STATE \hspace{0.5cm}{occ$\leftarrow$ \textsc{next\_int}( $x_r$ ) }
\STATE \hspace{0.5cm}{\textsc{write\_int}( occ )}
\STATE \hspace{0.25cm}{\textbf{while} ( occ-{}- ) }
\STATE \hspace{0.5cm}{next\_diff $\leftarrow$ diff[ $b$++ ] }
\STATE \hspace{0.5cm}{\textbf{if} ( has\_diff ) \textsc{write\_bit}( next\_diff ) }
\STATE \hspace{0.5cm}{\textbf{if} ( $p$.left.atomic ) }
\STATE \hspace{0.75cm}{\textbf{if} ( next\_diff ) \textsc{write\_int}( delta[ $d$++ ] ) }
\STATE \hspace{0.5cm}{\textbf{else} }
\STATE \hspace{0.75cm}{\textsc{enc\_patched}( $x_r$, $s$, diff, $b$, delta, $d$, next\_diff )}
\STATE \hspace{0.25cm}{\textbf{if} ( $\lnot$ $p$.regular ) \textbf{return} }
\STATE 
\STATE \hspace{0.25cm}{occ $\leftarrow$ 1 }
\STATE \hspace{0.25cm}{\textbf{if} ( $p$.right.option ) }
\STATE \hspace{0.5cm}{occ$\leftarrow$ \textsc{next\_bit}( $x_r$ ) }
\STATE \hspace{0.5cm}{\textsc{write\_bit}( occ )}
\STATE \hspace{0.25cm}{\textbf{if} ( $p$.right.repetition ) }
\STATE \hspace{0.5cm}{occ$\leftarrow$ \textsc{next\_int}( $x_r$ ) }
\STATE \hspace{0.5cm}{\textsc{write\_int}( occ )}
\STATE \hspace{0.25cm}{\textbf{while} ( occ-{}- ) }
\STATE \hspace{0.5cm}{next\_diff $\leftarrow$ diff[ $b$++ ] }
\STATE \hspace{0.5cm}{\textbf{if} ( has\_diff ) \textsc{write\_bit}( next\_diff ) }
\STATE \hspace{0.5cm}{\textbf{if} ( $p$.right.atomic ) }
\STATE \hspace{0.75cm}{\textbf{if} ( next\_diff ) \textsc{write\_int}( delta[ $d$++ ] ) }
\STATE \hspace{0.5cm}{\textbf{else} }
\STATE \hspace{0.75cm}{\textsc{enc\_patched}( $x_r$, $s$, diff, $b$, delta, $d$, next\_diff )}
\STATE \hspace{0.25cm}{\textbf{return} }
\end{algorithmic}
\end{algorithm}

\begin{algorithm}[!ht]
\caption{Writing compressed $x_r\mid\mathcal{D}_r$ with $\underline{x}\mid x_r$ patch, main driver.}\label{alg:write:patched}
\begin{algorithmic}[1]
\STATE {\textsc{write\_patched}( $x_r$, diffs, deltas, $n$ ) }
\STATE \hspace{0.25cm}{$s$ $\leftarrow$ \textsc{next\_ref}( $x_r$ ) }
\STATE \hspace{0.25cm}{\textsc{write\_ref}( $s$ )}
\STATE \hspace{0.25cm}{\textbf{if} ( $s$.EndOfString )}
\STATE \hspace{0.5cm}{\textbf{return} }
\STATE 
\STATE \hspace{0.25cm}{diff $\leftarrow$ diffs[ $n$ ] }
\STATE \hspace{0.25cm}{$b$ $\leftarrow$ 0 }
\STATE \hspace{0.25cm}{delta $\leftarrow$ deltas[ $n$ ] }
\STATE \hspace{0.25cm}{$d$ $\leftarrow$ 0 }
\STATE \hspace{0.25cm}{occ $\leftarrow$ 1 }
\STATE \hspace{0.25cm}{\textbf{if} ( $s$.option ) }
\STATE \hspace{0.5cm}{occ$\leftarrow$ \textsc{next\_bit}( $x_r$ ) }
\STATE \hspace{0.5cm}{\textsc{write\_bit}( occ )}
\STATE \hspace{0.25cm}{\textbf{if} ( $s$.repetition ) }
\STATE \hspace{0.5cm}{occ$\leftarrow$ \textsc{next\_int}( $x_r$ ) }
\STATE \hspace{0.5cm}{\textsc{write\_int}( occ )}
\STATE \hspace{0.25cm}{\textbf{while} ( occ-{}- ) }
\STATE \hspace{0.5cm}{next\_diff $\leftarrow$ diff[ $b$++ ] }
\STATE \hspace{0.5cm}{\textsc{write\_bit}( next\_diff ) }
\STATE \hspace{0.5cm}{\textbf{if} ( $s$.atomic ) }
\STATE \hspace{0.75cm}{\textbf{if} ( next\_diff ) \textsc{write\_int}( delta[ $d$++ ] ) }
\STATE \hspace{0.5cm}{\textbf{else} }
\STATE \hspace{0.75cm}{\textsc{enc\_patched}( $x_r$, $s$, diff, $b$, delta, $d$, next\_diff )}
\STATE \hspace{0.25cm}{\textsc{write\_patched}( $x_r$, diffs, deltas, ++$n$ )}
\end{algorithmic}
\end{algorithm}

\subsection{Coding Marginal Common Strings (MCS)} \label{app:serialize:marginal}

By Eq.~\ref{eq:mset:union}, this section only stands for the sake of completeness. We show how to achieve bit-exact symmetry of information for actual storage of string multisets. 

Since the $(\hat{\mathcal{D}}_{x_i},\hat{x}_i)$ have been computed separately, common substrings may still appear among strings in $\hat{X}$. They are called {\em marginal} common strings (MCS). However, because our underlying representation is patch-based, the same chunks of compressed data may be used to encode different chunks of uncompressed data. While the converse need not be true, a match in compressed form shall be a match in uncompressed form. Hence, we must use the patching information saved from Sec.~\ref{app:serialize:patching} to ensure exact matches between MCS in {\em simultaneous} un/compressed forms. Not all repetitions in $\hat{X}$ will be factorized, only those that map to the same chunks of uncompressed data. This way, {\em (i)} MCS are encoded just like an input string, {\em (ii)} there is no need to extend the representation to also abstract patching operations, and {\em (iii)} the search for MCS occurs in the patched, compressed representation. 

We now use the same index data structure for all the $\hat{X}=[\hat{x}_i]$. This allows to extract MCS from longest to shortest, creating each time an entry in $\hat{\mathcal{D}}_X$ and updating string slots, much like we did in Alg.~\ref{alg:deflate} except this time regexp parameter string slots are also emptied and the string slot containing the reference to the new MCS is written with a zero diff bit for patching (storing an MCS in the dictionary ``steals'' the local patching information, that is guaranteed to be the same among all instances of the same MCS). MCS creation is controlled by ensuring it does not expand the bitstream size

{\em Known limitation of} \texttt{mpx}: The search for MCS in mixed compressed/patched form is surprisingly tedious to implement, and currently only leads to code far below basic programming literacy. For the time being, we only search for MCS that are the sizes of the input strings. This still formally captures sets of patched strings, but more work is needed to make it fully operational and less cosmetic, although in essence it {\em is} cosmetic, as the reader now may see. This is the only discrepancy of \texttt{mpx} with what we have described. Of course, this has no influence on the results we report and it correctly illustrates the limiting case where collections of objects are indeed multisets.

MCS are written just like input strings along the lines of Alg.~\ref{alg:write:patched}. Only after they are added to the dictionary, we set the EndOfString dummy identifier to $|\hat{\mathcal{D}}_X|$. The decoder is trivially extended to support references to MCS. This eventually captures the case $x=y$ so support for collections of objects as string multisets is native. 

We can now enforce symmetry of information for lossless reconstruction of string multisets, for free. That is: $C(\mathcal{D}_{\hat{x},\hat{y}},\{\hat{x},\hat{y}\}, \{x, y\})=C(\mathcal{D}_{\hat{y},\hat{x}},\{\hat{y},\hat{x}\}, \{y, x\})$. We have $\mathcal{D}_{\hat{x},\hat{y}}=\mathcal{D}_{\hat{y},\hat{x}}$ from Eq.~\ref{eq:mset:union} after Alg.~\ref{alg:kompress}, and because dictionaries are written in canonical order, $C(\mathcal{D}_{\hat{x},\hat{y}})=C(\mathcal{D}_{\hat{y},\hat{x}})$. Now, just by resetting the entropy-coding stage downstream upon writing the next compressed string without padding bits, we ensure symmetry of information no matter the order of $x$ and $y$ inside the bitstream, {\em down to the bit level on disk}.

\subsection{Illustration: joint compression, multiset support}

Focusing on the core approximation of joint Kolmogorov complexity, we compare string (multi)set support for \texttt{mpx}, \texttt{gzip} and \texttt{xz}. For off-the-shelf compressors, we must construct the concatenation of the input files beforehand, in the order in which files are specified. For consistent comparison, we use the lossless models.

Tab.~\ref{tab:set:support} illustrates one example of spurious discrepancy reported in~\cite{cebrian:2005}. The LZ77 sliding window of \texttt{xz} is much larger (1MiB) than that of \texttt{gzip} (32KiB), so using \texttt{gzip} will incur more and more numerical discrepancy as the concatenated data overflows its internal buffer (which can accomodate three texts but only two DNA samples). On the other hand, \texttt{mpx} implements bit-exact symmetry of information. 

\begin{table}[!ht]
  \caption{(Multi)set support: \texttt{gzip} {\em vs.} \texttt{xz} {\em vs.} \texttt{mpx} (lossless models). File sizes in bytes as reported by the filesystem. \texttt{mpx} ensures symmetry of information on (multi)sets of strings down to the bit level by design.\label{tab:set:support}}
  \centering
  \begin{tabular}{|l||c|c||c|}\hline
    Joint Kolmorogov Complexity         & \texttt{gzip} & \texttt{xz} & \texttt{mpx} \\\hline\hline
    \texttt{english,english,french}     &          8231 &        7808 &        10168 \\\hline
    \texttt{english,french,english}     &          8260 &        7812 &        10168 \\\hline\hline
    \texttt{cat,cat,mouse}              &         10754 &        8492 &         9404 \\\hline
    \texttt{cat,mouse,cat}              &         14487 &        8424 &         9404 \\\hline
  \end{tabular}
\end{table}

Fig.~\ref{fig:dump:archive} outlines the internal structure of a Solomonoff archive that is produced in the process (\texttt{english-french-english.mpz}, thereby demonstrating multiset support). The dump of \texttt{english-english-french.mpz} is not shown, as basically only the information for the last two strings are swapped. Most values may only be printed after decompressing the whole archive. We did not even care to implement a table of file names, as it would only add even more precious bloat in the implementation. The reasons are that Eq.~\ref{eq:mset:union} and caching only need writing single string archives, and the file name may be recovered from the name of the cached file in practice.

\begin{figure}[!ht]
  \tiny
  \begin{verbatim}
<Model>

   K*  = 499 parselets (flushed: +17)
   K_D = 9844 bits (flushed: +236 bits)

 <#0::String>

   Decompr. = 10843 bytes
   K_2      = 9883 bits + 33688 MCS bits -> 4.02 bpl
     LDepth = 17456 steps
   Patch    = 0 bits

 <#1::String>

   Decompr. = 12391 bytes
   K_2      = 47271 bits -> 3.81 bpl
     LDepth = 20306 steps
   Patch    = 0 bits

 <#2::String>

   Decompr. = 10843 bytes
   K_2      = 9883 bits + 33688 MCS bits -> 4.02 bpl
     LDepth = 17456 steps
   Patch    = 0 bits

 <Total>

   Decompr. = 34077 bytes
   K_2      = 81511 bits -> 2.39 bpl
     LDepth = 55218 steps
   Patch    = 0 bits
\end{verbatim}
\caption{Internal structure dump of Solomonoff archive \texttt{english-french-english.mpz} (lossless models).}
  \label{fig:dump:archive}
\end{figure}

\section{Implementation notes} \label{app:data:structures}

The following is a possible implementation for the data structures of interest in this work: parselets, dicionary, string slots, string data, and the index.

This implementation imposes $|\mathcal{D}_{\hat{X}}|<2^{29}-|\alphabet|-1$ and a maximum finite number of repetitions of $2^{29}-2$, both of which are more than enough for any conceivably useful scenario. Ultimately, the index causes more pressure on the memory bus than off-the-shelf compressors. This is because ours has to be offline. 

As for elementary data structures (dynamic arrays and hashtables with amortized allocations), they are macro-templated so as to avoid constant dereferencement of generic pointers. Parallel (\textbf{spawn}) computations use a thread pool internally. 

\subsection{Parselets}

Fig.~\ref{fig:parselet:compact} depicts a compact representation of a parselet that is used to describe each element of a dictionary $\mathcal{D}$ of parselets in memory. 

\begin{figure}[!ht]
  \centering
  \begin{bytefield}[]{32}
    \bitheader[endianness=big]{31,29,0}\\
    \begin{rightwordgroup}[rightcurly=.]{\texttt{left}}
      \bitbox{1}{\rotatebox{90}{\tiny reg}} & \bitbox{2}{\texttt{op}} & \bitbox{29}{\texttt{id0}} 
    \end{rightwordgroup}\\
    \begin{rightwordgroup}[rightcurly=.]{\texttt{right}}
      \bitbox{1}{\rotatebox{90}{\tiny alt}} & \bitbox{2}{\texttt{op}} & \bitbox{29}{\texttt{id1}} 
    \end{rightwordgroup}
  \end{bytefield}
  \caption{Compact representation of a parselet on 64 bits.\label{fig:parselet:compact}}
\end{figure}

Following Fig.~\ref{fig:parselet:compact}, if the bit \texttt{reg} is set, the parselet has a right part and is regular. Otherwise, it is irregular. A regular parselet is a disjunction if the bit \texttt{alt} is set, a conjunction otherwise. Regexp operators are encoded like \texttt{00} for one occurrence of the parselet, \texttt{01} for a positive number of occurrences (\texttt{*}), and \texttt{10} for an optional occurrence (\texttt{?}). This implements all core predicates. The value \texttt{11} is unused so as to accomodate the definition of a string slot {\em infra}.

\subsection{Dictionary}\label{sec:impl:dictionary}

The dictionary is basically a dynamic array of parselets (and MCS) with counters and a hashtable~\cite{celis:1985} for fast set-theoretic operations (set-membership, unions, {\em etc.})

Counters of parselet types ({\em resp.} activations) are updated upon parselet insertion ({\em resp.} reading/writing, see Algs.~\ref{alg:expand}-\ref{alg:deflate}-\ref{alg:cond:deflate}).

Fast implementation of the $\sim$ predicate is done by associating to each parselet a string that encodes its syntactic structure. This string is computed with Alg.~\ref{alg:canonical} (parentheses and regexp operators are written down, letters are replaced by a dot placeholder).

Alg.~\ref{alg:canonical} is also used to compute for each parselet a string that encodes its syntactic expansion (letters are written down this time). The hashtable that enables fast implementation of Alg.~\ref{alg:merge} uses these strings for parselet indexing. 

\subsection{String slot}

In memory, a string is described as a sequence of {\em string slots} that borrow the same layout used in parselets: this makes a string slot containing a reference to $\mathcal{D}$ functionally indistinguishable from an irregular parselet. Hence, this models a string as a conjunction of irregular parselet sets. 

\begin{definition}[String slot]
  A string slot in memory is described with 32 bits like:\\

  \begin{center}

    \begin{bytefield}[]{32}
      \bitheader[endianness=big]{31,29,0}\\
      \bitbox{1}{0} & \bitbox{2}{\texttt{op}} & \bitbox{29}{\texttt{id}/\texttt{val}/\texttt{bit}} \\
    \end{bytefield}

  \end{center}

\end{definition}

String slots implement dynamic typing using the following boxing scheme:
\begin{itemize}
\item If \texttt{op} is not \texttt{11}, then it encodes the operator associated to the identifier \texttt{id} in the other bits (which must be less than $2^{29}-1$);
\item If \texttt{op}=\texttt{11}, two cases occur:
  \begin{itemize}
  \item If \texttt{val} has all bits set ($2^{29}-1$), the string slot is {\em empty} (so its content may be {\em erased});
  \item Otherwise, \texttt{val} does not encode a reference to $\mathcal{D}$, but a regexp operator parameter value. Special values $2^{29}-2$ and $2^{29}-3$ are reserved to encode bits (for selecting an alternative or a \texttt{?} regexp operator option). Values less than $2^{29}-4$ encode an unsigned number of repetitions associated to a \texttt{*} regexp operator.
  \end{itemize}
\end{itemize}

\subsection{String data}

\begin{figure}[!ht]
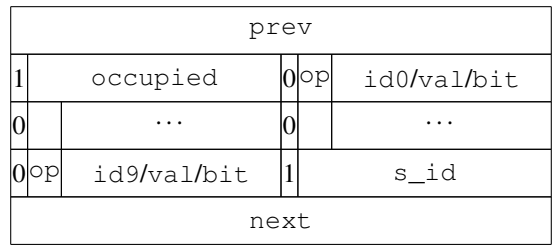

  \centering
  \begin{bytefield}[]{32}
    \bitbox{32}{\texttt{prev}}
  \end{bytefield}\\
  \begin{bytefield}[]{16}
    \bitbox{1}{1} & \bitbox{15}{\texttt{occupied}}
    \bitbox{1}{0} & \bitbox{2}{\texttt{op}} & \bitbox{13}{\texttt{id0}/\texttt{val}/\texttt{bit}}
  \end{bytefield}\\
  \begin{bytefield}[]{16}
    \bitbox{1}{0} & \bitbox{2}{} & \bitbox{13}{\dots} 
    \bitbox{1}{0} & \bitbox{2}{} & \bitbox{13}{\dots} 
  \end{bytefield}\\
  \begin{bytefield}[]{16}
    \bitbox{1}{0} & \bitbox{2}{\texttt{op}} & \bitbox{13}{\texttt{id9}/\texttt{val}/\texttt{bit}} 
    \bitbox{1}{1} & \bitbox{15}{\texttt{s\_id}} 
  \end{bytefield}\\
  \begin{bytefield}[]{32}
    \bitbox{32}{\texttt{next}} 
  \end{bytefield}
  \caption{Memory layout of a string slot array fitting a 64-byte L2 cache line, implementing a bidirectional sequence of string slots to represent string \texttt{s\_id} in either un/compressed form. This helps maintain improved memory locality, as chunks of 10 consecutive string slots are pushed at once from main memory to the cache, excluding the two pointers and additional fields. These arrays are allocated using \texttt{posix\_memalign(3)} to enforce cache line size alignment. 
    \label{fig:sequence}}
\end{figure}

In memory, the {\em sequence} of string slots must support efficient deletion and insertion of string slots from arbitrary locations in the string. Our sequence is implemented as a doubly-linked list of arrays of string slots that fit an L2 cache line. Such a string slot array is depicted on Fig.~\ref{fig:sequence}. 

The crux of the design of a string slot array is, given any pointer to a string slot it contains (our compressor is offline), to be able to evade the array so as to find the previous or the next in the sequence. This is done by surrounding actual string slots with 32-bit additional fields with their msb set, so we can eventually access the previous/next pointers located outside these fields. The list of string slots array is not circular because we need to know when we reach the start or end of the string. 

To support fast compaction of empty slots that shall accumulate during compression (and help maintain good locality of memory), we store the bitmap of occupied string slots in the first additional field, and empty string slot arrays are eventually freed. The second additional field stores the identifier of the string in its multiset $X$, which must be less than $2^{31}-1$. 

Hence, given the address of any string slot in memory, we are able to navigate the string forward/backward, from any arbitrary location to start or end, to skip (or not) regexp parameter values (because of dynamic typing), and generally to skip/insert/remove/replace a string slot in amortized constant time. When we denote a string with $x$, the reader really should think of $x$ as a set of pointers in string data that are references to $\mathcal{D}$ (and their possible regexp parameter values $v$ are immediately accessible from there). 

\subsection{Index}

The compressor will need an index to quickly locate pairs of common consecutive string slots \texttt{foo}op? \texttt{bar}op?. This is implemented as a two-level hashtable: the first level indexes the left string slot identifier in a pair (using the regexp operator as secondary key), and the second level indexes sets of pair locations (as pointers to the right string slot of the pair occurrences) ordered by the same right string slot identifier (and regexp operator) they share.

These sets are indexed by their size in an array of pointers, so the index also dynamically maintains the number of common pairs of consecutive string slots: this allows to find the most frequent pair of common string slots, and their locations, in $O(|\mathcal{D}|^2)$ worst-case time. The index is updated upon insertion/deletion of string slots. This likely makes our compressor the most memory-hungry to date. The index provides the primitives \textsc{most\_*} that return both the next best parselet and the set of locations where it was found.

Fig.~\ref{fig:mem:init} depicts the state of the above data structures after loading $x=\mbox{\texttt{aabcaaabaac}}$ ($\mathcal{A}=\left[\mbox{\texttt{a}}, \mbox{\texttt{b}}, \mbox{\texttt{c}}\right]$, so the first parselet has identifier \texttt{0x3}) and just before parselet creation. Fig.~\ref{fig:factorize} depicts the same data structures as in Fig.~\ref{fig:mem:init} after creation of the parselet \texttt{0x3} $\rightarrow$ \texttt{a* b}.

\begin{figure}[!ht]
  \begin{center}

    \begin{tikzpicture}
      \node[anchor=north] (dict_lbl) at (-2,0) {$\mathcal{D}$};
      \node[text width=2em] (dict) at (-.8,-.25) {$\emptyset$};
      \draw[ref] (dict_lbl.east) -- (dict);

      \node[anchor=north] (idx_lbl) at (-4,-1) {Index};
      \node[bnode,rectangle split parts=9] (idx) at (0,-1) {\texttt{a} \nodepart{two} \texttt{a?} \nodepart{three} \texttt{a*} \nodepart{four} \texttt{b} \nodepart{five} \texttt{b?} \nodepart{six} \texttt{b*} \nodepart{seven} \texttt{c} \nodepart{eight} \texttt{c?} \nodepart{nine} \texttt{c*}};
      \draw[ref] (idx_lbl) -- (idx);

      \node[bnode,rectangle split parts=6] (ha) at (-3,-2) {2 \nodepart{four} 1 \nodepart{six} 0 };
      \draw[ref] (idx.three south) -- (ha.head north);

      \node[bnode,rectangle split parts=5] (hb) at (.5,-2) {1 \nodepart{three} 1 \nodepart{five} 0};
      \draw[ref] (idx.four south) -- (hb.head north);
      \node[bnode,rectangle split parts=3] (hc) at (3,-2) { 1 \nodepart{three} 0 }; 
      \draw[ref] (idx.seven south) -- (hc.head north);

      \node[anchor=north] (string_lbl) at (-4,-3.845) {\shortstack[l]{String \\ data}};
      \node[bnode,rectangle split parts=10] (string) at (.5,-4) {\texttt{a*} \nodepart{two} 2 \nodepart{three} \texttt{b} \nodepart{four} \texttt{c} \nodepart{five} \texttt{a*} \nodepart{six} 3 \nodepart{seven} \texttt{b} \nodepart{eight} \texttt{a*} \nodepart{nine} 2 \nodepart{ten} \texttt{c} };
      \draw[ref] (string_lbl.east) -- (string.one west);

      \draw[ref] (ha.two south) -- (string.three north);
      \draw[ref] (ha.three south) -- (string.seven north);

      \draw[ref] (ha.five south) -- (string.ten north);

      \draw[ref] (hb.two south)    -- (string.eight north);
      \draw[ref] (hb.four south)   -- (string.four north);

      \draw[ref] (hc.two south) -- (string.five north);

    \end{tikzpicture}

  \end{center}
  \caption{Data structures after loading $x=\mbox{\texttt{aabcaaabaac}}$ ($\mathcal{A}=\left[\mbox{\texttt{a}}, \mbox{\texttt{b}}, \mbox{\texttt{c}}\right]$) and the first call to \textsc{rlc}. The second level of the hashtable maintains the number of pairs sharing the same string slot values, which allows fast scanning of these arrays.\label{fig:mem:init}}
\end{figure}
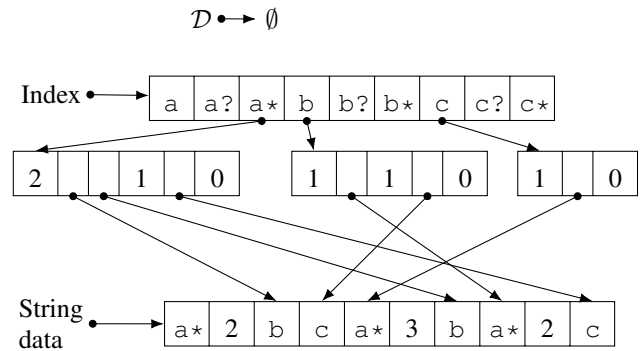

\begin{figure}[!ht]
  \begin{center}

    \begin{tikzpicture}
      \node[anchor=north] (dict_lbl) at (-4,0) {$\mathcal{D}$};
      \node[text width=10em] (dict) at (-1.3,-.25) {{\tt 0x3} $\rightarrow$ \texttt{a* b} };
      \draw[ref] (dict_lbl.east) -- (dict);

      \node[anchor=north] (idx_lbl) at (-5.38,-1) {Index};
      \node[bnode,text width=1.5em,rectangle split parts=11] (idx) at (-1.5,-2) {\dots \nodepart{two} \texttt{a*} \nodepart{three} \texttt{b} \nodepart{four} \texttt{b?} \nodepart{five} \texttt{b*} \nodepart{six} \texttt{c} \nodepart{seven} \texttt{c?} \nodepart{eight} \texttt{c*} \nodepart{nine} \texttt{\scriptsize 0x3} \nodepart{ten} \texttt{\scriptsize 0x3?} \nodepart{eleven} \dots};
      \draw[ref] (idx_lbl.south) -- (idx.one north);

      \node[bnode,rectangle split parts=3] (ha) at (-4,-3) {1 \nodepart{three} 0 };
      \draw[ref] (idx.two south) -- (ha.head north);

      \node[bnode,rectangle split parts=3] (hc) at (-2,-3) { 1 \nodepart{three} 0 }; 
      \draw[ref] (idx.six south) -- (hc.head north);

      \node[bnode,rectangle split parts=5] (hn) at (1,-3) {1 \nodepart{three} 1 \nodepart{five} 0 };
      \draw[ref] (idx.nine south) -- (hn.head north);

      \node[anchor=north] (string_lbl) at (-5,-4.845) {\shortstack[l]{String \\ data}};
      \node[bnode,rectangle split parts=10] (string) at (-1,-5) {\texttt{\scriptsize 0x3} \nodepart{two} 2 \nodepart{four} \texttt{c} \nodepart{five} \texttt{\scriptsize 0x3} \nodepart{six} 3 \nodepart{eight} \texttt{a*} \nodepart{nine} 2 \nodepart{ten} \texttt{c} };
      \draw[ref] (string_lbl.east) -- (string.one west);

      \draw[ref] (ha.two south) -- (string.ten north);

      \draw[ref] (hc.two south) -- (string.five north);

      \draw[ref] (hn.two south) -- (string.eight north);
      \draw[ref] (hn.four south) -- (string.four north);

    \end{tikzpicture}

  \end{center}
  \caption{Data structures of Fig.~\ref{fig:mem:init} after creation of the parselet \texttt{a* b}, which has identifier \texttt{0x3} in $\mathcal{D}$. The left string slot of the parselet occurrences are replaced with the identifier of the new parselet (local compilation), and the right string slots are emptied. Observe that doing so, we preserve the order in which operator parameter values should appear in the data stream. Compaction of empty string slots occur periodically.\label{fig:factorize}}
\end{figure}
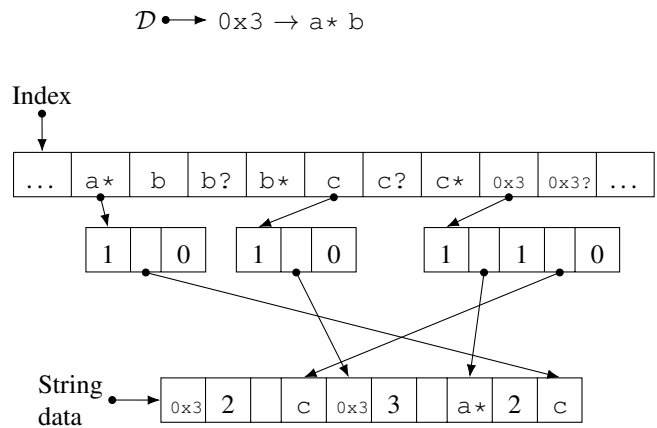

\end{document}